\documentclass{emulateapj}
\usepackage{graphicx}
\usepackage{amsmath}
\usepackage{epstopdf}
\usepackage{verbatim}
\usepackage{boxhandler}

\newcommand{\be}{\begin{equation}}
\newcommand{\ee}{\end{equation}}
\newcommand{\ba}{\begin{eqnarray}}
\newcommand{\ea}{\end{eqnarray}}

\newcommand{\nn}{\mbox{} \nonumber \\ \mbox{} }

\newcommand{\K}{{\rm K}}

\begin{document} 

\title{Rotation and Magnetism of Massive Stellar Cores}
\author{Yevgeni Kissin}
\affil{Department of Astronomy and Astrophysics, University of Toronto, 50 St. George St., Toronto, ON M5S 3H4, Canada}
\author{Christopher Thompson}
\affil{Canadian Institute for Theoretical Astrophysics, 60 St. George St., Toronto, ON M5H 3H8, Canada}

\begin{abstract}
The internal rotation and magnetism of massive stars are considered in response to i) the inward pumping
of angular momentum through deep and slowly rotating convective layers; and ii) the 
winding up of a helical magnetic field in radiative layers.  Field winding can transport
angular momentum effectively even when the toroidal field is limited by kinking.  Magnetic helicity is pumped into
a growing radiative layer from an adjacent convective envelope (or core).  The receding convective
envelope that forms during the early accretion phase of a massive star is the dominant source of helicity
in its core, yielding a $\sim 10^{13}$ G polar magnetic field in a collapsed neutron star (NS) remnant.
Using MESA models of various masses, we find that the NS rotation varies significantly,
from $P_{\rm NS} \sim 0.1-1$ s in a 13$\,M_\odot$ model to $P_{\rm NS} \sim 2$ ms in a $25\,M_\odot$ model
with an extended core.  Stronger inward pumping of angular momentum is found in more massive
stars, due to the growing thickness of the convective shells that form during the later stages of 
thermonuclear burning.  On the other hand, stars that lose enough mass to form blue supergiants in 
isolation end up as very slow rotators.  The tidal spin-up of a 40$\,M_\odot$ star by a massive binary companion
is found to dramatically increase the spin of the remnant black hole, allowing a rotationally supported torus
to form during the collapse.  The implications for post-collapse decay or amplification of the magnetic field
are also considered.
\end{abstract}

\section{Introduction}\label{s:intro}

This paper addresses some gaps in the current understanding of the rotation and magnetism of massive stars as
they evolve toward core collapse.   Our focus is on the transport of angular momentum by convective and magnetic stresses,
and on the genesis of stable magnetic fields in radiative layers of a star.  We note in particular the following:

\vskip .05in \noindent
1. Existing stellar evolution codes do not allow for the advective transport of angular momentum by extended convective plumes.
This may not result in significant inaccuracies for solar-type stars with shallow convective envelopes,
or for rapidly rotating stars that support vigorous magnetic dynamos (see, e.g., \citealt{auguston16}).  
But we suggest that the situation is different for supergiants with deep and slowly rotating convective layers:
in this case, the inward advection of a small fraction of the stellar angular momentum toward the core will have a
dramatic effect on the inner rotation.  The presence of deep convective plumes is supported by fits of mixing-length
models to giant stars \citep{SackB1991}, and by anelastic calculations of non-magnetized convection \citep{BrunP2009}.
The consequences for the rotation of stars of solar and intermediate mass were previously considered by \cite{KissT2015a,KissT2015b},
where the largest effect was seen near the tips of the red giant branch (RGB) and asymptotic giant branch (AGB).

Consider, for example, stars more massive than about $20\,M_\odot$ which also retain a hydrogen-rich envelope 
at the moment of core collapse.   The envelope and burning shells of these stars form a broad sequence of convective 
layers.  A key result of this paper is that strong angular momentum pumping within deep convective layers can significantly
compensate the rotational braking of a contracting stellar core by internal stresses, thereby allowing for more uniform rotation
across radiative-convective boundaries than is seen in preceding rotational models.  
A similar, but overall weaker, effect will be encountered in massive helium stars.

\vskip .05in \noindent
2. The radiative layers of an evolving star tend toward a state of differential rotation, which is
resisted by the winding up of an embedded magnetic field.  One popular approach to magnetic field
growth in radiative layers starts with a weak seed polar magnetic field, and then relies on a kink
instability of the wound-up toroidal field \citep{Tayler73} to feed back on the poloidal field \citep{Spru2002}. 
\cite{HegeWS2005} incorporated this formalism into evolutionary models of massive stars.
On the other hand, the magnetic fields of radio pulsars may indicate the presence of a stronger embedded magnetic flux
which could significantly redistribute angular momentum within the contracting stellar core (e.g. \citealt{SpruP1998,MM14}).
We re-examine this process, taking into account the kinking of the wound-up poloidal flux, and argue that kinking
does not much reduce its effectiveness. 
The competing effect of the magnetorotational instability (MRI) is much more localized
near radiative-convective boundaries, and involves small-scale fluid motions that induce strong compositional mixing 
and significant structural changes in the star \citep{wheeler15}.  These effects are greatly suppressed in our model
by the smoothing action of large-scale Maxwell stresses.

\vskip .05in \noindent
3. Stability of the polar magnetic field threading a radiative layer of the star depends on the presence of
a buried toroidal flux that carries a net twist \citep{BraiS2004}.  The accumulation of magnetic helicity within
a radiative layer cannot simply result from the freezing of a large-scale fossil magnetic field, as is sometimes
invoked to explain the magnetic moments of white dwarfs and neutron stars (NSs).  Instead it depends on the transport
of magnetic helicity across the evolving convective-radiative boundary that, in most cases, preceded the formation
of the radiative layer \citep{KissT2015b}.  There is a simple connection between this helicity flux and
the large-scale Maxwell stress that is needed to compensate a persistent, inhomogeneous Reynolds stress
that is imposed at a convective boundary.  The required Maxwell stress is a natural outcome of even a mild dynamo
instability.

\vskip .05in
The method just outlined allows us to identify the evolutionary stage that contributes most to
the seed magnetic field in a collapsing stellar core, and to investigate how the spin rate of the collapsed
core depends on the progenitor mass and spectral type.  
We find that the pre-MS accretion phase is the dominant source of magnetic helicity in the core of a massive star.
The outer part of a growing massive star transitions from a convective to a radiative state while the mass 
still below $\sim 10\,M_\odot$.  A significant merger could create a brief convective state that re-ignites this process,
but would not necessarily generate much stronger helicity in the {\it inner} part of a massive star that had
already reached the main sequence.  Existing treatments of magnetic field amplification during a binary 
stellar merger (e.g. \citealt{WickTF2014}) have not so far accounted for the evolution of the magnetic helicity.

The rotation of radio pulsars and
stellar-mass BHs is much faster than would be expected if their progenitors were able to relax continuously
to a state of solid rotation.   This is commonly taken to imply that the transport of angular 
momentum must freeze out in some parts of the progenitor during its evolution toward core collapse.  

For example, \cite{HegeWS2005} obtain pulsar rotation periods $P_{\rm NS} \sim 3$-15 ms from 
models of initial (zero-age main sequence) mass $M_{\rm ZAMS} = 12$-35$\,M_\odot$, and solar metallicity.   
These results are consistent with the fastest pulsar spins, but not with the substantial fraction of 
pulsars born spinning in the range $P_{\rm NS} \sim 0.1$-1 s (e.g. \citealt{PopoT2012}).  Although
the multipolar structure of the poloidal magnetic field is not clear in the formalism of \cite{Spru2002}, 
making an optimistic assumption of a dipolar radial field still implies a NS polar field $\sim 10^{10}$ G, 
substantially weaker than is typical of radio pulsars.

In comparison, the calculations presented here imply the presence of a stronger poloidal magnetic flux
within the cores of massive stars, whose large scale nature is directly tied to the presence of net
magnetic helicity.  We find only modest variations between progenitor models in the magnetic field entrained
in the collapsing core, corresponding to $B_p \sim 10^{12}$-$10^{13}$ G in the NS remnant.   
This demonstrates that the internal transport of magnetic helicity is an essential ingredient in 
a model of stellar rotation;  we incorporate it here for the first time in the case of massive stars.

Because the internal magnetic torques are stronger than in the calculations of \cite{HegeWS2005}, we obtain
a wider range of spin periods.  We find a significant dependence of $P_{\rm NS}$ on progenitor 
mass and spectral type.  In part, this is because more massive stars develop thicker convective burning shells
and experience stronger angular momentum pumping; and also because the angular moment lost by a supergiant
to a wind depends strongly on its effective temperature.   In combination with the relatively uniform
pre-collapse magnetic field, this suggests that the rotation of the post-collapse NS is an important 
variable determining how the seed poloidal magnetic field evolves after the collapse, a possibility 
that we discuss briefly.

\subsection{Additional Transport Processes and Simplified Approach}

Most stars that experience core collapse (those born with a mass $M_{\rm ZAMS} \gtrsim 8\,M_\odot$) 
rotate rapidly during the hydrogen burning phase (\citealt{HuanGM2010,Ramietal2013}).   Early 
numerical studies of the rotation of post-MS stars, which either neglected angular momentum transport or 
included it in an elementary way, found that the core angular velocity eventually surpassed the 
Keplerian rate (\citealt{KippMT1970,EndlS1976}).  This is obviously problematic and we now know unrealistic. 

A suite of mixing processes have been studied and implemented in one-dimensional evolutionary calculations
over the past few decades:  
see for example \cite{EndlS1978}, \cite{Zahn1992}, \cite{MaedZ1998}, \cite{MeynM2000}, \cite{HegeLW2000},
and \cite{wheeler15}.
Mechanisms of angular momentum transport can be divided into two main categories: dynamic and secular. 
Dynamic processes operate on the convective or rotational timescale, a familiar example being the MRI
\citep{BalbH1994}, which can enforce nearly solid rotational motion within spherical shells over a 
modest number of rotations.  Convection instead generates differential rotation in latitude (see \citealt{BrunP2009}
for simulations of deep envelopes).  Because our leading
concern is the redistribution of angular momentum in radius, we adopt a simplified description of
the rotation profile, taking the angular velocity to be a function only of spherical radius, 
$\Omega = \Omega(r)$.  Such a rotation profile is sometimes described as `shellular'.  

Secular processes include i) radial mixing that is driven by an angular velocity gradient and 
facilitated by thermal diffusion \citep{Townsend58,GoldS1967,Fric1968,Zahn1992,MaedZ1998,MeynM2000};
and ii) the transport of angular momentum across convective-radiative boundaries by gravity waves, which are sourced
by convective motions and then damp within radiative layers \citep{gk90,kq97,ztm97,tkz02}.  We find that 
turbulent hydrodynamic stresses are subdominant to the Maxwell stresses that emerge from our model for magnetic
helicity growth; and, furthermore,
the mean polar flux produced is consistent with the post-collapse magnetic fields measured in radio pulsars.  The
estimates of internal Maxwell stresses made by \cite{MM14}, which are based on outward extrapolations of pulsar 
magnetic fields (and may therefore underestimate the large-scale poloidal field in the outer stellar core)
are consisent with this conclusion.  The MRI feeds off negative $d\Omega/dr$
in thin and weakly stratified layers near radiative-convective boundaries \citep{menou04}, but combining its
effects with those of large-scale Maxwell stresses has not yet been attempted in stellar dynamo models and
is beyond the scope of this work.  Finally, we note that processes that operate on the Kelvin-Helmholtz
timescale $\sim GM^2/RL$, such as meridional circulation in a 
massive and rapidly rotating star, are not generally competitive with those considered here.  

Our rotational models combine two novel effects:  large-scale helical Maxwell stresses, which dominate the 
transport of angular momentum in radiative layers of the star, and the inward advection of angular momentum in
deep convective layers, which enforces strong radial differential rotational.  For this reason, we take an otherwise
simplified approach, turning off transport by other processes, especially rotationally induced mixing and internal
gravity waves.  In the latter case, predictions of the sign of the angular momentum transport have varied between 
different authors.  However, it is possible that pumping of angular momentum into radiative layers by gravity waves
sometimes overwhelms the smoothing effect of the large-scale Maxwell stress, especially if the exciting convective motions
have a high Mach number.  
The rotation periods of collapsed stellar cores estimated by \cite{FullCLQ2015} are competitive with those obtained here
for the lowest mass stellar models, but not for higher masses.  Although the neglect of gravity wave transport may
introduce the strongest systematic bias in our results, the influence of hydrodynamic instabilities on core properties
can be more straightfowardly quantified, and is examined briefly.

A recent numerical simulation of core collapse starting from a dynamic state of oxygen burning 
in a $18\,M_\odot$ progenitor \citep{muller17} finds that a fairly rapid ($P_{\rm NS} \sim 20$ ms)
neutron star spin results from an asymmetric collapse and explosion.   Velocity perturbations seeded by convection
in the oxygen shell have been argued to facilitate an explosion \citep{co13,muller15}.  This result appears to depend
on a combination of physical processes:  first, the growth of non-spherical velocity perturbations in the collapsing
material (due essentially to conservation of angular momentum:  \citealt{lg00}), which then seeds a global buoyancy 
instability of the hot shocked material \citep{thompson00}.  Such dynamic effects fall beyond the scope of the
approach advanced here.

Finally, rotational support during the collapse provides additional channels for angular momentum transport.  The formation
of a quasi-Keplerian torus is a key ingredient in the collapsar model of GRBs \citep{woosley93}.  In the most massive
models that form BHs, we are able to calculate the mass that is directly incorporated into the BH, and the mass which
may be expelled in an outflow.   More generally, we divide our stellar models into those which form a NS or a BH based
on estimates of the compactness of the progenitor core that emerges from the MESA calculation.  Previous 
calculations of massive stellar evolution with different codes, which were used to estimate threshold conditions for
an explosion, imply slightly different relations between ZAMS mass and the compactness and iron mass of the evolved 
core \citep{OConO2011,ErtlJWSU2016,muller16a}.  This implies an intrinsic theoretical `fuzziness' in the progenitor
mass that will evolve to a given set of pre-collapse core conditions.  For this reason, implications for the
post-collapse rotation of the most massive models are considered with both outcomes in mind.

\subsection{Plan of the Paper}

The set-up of the MESA models used in this paper is described in Section \ref{sec:stellar_models}.
Then in Section \ref{sec:angular_momentum_transport} we explain our prescription for angular momentum
transport in convective and radiative layers, and how the rotation profiles of successive MESA snapshots are connected
to each other.  The role of the Coriolis force in limiting angular momentum pumping by convection is discussed
in Section \ref{sec:convective_transport}, and of kinking in limiting transport by the Maxwell stress in Section
\ref{sec:J_transport_kink}.  We outline how the different parts of the model stars divide into transporting and
non-transporting layers, and demonstrate the effectiveness of the adopted angular momentum transport mechanisms
in comparison with more popular prescriptions for rotationally driven mixing.  The accumulation of magnetic helicity
in growing radiative layers is described in Section \ref{sec:helicity}.  The results for the overall evolution
of the rotation and magnetic helicity in isolated model stars are presented in Section \ref{sec:combined_evol},
with a focus on the rotation and polar magnetic flux of the remnant NS or BH.  Spin-up of the stellar core by
tidal angular momentum exchange with a massive stellar companion is explored in Section \ref{sec:binary_interaction}.
Our main results and conclusions are summarized in Section \ref{s:conclusions}.  Appendix A shows our
MESA model parameter lists, and Appendix B describes how kinking limits the growth of the Maxwell stress
in radiative layers of a star containing a helical magnetic field.

\section{Stellar Models} \label{sec:stellar_models}

Using the one-dimensional stellar evolution code MESA (Modules for Experiments in Stellar Astrophysics:
\citealt{Paxtetal2011}, version 8118), we created evolved models of several masses, beginning with pre-MS
accretion onto a low-mass core and extending to core-collapse.   We aimed to include models that i) produce
both NS and BH remnants; ii) have a range of effective temperatures and convective penetration during the
supergiant phase iii) have a wide range of peak luminosities, implying a range of convective depths within
burning shells; and iv) experience a range of peak mass loss rates.  We also allow for tidal interaction
with a binary stellar companion by adding a source term to the stellar angular momentum. 

Models of mass $M_{\rm ZAMS}$ = 13$\,M_\odot$ and 40$\,M_\odot$ were chosen as progenitors
of NSs and BHs respectively, with 25$\,M_\odot$ as an intermediate case.
Each of the stellar models has solar metallicity, with
the 40$\,M_\odot$ model duplicated at 30\% solar metallicity.  
These choices were guided by qualitative predictors of the outcome of
core collapse.  The first of these is the compactness parameter at core bounce,
\be
\xi_M \equiv \frac{M/M_\odot}{r(M)/1000~{\rm km}}\bigg|_{t=t_{\rm bounce}}.
\ee
\cite{OConO2011} found that $\xi_M$ evaluated at enclosed baryonic mass $M = 2.5\,M_\odot$ implies an explosion 
for $\xi_{2.5} < 0.4$.  

\begin{figure}
\epsscale{1.25}
\plotone{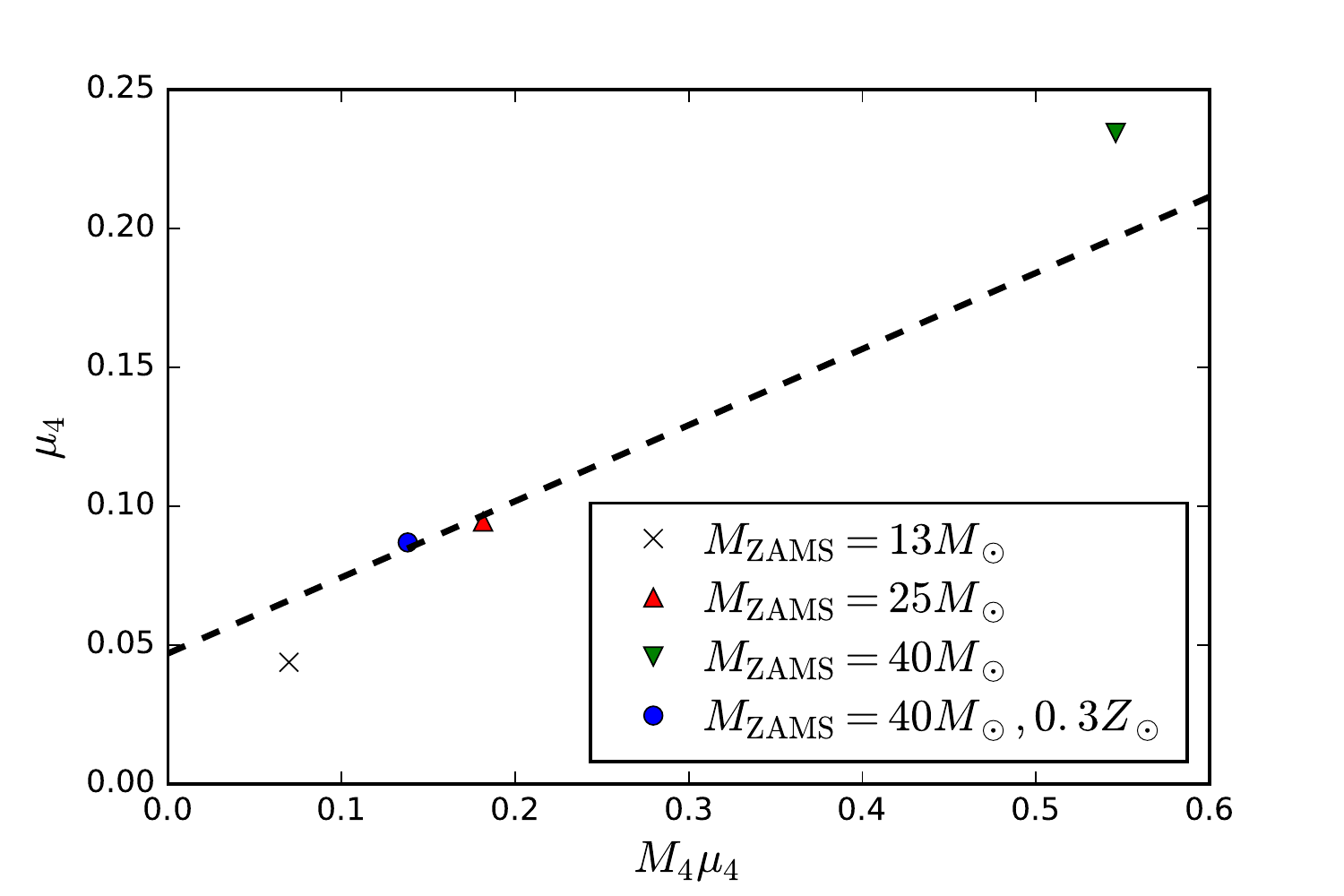}
\caption{Positions of the 4 MESA models considered in this paper, in the plane defined by $M_4\mu_4$
and $\mu_4$, where $M_4$ is the enclosed mass in solar units and $\mu_4 = d\ln M/d\ln r|_{s=4}$,
both evaluated at the radius where the specific entropy $s=4$.  The domain below the diagonal line comprises
models which are predicted to explode and leave behind NS remnants; whereas the models above the line
are expected to collapse to form a BH \citep{ErtlJWSU2016}.   The 25$\,M_\odot$ and low-metallicity 40$\,M_\odot$ models
lie near the boundary between explosive success and failure, meaning that additional physical
processes such as magnetorotational feedback could have a significant influence on the result. The diagonal boundary was calibrated using
the 19.8$M_\odot$ model of \cite{UgliJMA2012}.}
\vskip .2in
\label{fig:explodability}
\end{figure}

More recently, \cite{ErtlJWSU2016} found a tighter relation between pre-collapse mass
profile and `explodability', which is the one we used.  The division between NS and BH remnant corresponds to a
linear curve in the plane defined by the enclosed mass and the radial mass derivative, both evaluated where the
dimensionless entropy per baryon $s = 4$.  Figure \ref{fig:explodability} shows the four MESA models in this
plane.  The 13$\,M_\odot$ model appears to explode easily, and the solar-metallicity 40$\,M_\odot$ model
appears to fail.  

The 25$\,M_\odot$ model was drawn from a narrow mass range over which the core compactness
drops significantly, suggesting the possibility of a successful explosion.   This model lies close
to the diagonal line in Figure \ref{fig:explodability} separating failed from successful explosions.  
The same is true of the low-metallicity 40$\,M_\odot$ model.  This means that additional physical processes
than neutrino heating, such as magnetorotational feedback, could play a significant role in determining
the outcome of the core collapse.   Interestingly, we find that one of these models does produce a rapidly
rotating neutron star, whereas the other does not.

\begin{figure}
\epsscale{1.25}
\plotone{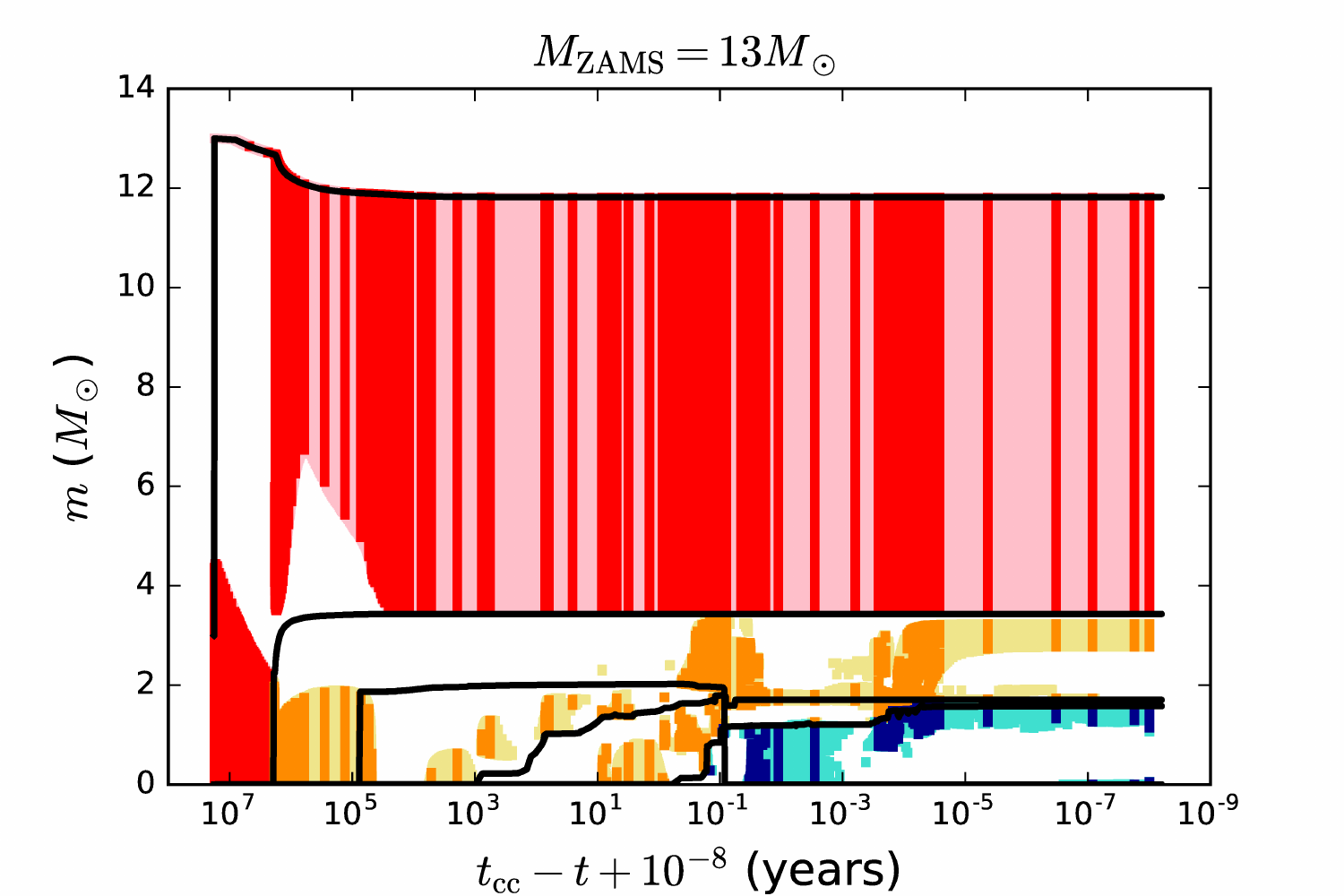}
\plotone{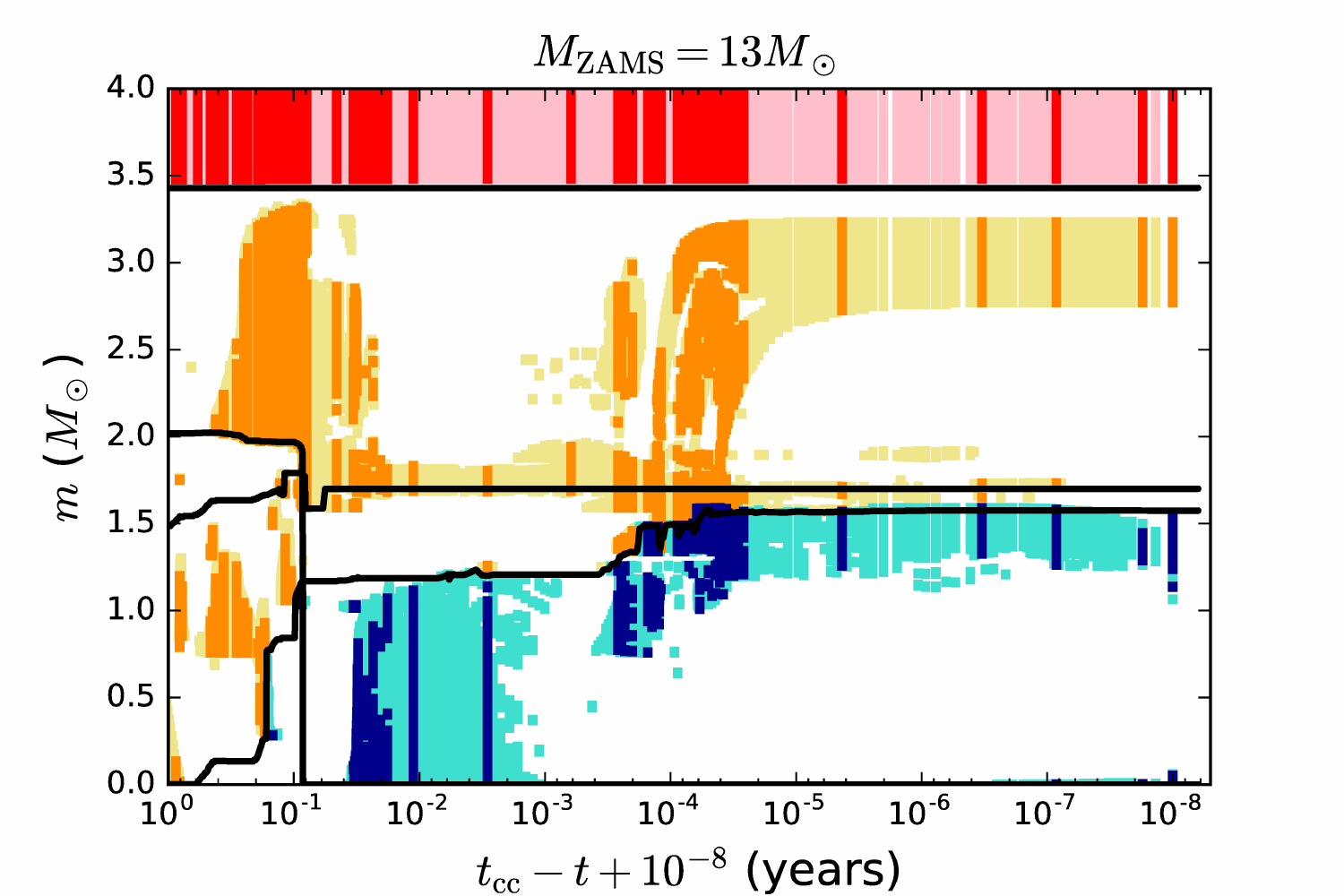}
\caption{{\it Top panel:}  Convective structure of the 13$\,M_\odot$ model, versus time to
core collapse. Only regions that are fully convective according to the Ledoux criterion are highlighted.
Light colored regions are constructed with a frequent MESA model output, and the solid lines are the
models we analyzed in detail.  Red:  hydrogen-rich convection zones;  orange:  helium-, carbon- and
oxygen-rich zones; blue:  silicon- and iron-rich zones.   The horizontal black lines mark (from the top)
the surface, and the upper boundaries of the helium, carbon, oxygen, and silicon cores and shells.
At $\sim 10^{-1}$ yr before core collapse, a distinct carbon-rich layer disappears.
{\it Bottom panel:}  Final year of the inner 4$\,M_\odot$.}
\vskip .2in
\label{fig:conv_zones_complete_13M_o}
\end{figure}

\begin{figure}
\epsscale{1.25}
\plotone{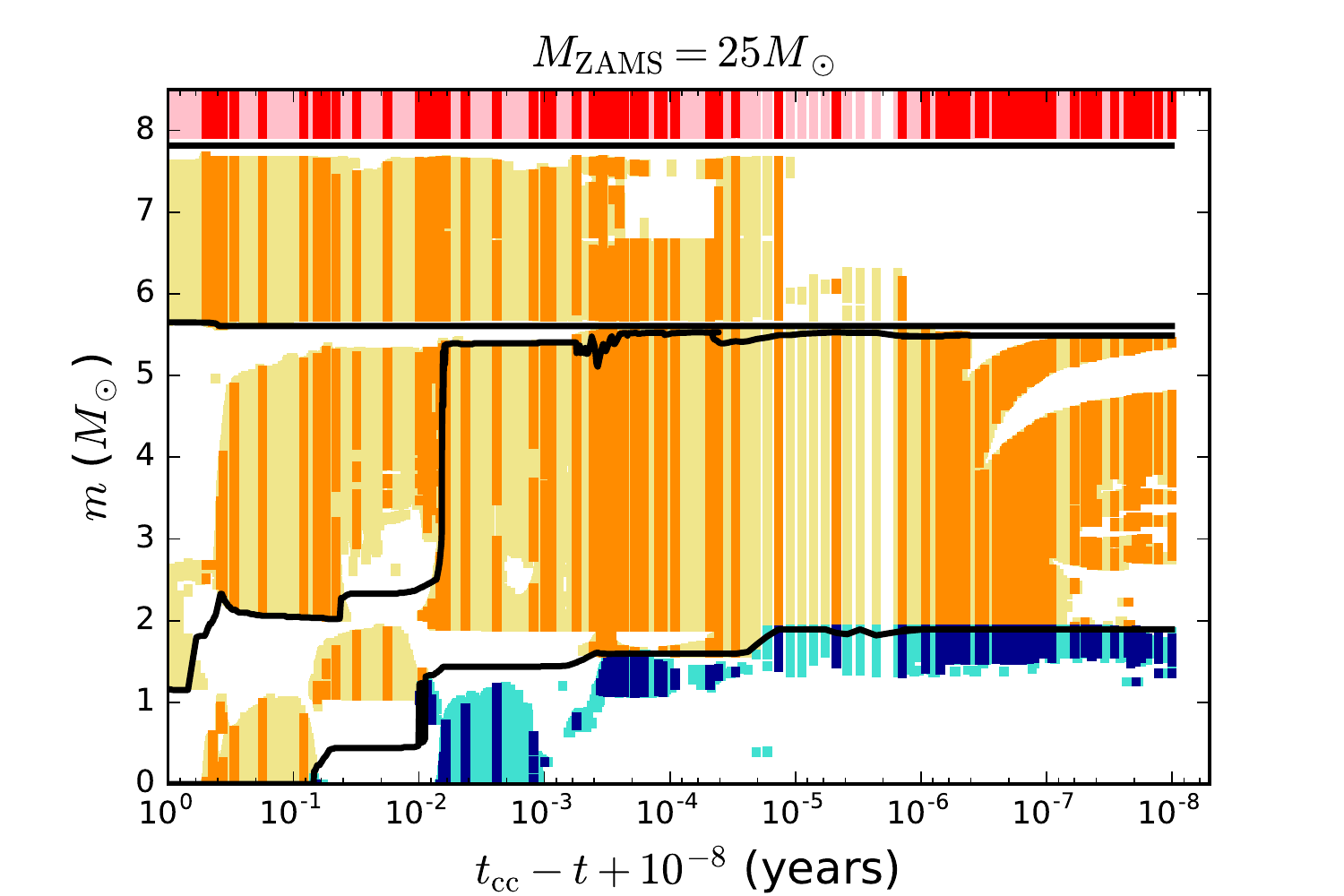}
\caption{Convective structure of the core of the 25$\,M_\odot$ model during the later
burning stages.  Comparison with the 13$\,M_\odot$ model (Figure \ref{fig:conv_zones_complete_13M_o})
shows a deeper and more connected sequence of convective layers during the $\sim 0.1$ yr before core
collapse.}
\vskip .2in
\label{fig:conv_zones_25M_o}
\end{figure}

The 13$\,M_\odot$ and 25$\,M_\odot$ models expand to 
become red supergiants, whereas the two 40$\,M_\odot$ models remain blue supergiants. 
We find that varying the effective temperature during peak mass loss has a strong influence on the angular
momentum carried off by the stellar wind, because convective envelopes of varying depths pump different amounts
of angular momentum into the stellar interior.  The same effect is encountered within the inner convective shells.
At the moment of core collapse, the inner core stores more angular momentum in more massive
progenitors, because they have deeper convective shells: compare the later stages of the 13$\,M_\odot$ model of Figure
\ref{fig:conv_zones_complete_13M_o} with the 25$\,M_\odot$ model of Figure \ref{fig:conv_zones_25M_o}.   

\subsection{Building the Star}

Each model star is assembled by gradual accretion, starting from a 3$\,M_\odot$ core.
Growth by mergers with other stars is harder to implement in a one-dimensional evolution code. 
The accretion rate is taken to grow with time, $\dot{M} \propto t$, as suggested by numerical 
simulations and analytic calculations \citep{PeteBKM2011,MurrC2015}.   The accretion rate is 
normalized so that the mass is fully assembled in $T = 10^5$ yr,
\be
\dot{M} = \frac{2(M_{\rm ZAMS} - 3M_\odot)}{T^2} \cdot t \ \ M_\odot\; {\rm yr}^{-1}.
\ee
Thereafter accretion is shut off, and the only further change in mass is due to the stellar wind.
Since a massive star can reach the main sequence while still accreting, we define
the zero-age MS to coincide with the end of accretion.

\begin{figure}
\epsscale{1.25}
\plotone{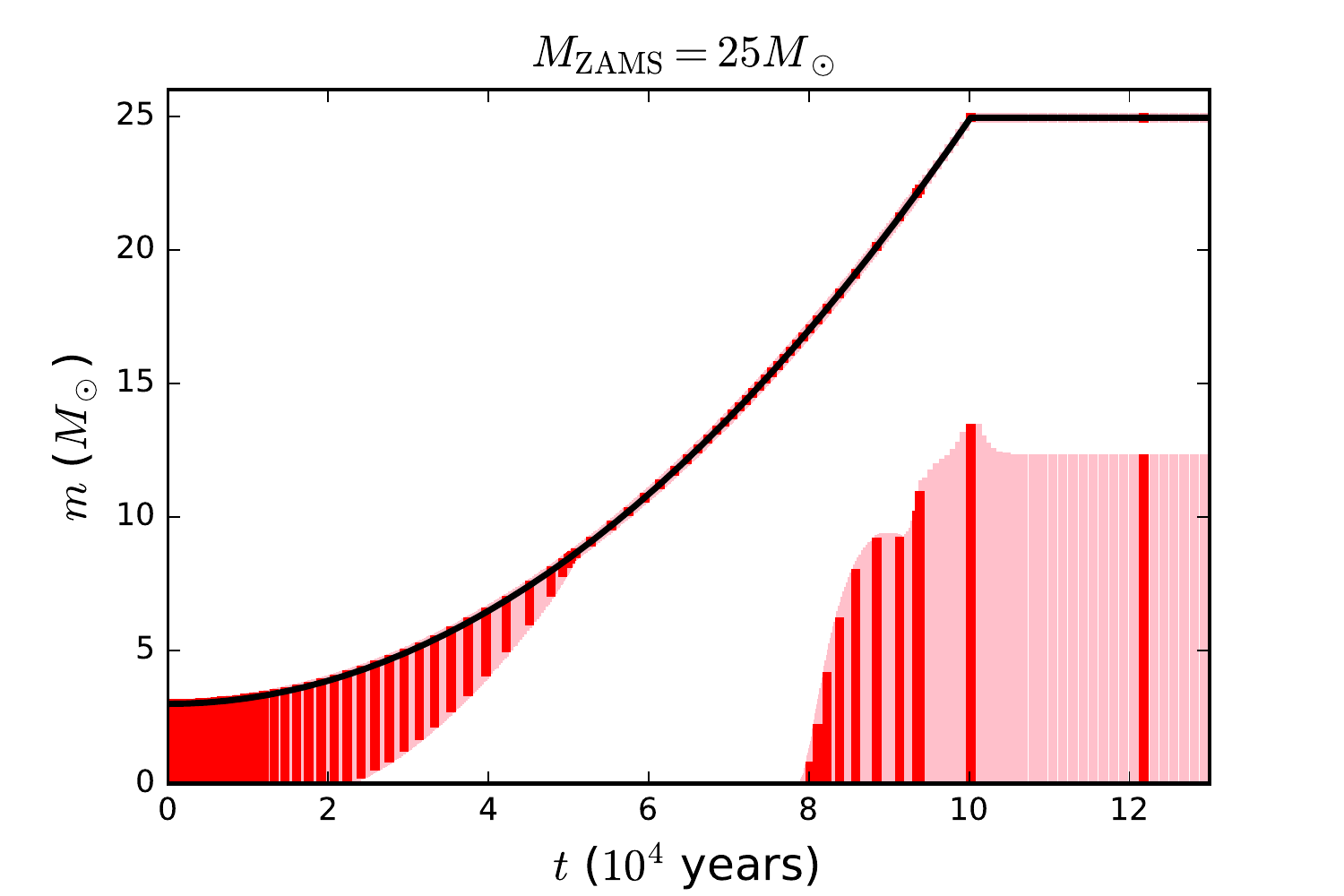}
\caption{Convective zones (marked red) during the pre-MS evolution of the 25$\,M_\odot$ model.  Black line:
surface of the star.  Accretion begins with a 3$\,M_\odot$ core, and is completed $T = 10^5$ yr later. 
Convective layers are unstable according to the Ledoux criterion.   Solid red lines
mark individual MESA profiles whose rotational evolution is analyzed in detail.
Recession of the outer convection zone is completed when the stellar mass has grown to $\sim 9\,M_\odot$.}
\vskip .2in
\label{fig:conv_zones_pre_MS_25M_o}
\end{figure}

The seed 3$\,M_\odot$ core is fully convective.  We follow changes in the early convective structure
because of its importance in magnetic helicity generation (Section \ref{sec:helicity}).  
This structure differs little between
the models.   When the accretion rate reaches $\dot{M} \sim 0.5$-$1 \times 10^{-4}\,M_\odot$ yr$^{-1}$,
the initial convection zone begins to retreat to the surface, and disappears when the stellar mass has
grown to $M \sim 7.7$-$9\,M_\odot$.  At the same time the convective core emerges. See Figure 
\ref{fig:conv_zones_pre_MS_25M_o}.

\subsection{Prescription for Numerical Evolution} \label{sec:model_params}

The MESA inlist file we used to generate the stellar models can be found in Appendix \ref{app:MESA_inlist}.
The free parameters with the greatest impact on the model behavior are those describing convective overshoot and mass loss.
We now discuss each of these briefly.

\vfil\eject
\subsubsection{Overshoot}

The convective structure of massive model stars is sensitive to the prescription for convective overshoot.
Strengthening the overshoot causes increased mixing, and significant changes in core mass.   The amplitude
of the overshoot is still poorly constrained, with some evidence for a dependence on stellar mass and metallicity:
see \cite{Herw2000} as well as the discussion in Appendix B.7.2 of \cite{Paxtetal2013}.  We use an exponential 
parametrization of overshoot, meaning that the convective mixing coefficient drops on the radiative side of a
radiative-convective boundary over $f_{\rm ov} = 0.01$ times the pressure scaleheight $l_P$:  
$D_{\rm ov} = D_0 e^{-|\Delta r|/f_{\rm ov}l_P}$, where $\Delta r$ the distance from the boundary.  The amplitude
$D_0$ is calculated a distance $f_{\rm ov,0} = 5\times10^{-4}l_P$ into the convection zone.

The amplitude of overshoot may well depend on a second parameter such as the magnetization of
the convective material \citep{nordhaus08}, especially given that the buoyancy of magnetized material can be
enhanced by a high radiative energy flux \citep{KissT2015b}.  The latter effect is especially relevant during
later stages of stellar evolution, or in massive stars.

\subsubsection{Mass and Angular Momentum Loss}

Mass loss plays an important role in the evolution of massive stars:  our 40$\,M_\odot$ model loses about half
its ZAMS mass by the moment of core collapse.  In the absence of a surface magnetic field, as assumed here, 
the angular momentum lost is proportional to the ejected mass.  We use the `Dutch' mass loss prescription in MESA,
which combines fitting formulae appropriate to different ranges of effective temperature and surface hydrogen abundance
$X_s$, as described by \cite{Glebetal2009}.   

\subsubsection{Convective Stability and Semi-convection}

The Ledoux criterion for convective instability is enforced.  Semi-convective zones also develop
within each model;  although they influence the degree of mixing and the energy transport,
they have a negligible effect on the angular momentum transport that we calculate in post-processing. 
That is because the fluid motions are relatively slow in semi-convective regions, so 
that angular momentum transport is dominated by the large-scale Maxwell stress.

\subsection{Angular Momentum Transport and \\ Magnetic Field Evolution}

We disable the MESA modules that handle internal rotation, as well as angular momentum transport by rotationally
induced instabilities and the `Spruit-Tayler' magnetic feedback process.  We find that a large enough magnetic
helicity is deposited in the radiative zones of our model stars to invalidate the starting assumptions of the
`Spruit-Tayler' process.  In addition, radial angular momentum transport by the winding up of the pinned
poloidal magnetic field is rapid enough to overwhelm rotationally driven mixing processes.

Instead the rotation profile is handled in post-processing, working with a spherical stellar model.  One
limitation of this approach is that the growth of the magnetic helicity in radiative layers of the star
(Section \ref{sec:helicity}), and the transport of angular momentum by convection and magnetic torques
(Section \ref{sec:angular_momentum_transport}), are decoupled from the mixing processes that influence
the growth of the stellar core.

\section{Rotational Evolution} \label{sec:angular_momentum_transport}

We now describe the initialization of the stellar rotation, and explain
our handling of angular momentum transport by convective and magnetic stresses.  These processes operate
in combination with each other:  the effectiveness of magnetic stresses at enforcing nearly solid rotation
in radiative layers of a star depends on the previous convective history.   

\subsection{Initialization of the Rotation}

At the end of the accretion phase, we set the equatorial rotation speed to $v_{\rm rot,eq} = 200$ km s$^{-1}$,
consistent with the measured average projected velocity $\langle v_{\rm rot,eq} \sin i\rangle =
{1\over 2}v_{\rm rot,eq} \sim 100$ km s$^{-1}$ in O and B stars \citep{HuanGM2010,Ramietal2013}.  
The rotation speed during the accretion phase is set to the same (fixed) fraction of the break-up speed, 
as determined by the evolving mass and radius.  

\subsection{Inward Pumping of Angular Momentum by Convection} \label{sec:convective_transport}

We adopt the approach to convective angular momentum transport described in \cite{KissT2015a}.
We make the plausible assumption that convective transport is rapid, with the rotation approaching
its equilibrium profile over a modest multiple of the convective period $\tau_{\rm con} \sim l_P/v_{\rm con}$.
We keep track of how $\tau_{\rm con}$ compares with the evolutionary time of the stellar mass profile,
as defined by the Eulerian speed $v_r$ of the stellar material,
\be\label{eq:tev}
\tau_{\rm ev} = {\rm min}\left[{l_P\over |v_r|}, {t_{\rm cc}-t\over 3}\right].
\ee
Here $l_P = P/\rho g$ is the pressure scaleheight.  The second term in (\ref{eq:tev}) accounts for
the rapid evolution of the inner burning shells just before core collapse (which happens at stellar age
$t_{\rm cc}$).  The last stages of nuclear burning are rapid enough that the convection in the intermediate
and outer parts of the star effectively freezes out.   

We restrict consideration of convective angular momentum pumping to thick shells and envelopes.  These we
define as having an aspect ratio $R_{\rm con+}/R_{\rm con-} \geq 2$, where $R_{\rm con+/-}$ is the radius
of the top/bottom of the convective layer.   This choice is motivated mainly by the observation
of a small radial angular velocity gradient throughout most of the solar envelope 
($R_{\rm con+}/R_{\rm con-} \sim 1.4$).  

When the rotation is slow (as measured by the Coriolis parameter ${\rm Co} \equiv \Omega \tau_{con} \lesssim 1$),
we allow convective plumes to conserve angular momentum, resulting in a rotation profile
$\Omega \propto r^{-2}$.  In faster rotating layers (${\rm Co} > 1$) we account for the back reaction 
of the Coriolis force by considering the vorticity equation in the `thermal wind' approximation 
(see Section 2 of \citealt{KissT2015a}).  The rotation profile in convection zones is therefore described by
\be \label{eq:rotation_profile}
\Omega(r) \propto \left\{
     \begin{array}{lr}
       r^{-2} &  \quad [{\rm Co}(r) \leq 1] \\
       r^{-(1+\beta)/2} &  \quad [{\rm Co}(r) > 1].
     \end{array}
   \right.
\ee
Here $\beta$ is the local radial power law dependence of gravitational acceleration ($g(r) \propto r^{-\beta}$).
These scalings are connected to each other at radial shells where ${\rm Co} = 1$.  

Recently \cite{kq17} investigated core rotation in sub-giant and giant stars (radius $\sim (4-10)\,R_\odot$)
during the early part of the first dredge up, where $\beta \sim 1$ and $\Omega \sim r^{-1}$ is predicted by 
Equation (\ref{eq:rotation_profile}).  They considered the relative splittings of pressure- and 
gravity-dominated modes as a probe of the relative rotation rates in the stellar core and envelope.
Typical splittings were found to be marginally consistent with such an envelope rotation profile, and also
to be consistent with most of the angular velocity offset concentrated in the layer between the hydrogen burning shell
and the convective envelope.  In the case of Kepler-56 ($R \sim 4\,R_\odot$, $M \sim 1.3\,M_\odot$ and a convective
envelope extending a factor $\sim 3$ in radius), a $\sim r^{-1}$ convective angular velocity 
profile is not precluded; but this gradient must be extended all the way down to the burning shell,
contradicting the assumption of solid rotation in all radiative layers by \cite{KissT2015a}.  A recent application
of the \cite{kq17} method to other Kepler giants by \cite{triana17} did not yield clear results.  In this paper,
we take a more general approach to rotation in radiative layers by calculating the limiting magnetic torque,
as enforced by kinking of the wound up magnetic field (Section \ref{sec:J_transport_kink}).  

The feedback of magnetic fields on the rotation of a convective layer is an important outstanding issue.
The MRI must be activated at some level where ${\rm Co} > 1$ (e.g. \citealt{BalbBLW2009}),
which is the case during the subgiant phase probed by the Kepler asteroseismological data.
By contrast, the Coriolis parameter never becomes very large in our isolated star models
during a supergiant phase (or, indeed, in the fully expanded RGB/AGB models of \citealt{KissT2015a}).  
Since the convective Mach number is relatively high in the envelope near full expansion, the Coriolis parameter 
also remains modest within interior convective shells.  We expect that a reduction in peak ${\rm Co}$ due to the activation
of the MRI would, therefore, have a modest effect on the rotation rate of the inner core in a red supergiant.
The effect may be larger in more compact stars, especially those which gain angular momentum from a binary companion.

\subsection{Magnetic Angular Momentum Transport Limited by Kink Instability} \label{sec:J_transport_kink}

The growing concentration of mass toward the center of an evolving star generates negative $\partial\Omega/\partial r$,
but in the presence of strong stable stratification this does not trigger the MRI.
In radiative zones we focus on the linear winding of the embedded radial magnetic field, which generates a growing toroidal
magnetic field and $r\phi$ Maxwell stress.  Because the magnetic field quickly becomes tightly wound, it is
susceptible to a kink instability \citep{Tayler73}.   The growth of the kink is, however, impeded by the Coriolis
force  (\citealt{PittT1985}; Appendix \ref{app:kink_instability}).   

We find that the net effect of the kink instability is a modest increase in the radial magnetic field needed to
erase most of the radial angular velocity gradient.  The threshold radial Alfv\'en speed 
$v_{{\rm A},r} \equiv B_r/(4\pi\rho)^{1/2}$ rises from $\sim |v_r|$ to
\be \label{eq:v_A_reduced}
v_{{\rm A},r}\biggr|_{\rm min} = {\rm max}\left[{l_P\over \tau_{\rm ev}} 
              \left({r\over 2l_P}\Omega \tau_{\rm ev}\right)^{1/4}, \; {(rl_P)^{1/2}\over \tau_{\rm ev}}\right].
\ee
We take solid rotation to be established in a radiative mass shell if $v_{{\rm A},r} > v_{{\rm A},r}|_{\rm min}$ 
within that shell.  The second term in Equation (\ref{eq:v_A_reduced}) allows for a situation where $\tau_{\rm ev}$
becomes so small that the stellar material is not able to execute a full rotation, $\Omega\tau_{\rm ev} \lesssim 1$.

\begin{figure}
\epsscale{1.25}
\plotone{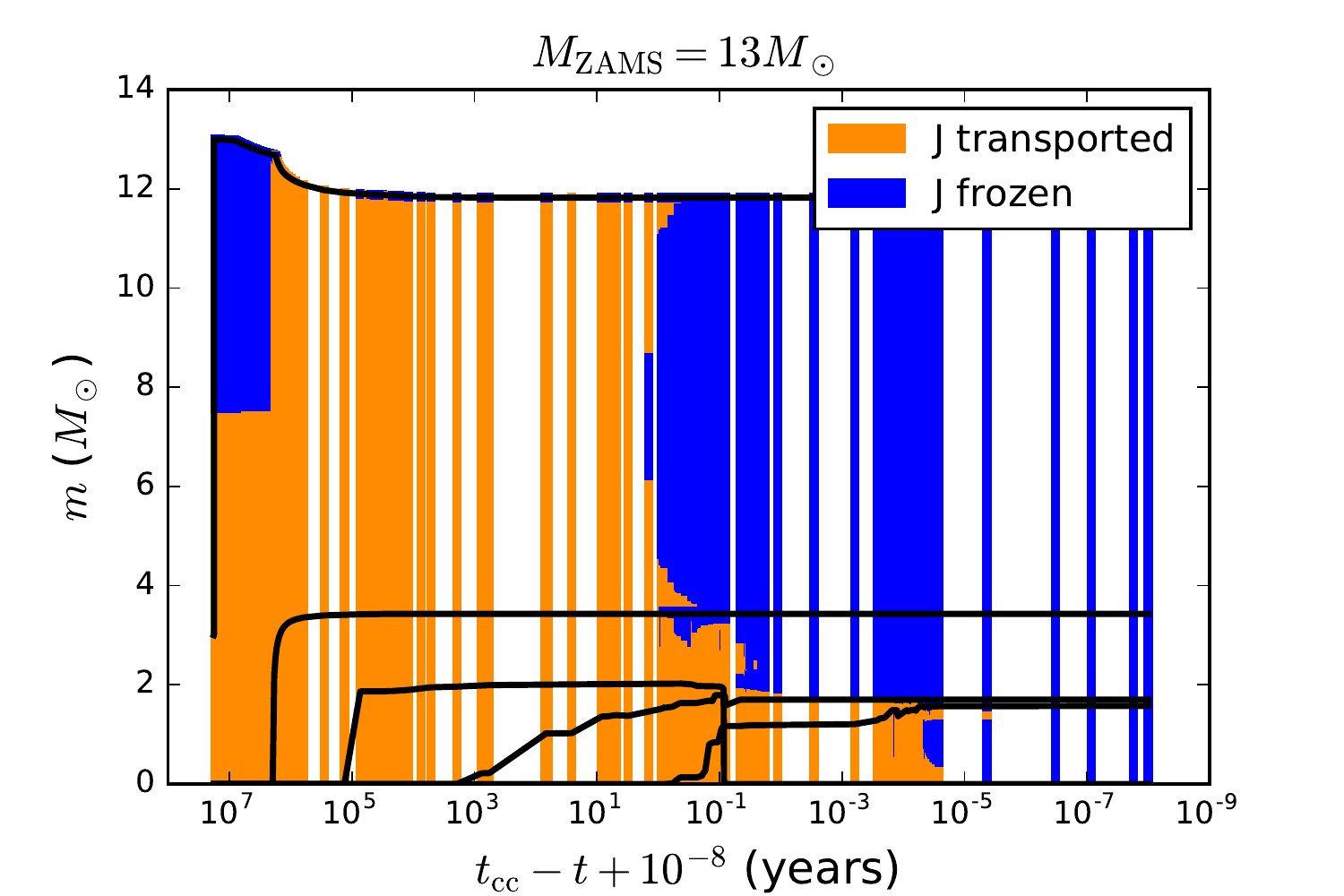}
\plotone{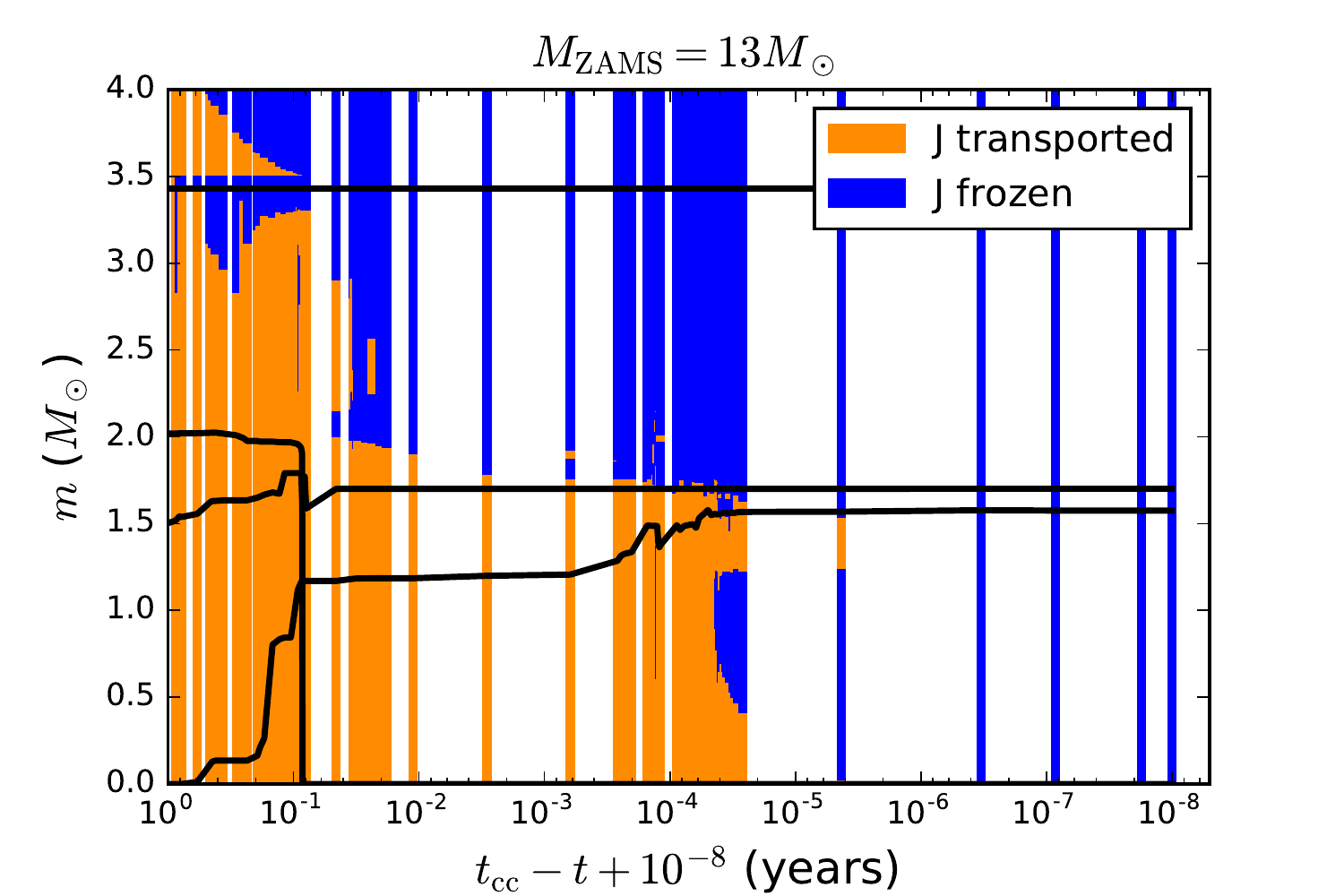}
\caption{The zones in the 13$\,M_\odot$ model which transport angular momentum (orange) and those in which angular
momentum is effectively frozen (blue).  {\it Upper panel:}  Full stellar model.  {\it Lower panel:} later evolution
of the inner 4$\,M_\odot$.  The outer blue zone appearing during the MS phase corresponds to the part of the star
that is not convective during the accretion phase, and so does not accumulate magnetic helicity at the evolving 
convective-radiative boundary.  This material may in fact be magnetized by the accretion disk.  This detail does not
influence the later post-MS rotational behavior, because the magnetized inner part of the star extends outside the
helium core.}
\vskip .2in
\label{fig:transport_and_non_transport_regions_complete_13M_o}
\end{figure}

\subsection{Rotational Evolution between Model Snapshots} \label{sec:J_transport}

The rotation profile $\Omega(r)$ is evolved as follows from one MESA model snapshot to the next:
 
\noindent\vskip .05in
1. In each radial zone we ask if $\tau_{\rm con} < \tau_{\rm ev}$,
or $v_{{\rm A},r} > v_{{\rm A},r}|_{\rm min}$.  If either inequality is satisfied then angular momentum is transported
between the interior and exterior zones; otherwise the angular momentum of the given zone is frozen.  
In this way, we divide each model snapshot into transporting and non-transporting layers.   Figure
\ref{fig:transport_and_non_transport_regions_complete_13M_o} shows the result over the full history of
the 13$\,M_\odot$ model.  Most of the star ceases to transport angular momentum effectively in the final 
$\sim 0.1$ yr before core collapse.   This conclusion also applies to the outer convective envelope, where
$\tau_{\rm con}$ becomes longer than the time to collapse.
\noindent\vskip .05in
2.  The update to the rotation rate in a non-transporting zone is obtained by matching the specific 
angular momentum $\frac{2}{3}\Omega r^2$ between corresponding mass shells in the successive snapshots.  
The new rotation profile in each {\it connected} transporting region (comprising multiple MESA mass shells)
is obtained by i) conserving the total angular momentum in that region; and ii) fitting the profile
(\ref{eq:rotation_profile}) in convective parts and solid rotation in radiative parts.  This fit is 
often non-linear, in the sense that both the normalization and the position of breaks in the slope of
the convective rotation profile (surfaces where ${\rm Co} = 1$) will depend on the net angular momentum
of the zone.  
\noindent\vskip .05in
3. The complex convective structure of massive stars generates circumstances in which 
thin non-transporting layers are sandwiched between two transporting zones.
In such a situation we envision that the two transporting regions would successfully communicate 
changes in angular velocity, if and only if the non-transporting layer is thinner than a pressure scaleheight 
$l_P$.

\subsection{Some Basic Results}

\begin{figure}
\epsscale{1.2}
\plotone{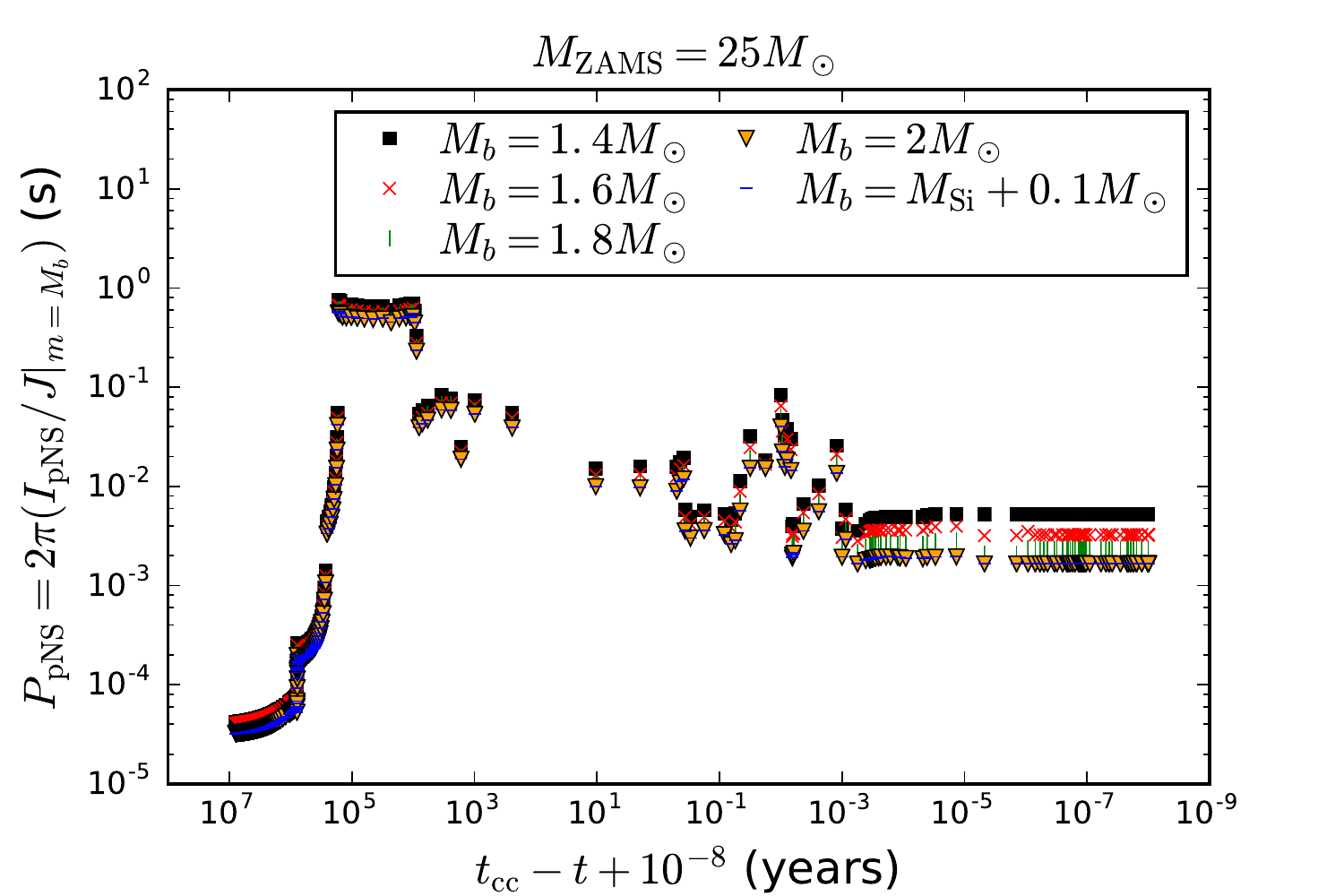}
\caption{Complete rotational evolution of the core of the 25$\,M_\odot$ model, as measured
by the spin period $P_{\rm NS}$ of a cold NS formed by the collapse of various baryonic masses $M_b$.
The radius of the NS is set to a uniform value 10 km.
The strong growth in $P_{\rm NS}$ around time $\bar{t} = 10^6$ yr before core collapse is associated
with the post-MS expansion.  Before a deep convective envelope begins to pump angular momentum inward, 
the core is coupled to the intermediate layers of the star by magnetic stresses.}
\vskip .2in
\label{fig:P_pNS_25M_o_complete}
\end{figure}

We now highlight some consequences of the rotational model just described, before presenting more detailed results
in Sections \ref{sec:combined_evol} and \ref{sec:binary_interaction}.   
Figure \ref{fig:P_pNS_25M_o_complete} summarizes the rotational evolution of the 25$\,M_\odot$ model.
Plotted is the rotation period $P_{\rm NS}$ that would result if the central 1.4, 1.6, 1.8, or 
2$\,M_\odot$ of stellar material were instantly incorporated into a neutron star, while conserving angular momentum. 

The result is shown as a function of time, to provide a measure of the changing angular momentum profile.
The first rise in (the effective value of) $P_{\rm NS}$ is driven by the post-MS expansion of the star, and is followed by a 
rapid drop as the outer convective envelope forms and deepens.   The pumping effect of the envelope is supplemented by 
the deep convective shells that form during the later stages of thermonuclear burning, resulting in a further
decrease in $P_{\rm NS}$.  

\begin{figure}
\epsscale{1.2}
\plotone{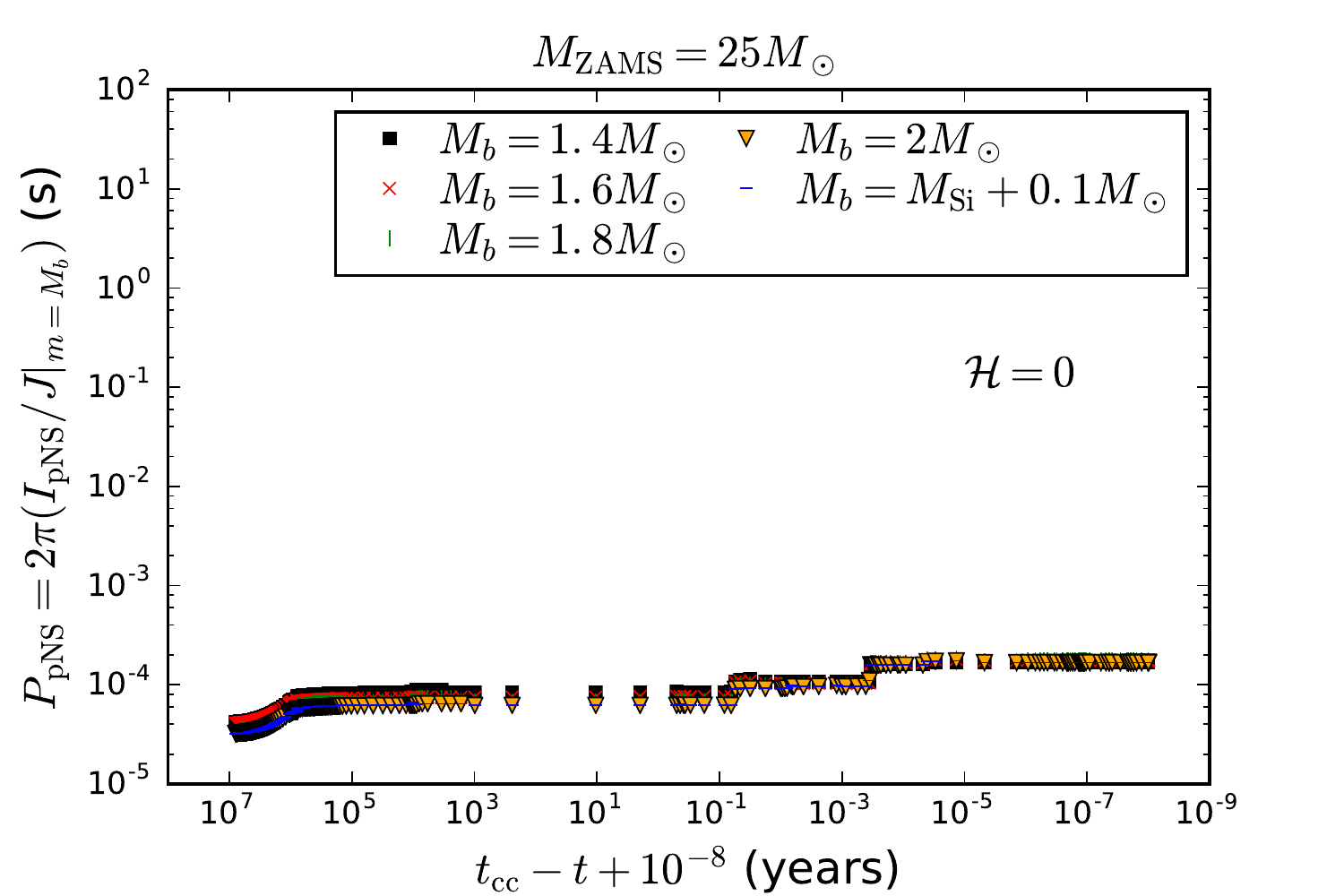}
\caption{Alternative rotational evolution of the 25$\,M_\odot$ model, with magnetic field set to zero
($\mathcal{H} = 0$) and rotationally driven mixing suppressed in radiative layers.  The core never couples
to the outer layers of the star, and conserves its angular momentum from the pre-MS phase.}
\vskip .2in
\label{fig:P_pNS_25M_o_complete_H_eq_0}
\end{figure}

Figure \ref{fig:P_pNS_25M_o_complete_H_eq_0} shows for comparison the effect of turning off angular momentum transport
by magnetic stresses in radiative layers, while maintaining the pumping effect of the convection.  Now the core 
does not couple to the outer layers of the star.  Conserving its angular momentum from the ZAMS, one sees
that the assumption of collapse to a hydrostatic NS leads to an inconsistency:   the implied $P_{\rm NS}$ 
is shorter than a millisecond.

\begin{figure}
\epsscale{1.25}
\plotone{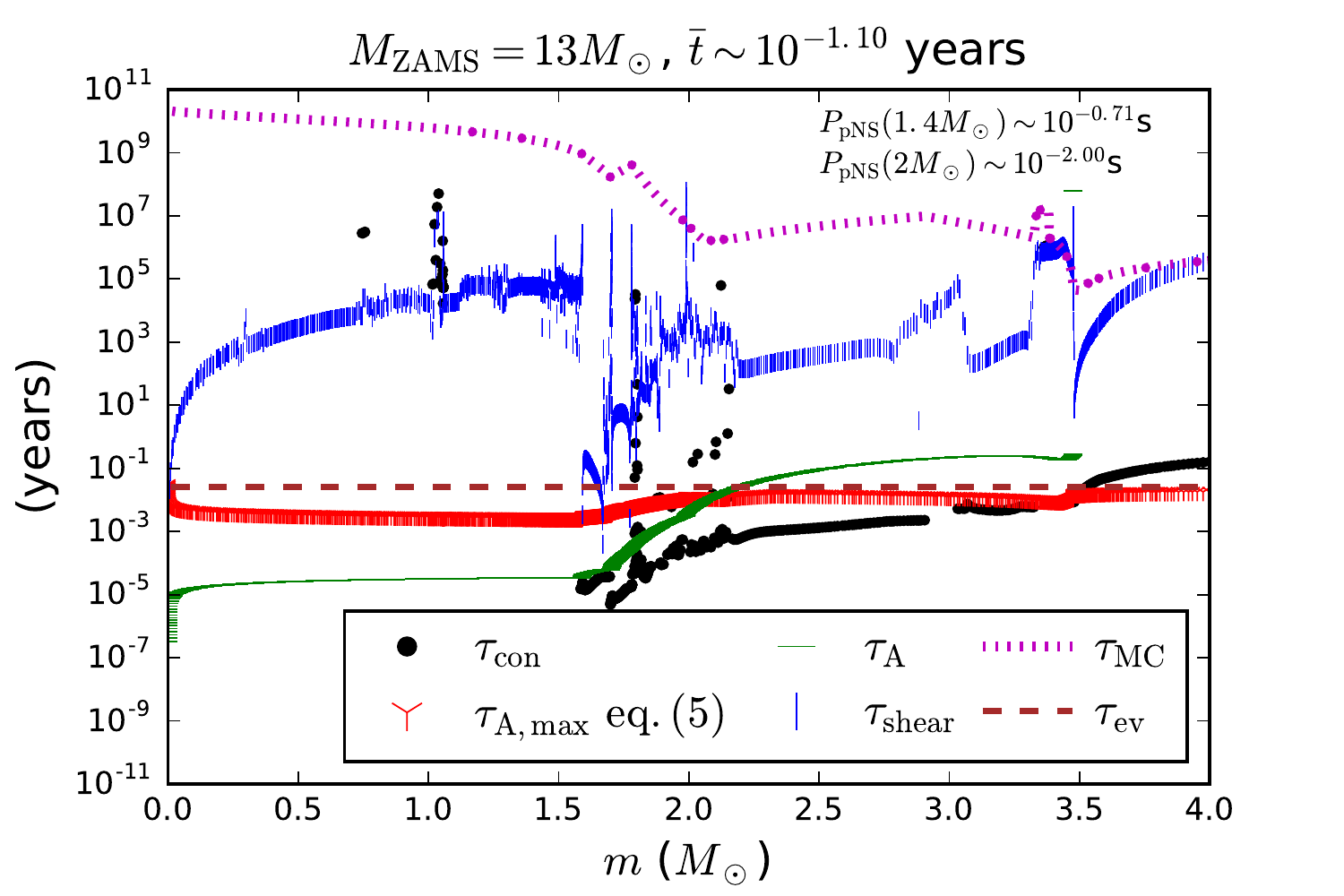}
\caption{Comparison of timescales for angular momentum transport by different mechanisms $\sim0.1$ yr before
core collapse. Horizontal dashed line: the evolutionary time of the stellar mass profile $\tau_{\rm ev}$,
which is limited by the time to core collapse (Equation (\ref{eq:tev})).  Black dots:  the convective timescale,
which is shorter than $\tau_{\rm ev}$ except in the outermost part of the envelope.
In the inner radiative parts of the star, 
the Alfv\'en time $l_P/v_{{\rm A},r}$ (green horizontal ticks) is comfortably below the critical value
that will allow solid rotation to be established (red tri-pointed stars) even after allowing for kinking
of the wound-up magnetic field.  This stands in strong contrast with the timescale (\ref{eq:tshear}) for
rotationally driven mixing (vertical blue ticks), which at this relatively advanced stage is much
longer than $\tau_{\rm ev}$ except for a narrow range of mass.  Meridional circulation
(mauve dotted curve) is everywhere slower than the other processes.}
\vskip .2in
\label{fig:rot_times}
\end{figure}

\subsection{Comparison with Rotationally Induced Mixing}

The efficiency of angular momentum transport by convective and Maxwell stresses, as implemented
here, can be compared with the rotationally induced mixing (RIM) as formulated by \cite{Zahn1992}.
The latter operates at low Prandtl number $\nu/\kappa_T$, where $\nu$ the kinematic viscosity
and $\kappa_T$ the thermal diffusivity, as is appropriate to the radiative zone of a star.  In this
situation, strong $\partial\Omega/\partial r$ combines with radiative diffusion to trigger a
linear axisymmetric instability \citep{GoldS1967,Fric1968}, as well as higher wavenumber
turbulent mixing \citep{Townsend58,Zahn1992}.  The corresponding timescale for
angular momentum transport is
\be\label{eq:tshear}
\tau_{\rm shear} = \frac{r^2}{D_{\rm shear}} = \frac{45}{2} {N^2 \kappa_T^{-1}} 
\left(\frac{\partial\Omega}{\partial r}\right)^{-2}.
\ee
Here $D_{\rm shear}$ is the diffusion coefficient, 
and $N$ is the Brunt-V\"ais\"al\"a frequency.  We also show for comparison the timescale for
meridional circulation,
\be
\tau_{\rm MC} \sim \tau_{\rm KH} \left(\frac{\Omega_{\rm K}}{\Omega}\right)^2,
\ee
where $\Omega_{\rm K}$ is the Keplerian angular frequency at the given radius and enclosed mass.

These RIM and circulation timescales are compared in Figure \ref{fig:rot_times} with 
i) the evolutionary time (\ref{eq:tev}); ii) the convective time;  iii) the radial 
Alfv\'en time $\tau_{{\rm A},r} = l_P/v_{{\rm A},r}$ as estimated from the hemispheric magnetic flux; 
and iv) the (maximum) Aflv\'en time $l_P/v_{{\rm A},r}|_{\rm min}$ that allows efficient
angular momentum transport after allowing for kinking of the wound-up field.
We choose a snapshot of the core of the 13$\,M_\odot$ model around the time (about 0.1 yr
before core collapse) when the core and envelope begin to decouple.  The RIM timescale
is consistently 3-4 orders of magnitude longer than the Alfv\'en and convective timescales, and 2-3 
orders larger than $\tau_{\rm ev}$;  the circulation
time even longer.  

Where the core material is radiative, the inequality $\tau_{\rm A,r} < l_P/v_{{\rm A},r}|_{\rm min}$ 
holds and the Maxwell stress is large enough to enforce solid rotation.  One sees from Figure
\ref{fig:rot_times} that this inequality is violated outside $\sim 2\,M_\odot$ enclosed mass, but
here the material is convective.   The convective timescale remains small enough to enforce the
equilibrium rotation profile, given here by Equation (\ref{eq:rotation_profile}), 
out to $\sim 3.5\,M_\odot$ enclosed mass.   Beyond that point, the convection has essentially
frozen out at this brief interval before core collapse.

We conclude that the addition of the shear instability would not {\it directly} change
the rotation profile significantly in our calculation.  The secondary effect of the influence
of rotationally-induced mixing on core mass and composition is addressed in the next section.

\begin{figure}
\epsscale{1.25}
\plotone{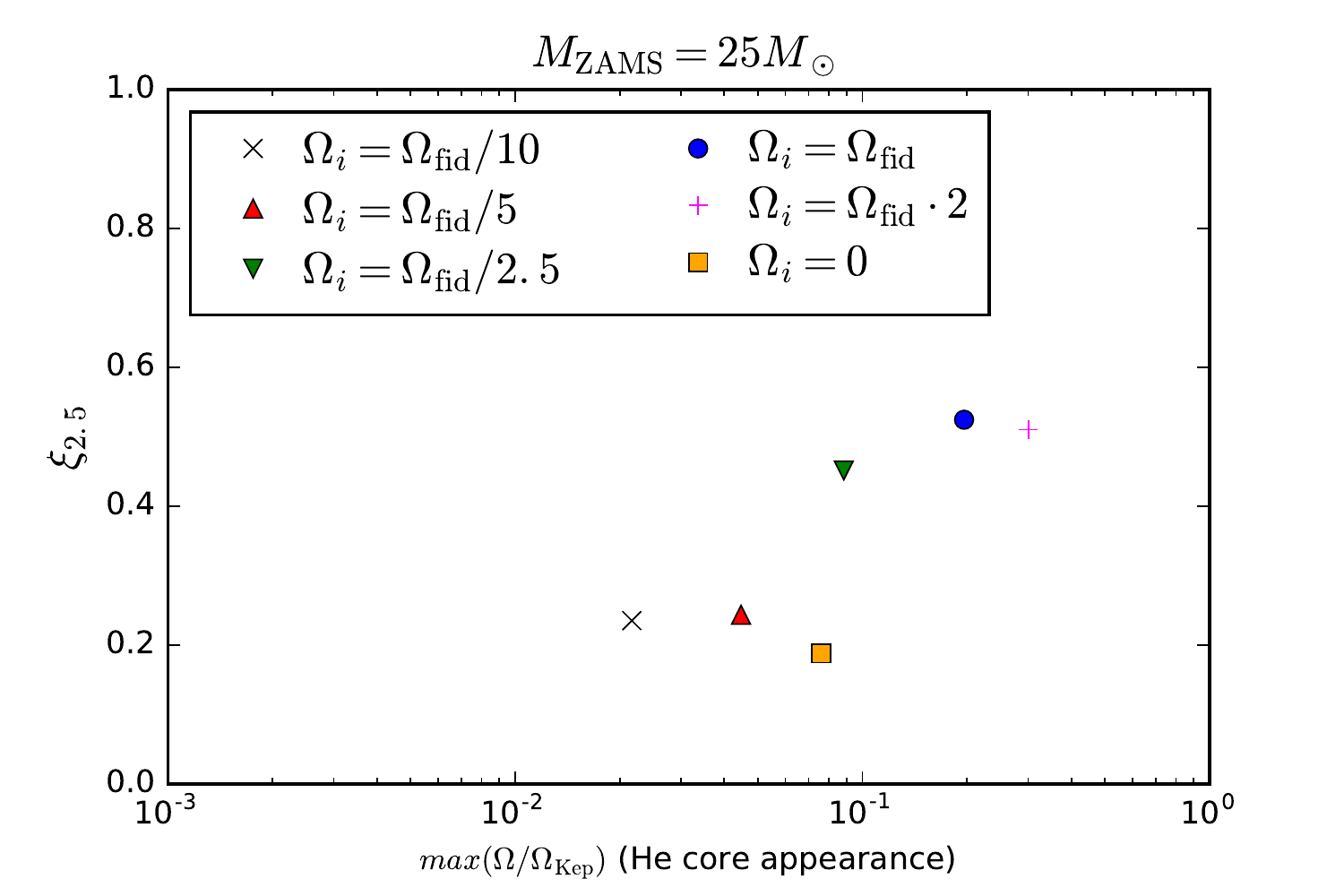}
\plotone{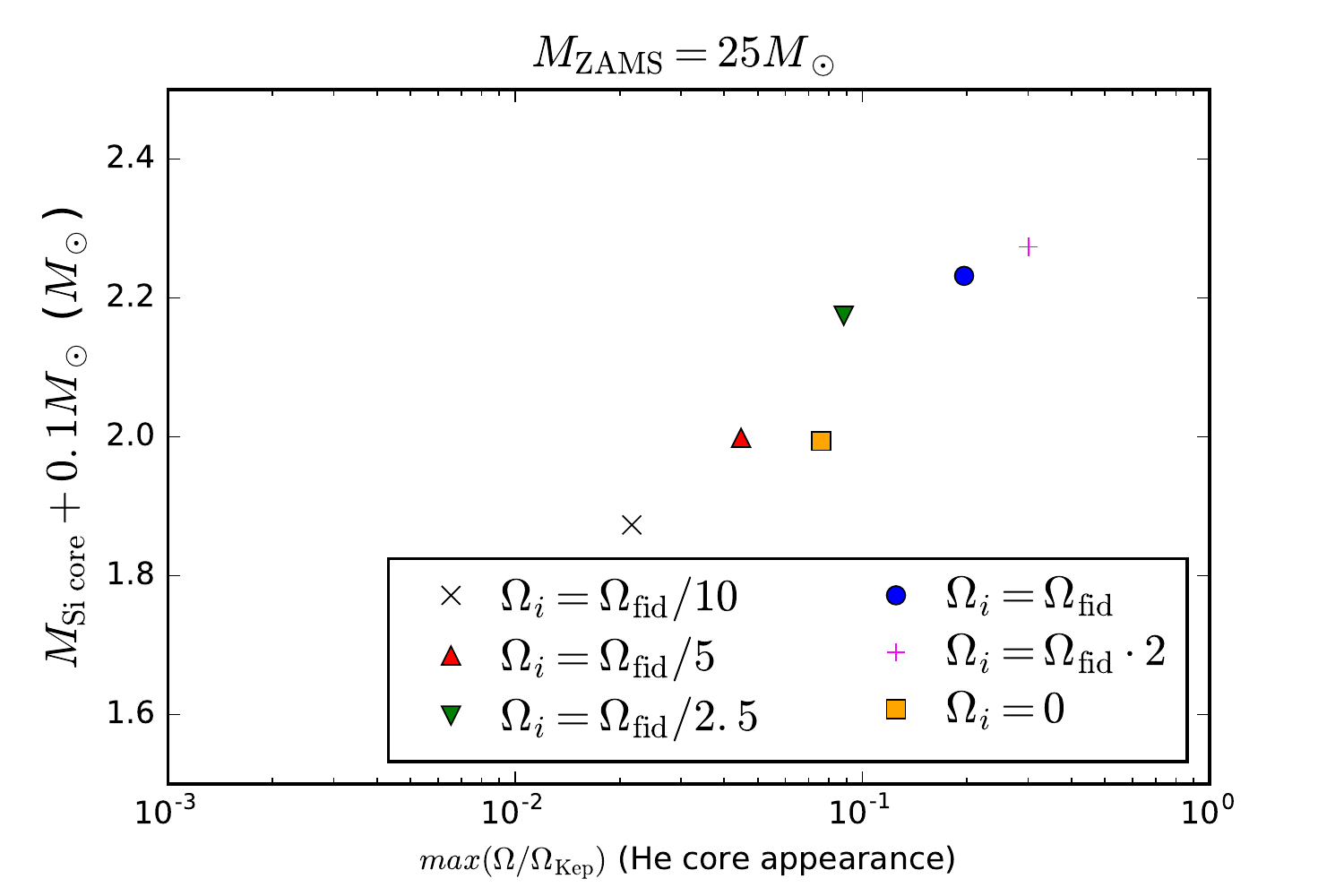}
\caption{Core compactness and total mass enclosed by the outer boundary of the Si layer just before core collapse
in a series of $25\,M_\odot$ models.  These are plotted versus the maximum value of $\Omega/\Omega_K$ within the
star at the formation of the helium core.  This maximum value is reached slightly interior to the inner boundary
of the convective envelope.  Five of the six points are obtained from MESA models with the built-in prescription
for angular momentum transport by rotationally induced mixing and the Spruit-Tayler dynamo turned on.  The rotation rate 
of the square yellow point represents a non-rotating MESA model with the angular velocity obtained in 
post-processing using the angular momentum transport prescriptions described in Section \ref{sec:angular_momentum_transport}. 
Here $\Omega_i$ is the initial angular velocity of the MESA model (with solid rotation on the ZAMS), and 
$\Omega_i = \Omega_{\rm fid}$ corresponds to ${1\over 2}$ of break-up at the stellar surface.}
\vskip .2in
\label{fig:hecorespin}
\end{figure}

\begin{figure}
\epsscale{1.25}
\plotone{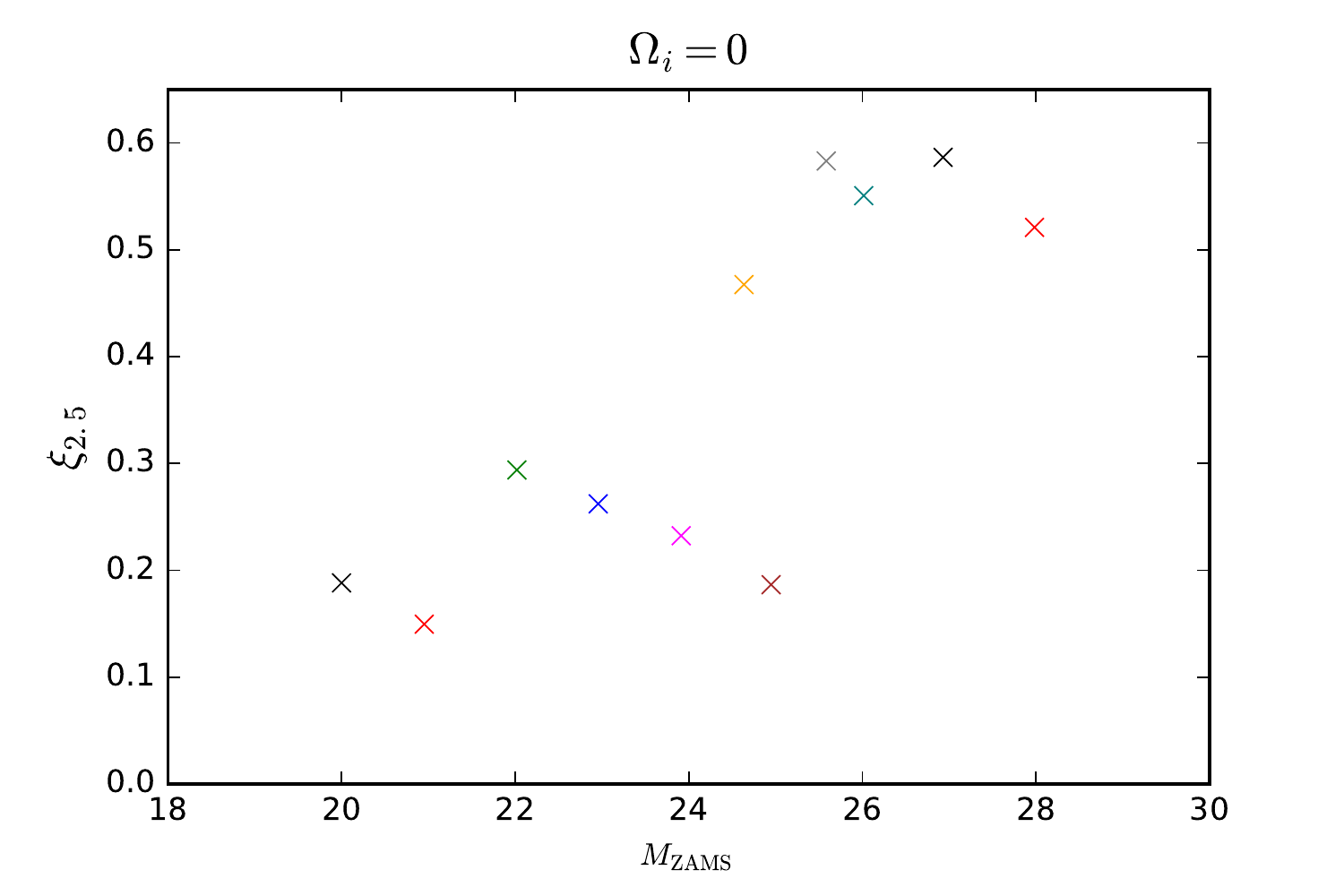}
\plotone{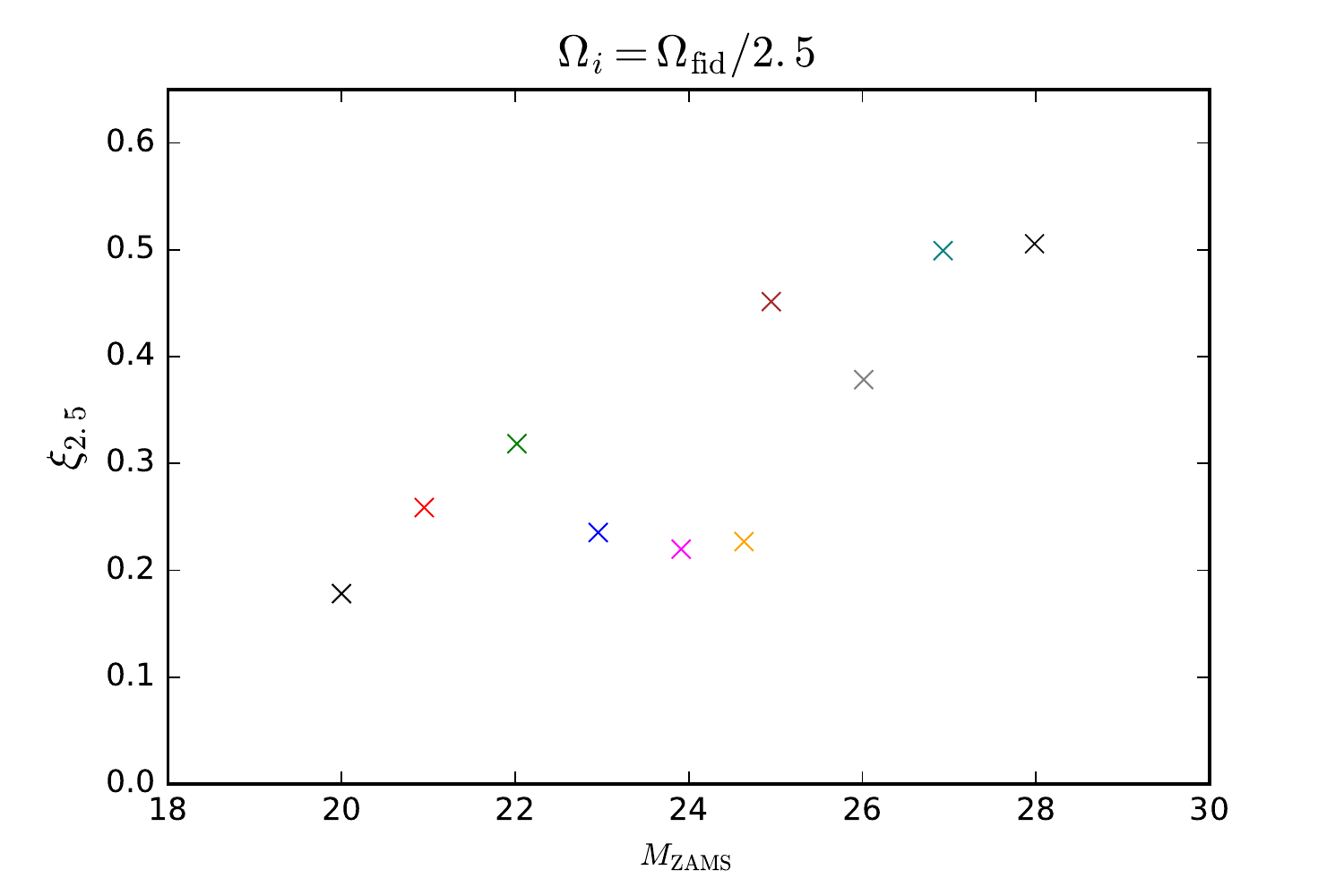}
\caption{Core compactness obtained for various progenitor masses from non-rotating MESA models (top panel),
and from MESA models with finite rotation and the effects of RIM and the Tayler-Spruit dynamo included in 
the evolution (bottom panel).  The initial rotation in the second case is chosen so that the peak value of 
$\Omega/\Omega_K$ at formation of the He core is comparable to that obtained by post-processing a non-rotating MESA 
model of the same $M_{\rm ZAMS}$.}
\vskip .2in
\label{fig:xi25}
\end{figure}

\vfil\eject
\subsection{Effect of RIM on Core Mass and Compactness}\label{s:rim}

The introduction of rotationally induced mixing into a one-dimensional stellar model, using
a simplified prescription \citep{Zahn1992,MaedZ1998} is known to modify the iron core mass and
the inner core compactness in models with rapid initial rotation.   For example, \cite{HegeLW2000} found
that when the ZAMS model rotates at about one-half of the breakup rate (as is typical of massive stars),
more than $\sim 10$\% changes in iron mass are possible in comparison with non-rotating models, even 
after allowing for the inhibition of mixing by mean molecular weight gradients.  Here we choose a similar
normalization of the mixing coefficients, with the parameters $f_c = 0.03$ and $f_\mu = 0.05$ in 
Equation (53) of \cite{HegeLW2000}.  In this case, in agreement with these previous results, the
total helium core mass is less sensitive to the initial rotation rate:  changes in $M_{\rm He}$
are limited to a few percent unless the rotation approaches breakup, or mean molecular weight gradients
are (unrealistically) ignored ($f_\mu = 0$).  We focus in this paper on massive stars that do not start 
as close binaries and develop extreme rotation from mergers.

The ZAMS rotation rate is too blunt a measure of the influence of rotation
on core properties.  We have checked systematically how the pre-collapse core properties correlate
with the strength of rotation at different stellar radii and at successive evolutionary stages.   By
far the strongest correlation turns out to be with the maximum rotation rate within the star at the
first appearance of the helium core.
This quantity is expressed as ${\rm max}(\Omega/\Omega_K)$, where $\Omega_K = [GM(<r)/r^3]^{1/2}$
and $M(<r)$ is the enclosed mass, in which case the maximum is situated typically around the base 
of the convective envelope.  We find ${\rm max}(\Omega/\Omega_K) \sim 0.08$ for an initial fiducial 
rotation rate $\Omega_{\rm fid}$ equal to ${1\over 2}$ the breakup rate, using the prescription 
for angular momentum transport given in Section \ref{sec:angular_momentum_transport}.   Performing
the same test during later burning stages (e.g. the appearance of the carbon core) shows a weaker
dependence on the local rotation rate, and so we focus on the helium core rotation rate as the 
second variable modulating the core properties near collapse.  

To explore the effect of variable core rotation, we construct a series of rotating MESA models with the 
built-in prescription for RIM and the Spruit-Tayler dynamo turned on, and the convective overshoot 
parameters normalized to the same values as in our non-rotating MESA models.  This allows us to indirectly
measure the error in pre-collapse core properties introduced by our neglect of RIM.
Figure \ref{fig:hecorespin} shows how the pre-collapse core compactness $\xi_{2.5}$ and the total
mass enclosed by the Si shell depend on ${\rm max}(\Omega/\Omega_K)$ at the first appearance of the
helium core, for $M_{\rm ZAMS} = 25\,M_\odot$.   The rotating MESA model with initial (solid-body)
rotation rate $\Omega_i = \Omega_{\rm fid}/2.5$ gives the closest correspondence with the post-processed, 
non-rotating model.  One finds that $\xi_{2.5}$ in our post-processed model is a factor $\sim 1.5$ smaller 
than is implied by the trend of $\xi_{2.5}$ with ${\rm max}(\Omega/\Omega_K)$.  

This rotation-dependent shift in core properties is combined with a strong sensitivity to progenitor mass.
The latter effect has mainly been explored so far by neglecting the effects of RIM
\citep{ErtlJWSU2016,muller16a}, and is also seen in our MESA models.   The implication here
is that RIM will {\it shift} the ZAMS mass that produces a given value of $\xi_{2.5}$.  
To gauge the magnitude of this shift, we plot in Figure \ref{fig:xi25} the pre-collapse core compactness for
a range of $M_{\rm ZAMS}$, and both $\Omega_i = 0$ and $\Omega_i = \Omega_{\rm fid}/2.5$.  One sees
that models producing the same $\xi_{2.5}$ are shifted in initial mass by 1-2$\,M_\odot$.  The rapid
variations seen in pre-collapse compactness as a function of $M_{\rm ZAMS}$,
when combined with the effects of RIM, imply an intrinsic fuzziness in the relation between $M_{\rm ZAMS}$ 
and post-collapse core properties of this order.

\section{Magnetic Helicity Accumulation} \label{sec:helicity}

The inner parts of all our stellar models pass through multiple convective phases, which leave behind
radiative material as they contract.  There are two main contributions to ${\cal H}$ in these growing
radiative layers \citep{KissT2015b}.  The flux of magnetic helicity across the convective-radiative boundary
is proportional to the magnetic torque which acts against growing differential rotation.  A minimal source
of differential rotation comes from the gradually changing mass profile of the star.  But a stronger source
is provided by inhomogeneous latitudinal rotation within the convective layer.  The latitudinal gradient in 
rotation sources a radial gradient on the opposing (radiative) side of the boundary, where the magnetic
field tends to enforce solid rotation.  This process generates a toroidal magnetic field, and is
a component of the hydromagnetic dynamo operating near the convective-radiative boundary.

The helicity is a conserved topological charge that, absent a boundary flux of
magnetic twist, can only decay on a long resistive timescale.   Net helicity is needed to stabilize
the magnetic field in a radiative layer.  The generic helical field configuration involves a finite open poloidal flux
$\Phi_r$ that is surrounded by a twisted toroidal loop carrying flux $\Phi_\phi \sim {\cal H}/\Phi_r$ \citep{BraiS2004}.  
The toroidal flux is confined by loops of poloidal field that close within the radiative material.  We assume, 
following \cite{KissT2015b}, that the twisted magnetic field isotropizes as the helicity accumulates, so that 
$\Phi_r \sim \Phi_\phi \sim {\cal H}^{1/2}$.   

The first convective phase is encountered during the pre-MS evolution, as the accreting massive star transitions
from a fully convective state to the MS configuration of a radiative envelope and a convective core
(Figure \ref{fig:conv_zones_pre_MS_25M_o}).  Net magnetic helicity is left behind in the inner $\sim 8$-9$\,M_\odot$
of material because the convective envelope offers an escape route for the compensating helicity, e.g. 
via flaring activity. The remainder of the star is assembled after it has developed a radiative envelope,
and its magnetization is determined by the interaction with the accretion disk.  We (somewhat arbitrarily)
set the magnetic field threading this outer material to zero.   

We now describe the process of magnetic helicity pumping in a general way, and then turn to describe 
specific episodes of convective retreat in more detail.  We first need an estimate of the helicity
that would be stored in a shell adjacent to the convective-radiative boundary (at radius $R_b$).  
The shell thickness is determined by the Lagrangian speed $v_{\rm b}$ of the boundary through the stellar material.
The helicity accumulated in a single hemisphere over a time $\delta t$ is then
\be
\delta{\cal H} \sim \pi B_r B_\phi R_{\rm b}^3 v_{\rm b}\delta t,
\ee
assuming that the magnetic field maintains a constant shape and strength over this interval.
See Equation (27) of \cite{KissT2015b}.   

The strength of the magnetic field in the dynamo layer is estimated by i) relating the toroidal field to
the poloidal field through the linear winding term in the induction equation; and ii) assuming that the poloidal
field is just strong enough to transfer angular momentum across a distance $\sim l_P$ over the dynamo period.  Then
\be\label{eq:Max}
{B_rB_\phi\over 4\pi} \sim {\rho l_P R_{\rm b} (\Delta\Omega)^2\over 2\pi N_{\rm dyn}} = \varepsilon_B \rho\Omega^2.
\ee
Here $\Delta\Omega$ is the mismatch between the (nearly uniform) angular velocity of the radiative layer,
and the angular velocity on the opposing convective side of the boundary.  We take the dynamo period to be 
$N_{\rm dyn}\cdot\Delta\Omega^{-1}$, with $N_{\rm dyn} = 10^2$.  Further assuming that 
$\Delta\Omega \sim \Omega$, as appropriate for a slowly rotating and deeply convective layer 
\citep{BrunP2009}, one finds $\varepsilon_B \sim 10^{-3}$.

The integral of the helicity flux over latitude does not generally vanish within a {\it single hemisphere},
and maintains a uniform sign as long as the pole-equator angular velocity difference also maintains a constant sign
\citep{KissT2015b}.  There is, however, a cancellation between hemispheres if the magnetic field
is reflection symmetric about the rotational equator.  This cancellation can only be approximate if the rotation
near the convective boundary is sustained by a modest number of deeply penetrating plumes.  The strength of the cancellation
is normalized here to $N_{\rm dyn}^{-1} \sim 0.01$ over a dynamo period, with a sign that varies stochastically over 
multiple cycles.  

Then the net helicity deposited over the dynamo period $P_{\rm dyn}$ is
\be\label{eq:netH}
\delta{\cal H}_{\rm dyn}  \sim {\varepsilon_B\over N_{\rm dyn}} \cdot\pi \delta M_{\rm dyn} \Omega^2(R_{\rm b}) R_{\rm b}^3.
\ee
The mass of radiative material added over the dynamo period is $\delta M_{\rm dyn} = 4\pi R_{\rm b}^2 v_{\rm b} 
P_{\rm dyn} \rho(R_b)$.
Given that the sign of the imbalance between hemispheres fluctuates randomly over multiple dynamo periods, the net
helicity accumulated over an interval $\delta t \gg P_{\rm dyn}$ is
\be
\delta({\cal H}^2) \sim {\delta t\over P_{\rm dyn}} (\delta {\cal H}_{\rm dyn})^2.
\ee
This expression is easily implemented in a sequence of MESA model snapshots (labelled $i$), each of age $t_i$:
\ba\label{eq:helicity}
\delta({\cal H}^2) &=& \sum_i {t_i-t_{i-1}\over P_{{\rm dyn},i}} (\delta {\cal H}_{\rm dyn})^2; \nn
       \delta {\cal H}_{\rm dyn} &\equiv& {\pi\varepsilon_B\over N_{\rm dyn}}
       \left[P_{{\rm dyn},i}{M_{{\rm rad},i}- M_{{\rm rad},i-1}\over t_i-t_{i-1}}\right] \Omega^2(R_{{\rm b},i}) R_{{\rm b},i}^3.\nn
\ea
This expression is independent of the time spacing between snapshots, as long and this spacing is fine enough to
resolve the large-scale changes in convective structures.

The normalization of the helicity flux given by Equation (\ref{eq:helicity}) is supplemented by a threshold condition
for the Coriolis parameter, measured close to the radiative-convective boundary.  When ${\rm Co} < {\rm Co}_{\rm crit}$,
the dynamo is shut off and the helicity flux vanishes.  In this paper we take ${\rm Co}_{\rm crit} = 0.1$.

We will need to ascribe a radial magnetic flux $\Phi_r$ to each radial zone in each snapshot.  As the radiative material
grows, we do this by summing $\delta({\cal H}^2)$ over the last shell of thickness $l_P$ added to the zone.  Thus
$\delta({\cal H}^2) \sim \Phi_r^4$ is the Lagrangian variable which follows each radiative mass shell, and which is modified by
successive convective structures.  The mass in this shell is $\delta M_{l_P} = 4\pi R_{\rm b}^2 \rho(R_{\rm b}) l_P$, meaning that
\be
{\delta({\cal H}^2)_{l_P}\over (\delta{\cal H}_{\rm dyn})^2} = {\delta M_{l_P,i}\over M_{{\rm rad},i}-M_{{\rm rad},i-1}}
      {t_i-t_{i-1}\over P_{{\rm dyn},i}}.
\ee
The sum (\ref{eq:helicity}) is recovered from the Lagrangian variable $\delta({\cal H}^2)_{l_P}$ by summing over
the mass shells (labelled $j$) of the radiative material:
\be
\delta({\cal H}^2) = \sum_j \delta({\cal H}^2)_{l_P,j} {r_j-r_{j-1}\over\ell_{P,j}}.
\ee
The hemispheric flux threading a single mass shell is estimated as
\be\label{eq:phirloc}
\Phi_r \sim \left[\delta({\cal H}^2)_{l_P}\right]^{1/4}.
\ee

Several pressure scale heights of material (number $N_P$) are mixed together when convection is excited in the core during
hydrogen and helium burning.  The total helicity contained in the core is ${\cal H}_{\rm con} \sim N_P^{1/2}
[{\cal H}^2_{l_P}]^{1/2}$, and the flux increases after mixing to
\be\label{eq:phicon}
\Phi_r \rightarrow {\cal H}_{\rm con}^{1/2} \sim N_P^{1/4}\left[\delta({\cal H}^2)_{l_P}\right]^{1/4}.
\ee

\begin{figure}
\epsscale{1.25}
\plotone{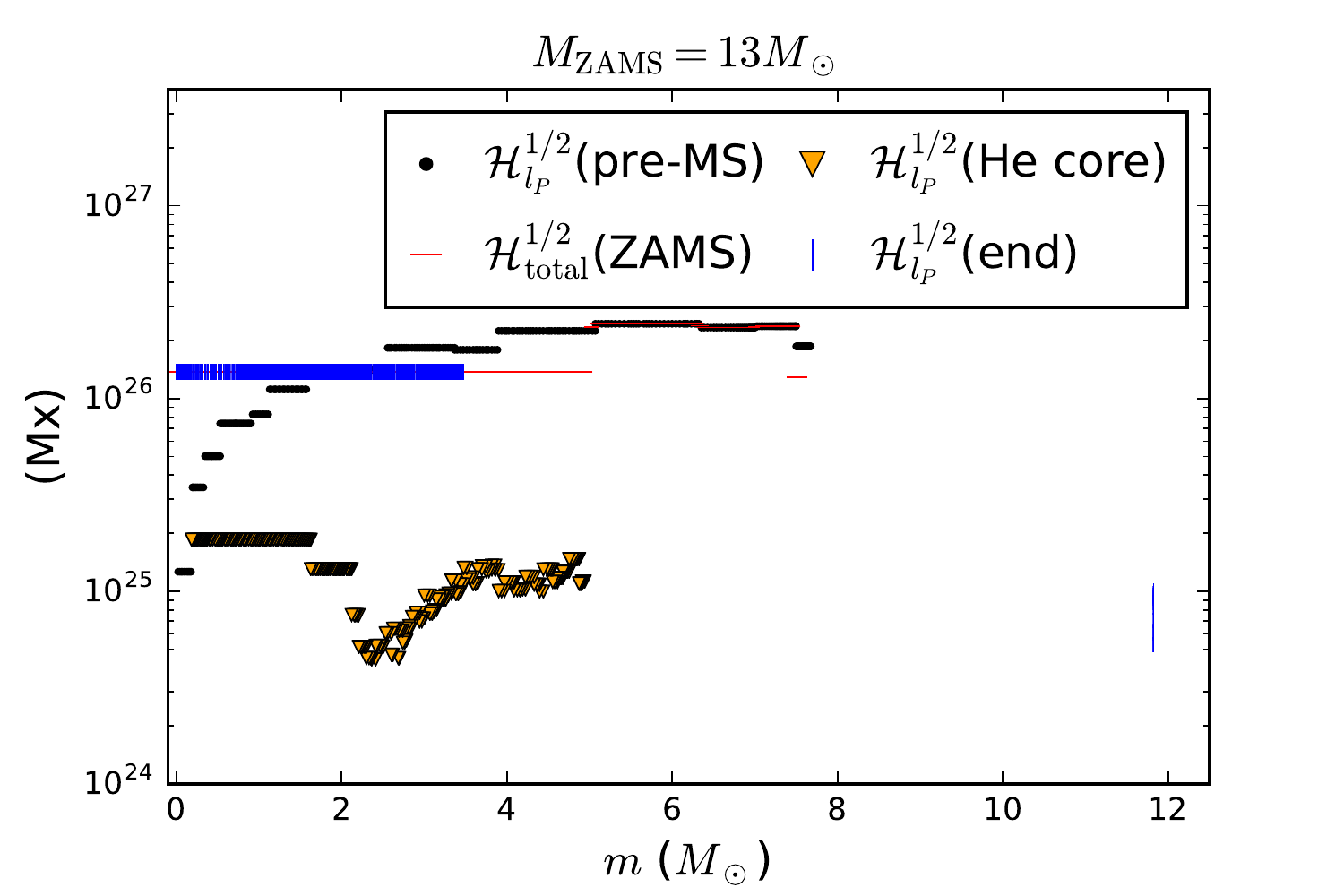}
\caption{Hemispheric poloidal magnetic flux, as measured by the square root of the magnetic helicity accumulated within
a pressure scale height during the transition from a convective to a radiative state (Equation (\ref{eq:phirloc})).  
Points show contributions from different evolutionary phases of our 13$\,M_\odot$ model.  
Black dots:  helicity left behind as the convective envelope recedes during the pre-MS contraction.  
Red horizontal dashes:  smoothed helicity profile in the convective ZAMS H-burning core
(Equation (\ref{eq:phicon})).  Green triangles: contribution from the receding He-rich 
convective core at the end of the MS.   Blue vertical ticks:  profile of 
${\cal H}^{1/2}$ right before core collapse.}
\vskip .2in
\label{fig:deposited_flux_13M_o}
\end{figure}

\begin{figure}
\epsscale{1.25}
\plotone{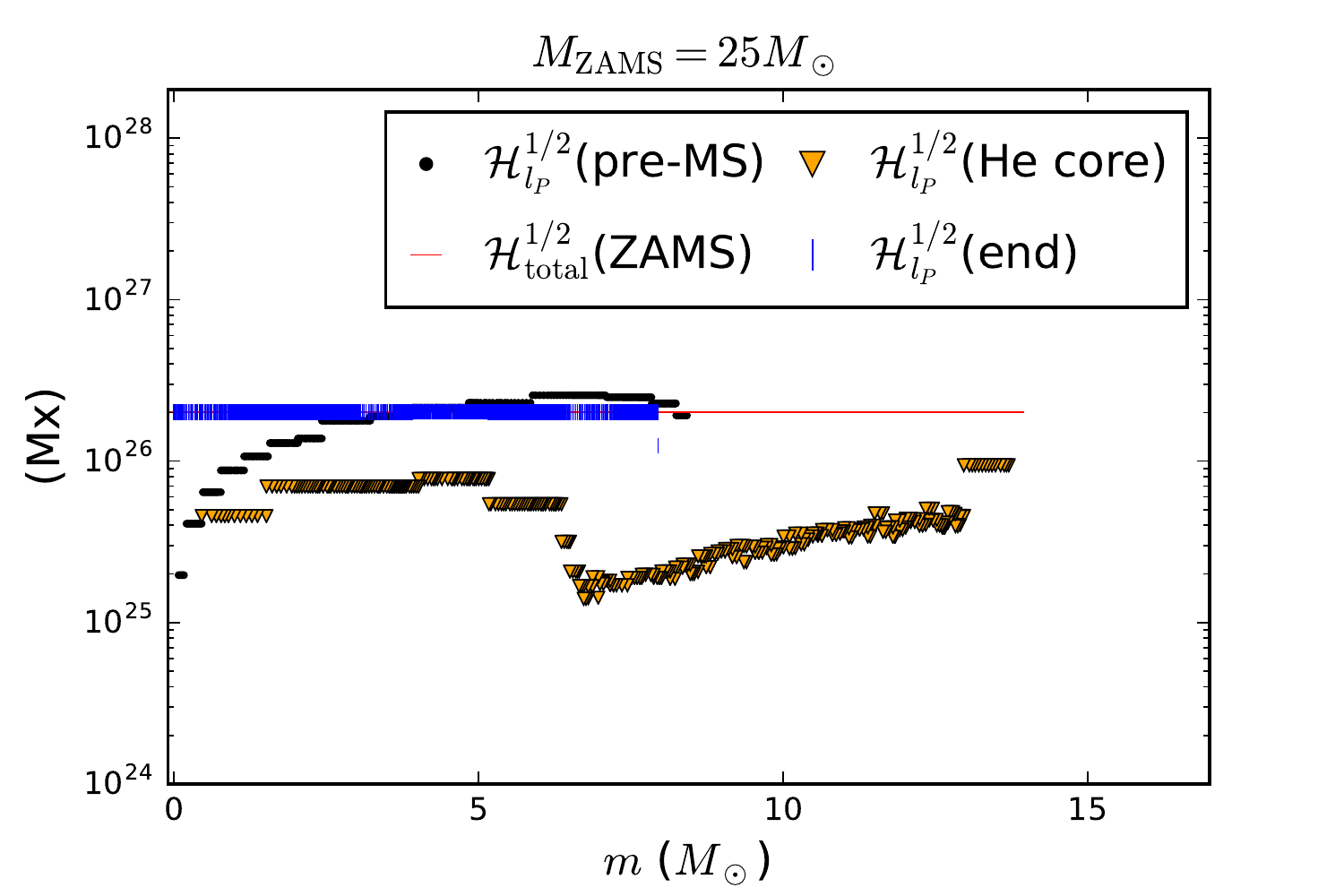}
\caption{Same as Figure \ref{fig:deposited_flux_13M_o} but for the 25$\,M_\odot$ model.}
\vskip .2in
\label{fig:deposited_flux_25M_o}
\end{figure}

\begin{figure}
\epsscale{1.25}
\plotone{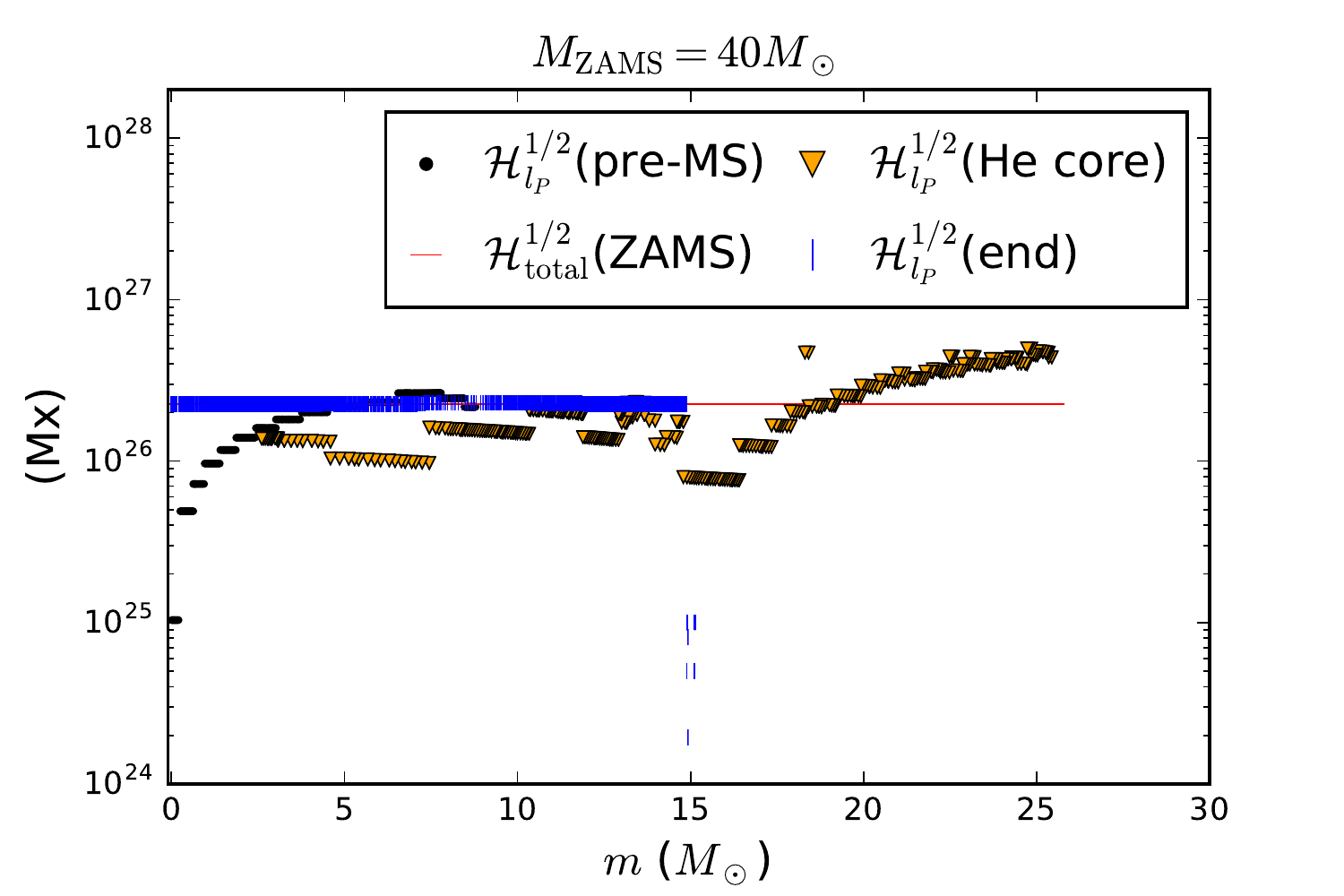}
\plotone{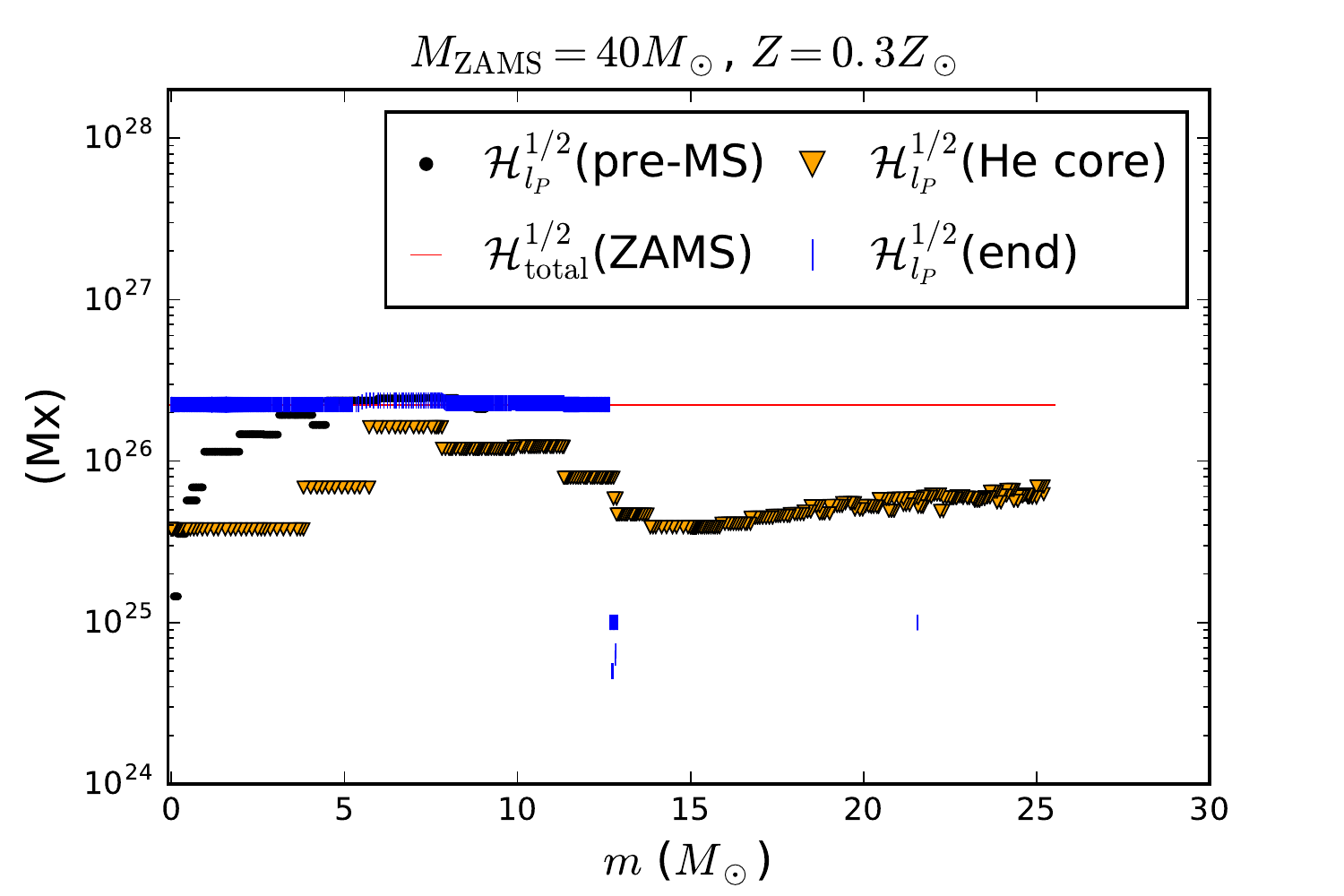}
\caption{Same as Figure \ref{fig:deposited_flux_25M_o} but for the 40$\,M_\odot$ model
with solar metallicity (top panel) and $Z = 0.3\,Z_\odot$ (bottom panel).}
\vskip .2in
\label{fig:deposited_flux_40M_o}
\end{figure}

\subsection{Successive Convective Structures}

We calculate the contribution to ${\cal H}$ from three successive convective episodes:  i) the transition from a convective
to a radiative envelope during the pre-MS phase; ii) the contraction of the He-rich convective core at the end of the MS; and
iii) the contraction of the C-rich convective core at the end of core He burning.  The first, pre-MS, contribution dominates in all the models
we consider.  Figures \ref{fig:deposited_flux_13M_o}-\ref{fig:deposited_flux_40M_o}
show the effective hemispheric magnetic flux, as corrected by convective mixing where appropriate (Equation \ref{eq:phicon}),
and the flux distribution at the moment of core collapse.

The relative importance of the pre-MS contribution to ${\cal H}$ arises from the fast recession of the envelope:
in the 25$\,M_\odot$ this takes place in $\Delta t \sim 3\times 10^4$ yr, as compared with 
$\sim 7 \times 10^6$ yr for the recession of the convective core at the end of the MS.  The cancellation factor
$\sim P_{\rm dyn}/\Delta t$ in ${\cal H}^2$ is much stronger in the later phase. 

We now summarize some of the finer details of this process.

{\it End of Core He Burning.}
Even though the recession of convective He burning core has the potential to add helicity
to the inner 2-12$\,M_\odot$ (depending on the ZAMS mass), in all of our models the Coriolis parameter during
this phase is $< 0.1$.  We therefore shut off this contribution.

{\it Multiple convective shells.}
The complex convective behavior of massive stars creates numerous instances of receding convection zones, which 
we cannot track individually.  To account for this contribution, we arbitrary add a term ($10^{25}\,{\rm Mx})^4$
to ${\cal H}^2$.  This turns out to have a negligible effect on the total helicity and angular momentum transport.

{\it Expulsion of helicity from a slowly rotating convective envelope.}
We sometimes find that the entire supergiant envelope is slowly rotating, meaning that it cannot sustain an
active dynamo but still can lose magnetic helicity through the stellar surface, especially during strong mass loss.
We allow this ejection process to occur if the radiative buffer at the surface is thinner than five pressure scaleheights.
The effects of surface helicity ejection can be seen in Figures \ref{fig:deposited_flux_13M_o}-\ref{fig:deposited_flux_40M_o},
which show a radial cutoff in helicity marking the deepest penetration of the slowly rotating convective envelope.

\vfil\eject
\section{Combined Magnetic and Rotational Evolution to Collapse} \label{sec:combined_evol}

The mechanisms of angular momentum transport described in Section \ref{sec:angular_momentum_transport} work
in concert with the mechanism of magnetic helicity deposition outlined in Section \ref{sec:helicity}.  
In this section, we contrast the behavior of the various stellar models, as manifested especially by the 
magnetization and rotation of the post-collapse remnants.

The rate of helicity deposition in growing radiative layers depends on the star's angular velocity profile and
history of angular momentum loss.   Conversely, the embedding of magnetic helicity facilitates a
rotational coupling between core and envelope, through intermediate radiative and semi-convective zones.

\begin{figure}
\epsscale{1.25}
\plotone{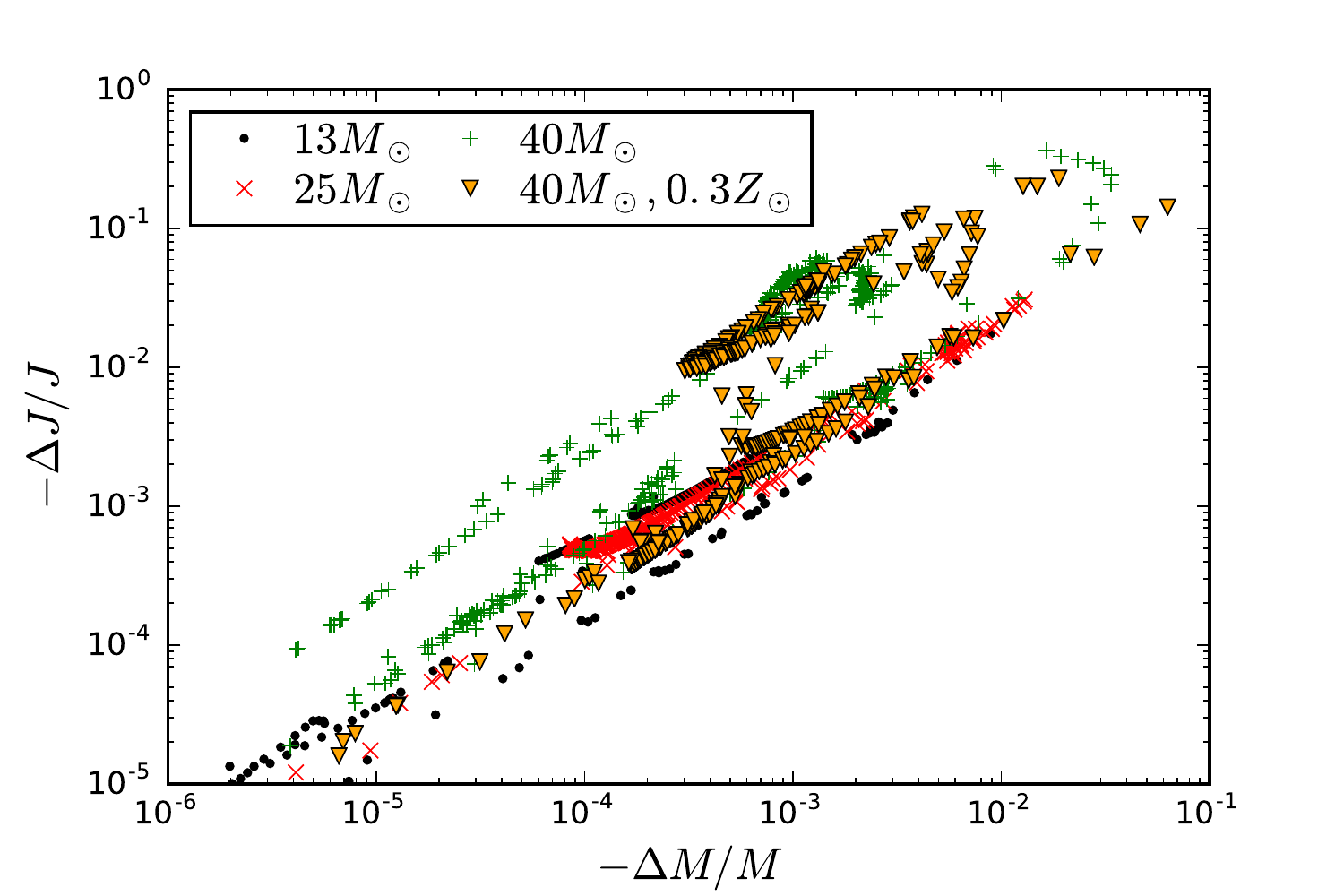}
\caption{Loss of angular momentum to a wind (as measured between successive MESA model snapshots) as a function of the wind
mass ejected from the star, for the four stellar models considered.  Stronger angular momentum loss
is found in the 40$\,M_\odot$ models, due to weaker inward pumping of angular momentum in a surface convective layer.}
\vskip .2in
\label{fig:delta_J_ratio_vs_delta_M_ratio}
\end{figure}

\subsection{Dependence on Effective Temperature} \label{sec:mass_T_dependence}

Both 40$\,M_\odot$ models become blue supergiants ($T_{\rm eff} \sim 8$-9000 $\K$), and we find that they end up with far
less angular momentum than the lower-mass models, which expand to become red supergiants ($T_{\rm eff} \sim 4000\,\K$).
Figure \ref{fig:delta_J_ratio_vs_delta_M_ratio} shows the relation between the angular momentum and mass
carried away by the stellar wind.  During the supergiant phase the rotation is slow enough that 
$\Omega(r) \propto r^{-2}$ throughout most of the envelope.  This means that the depth of the envelope has a strong
influence on the angular momentum stored close to the surface. The result is specific angular momentum several
times higher near the surfaces of blue supergiants.

It is worth emphasizing that this result does not depend on the detailed reasons why both 40$\,M_\odot$ models
fail to become a red supergiant. 
(See \citealt{WoosW1995} and \citealt{Ekstetal2012} for a summary of current thinking on this issue.)  
The stronger wind mass loss may be responsible.  Metallicity does not appear to be the key variable, and we are
not including rotationally driven mixing effects, or (in this section of the paper) allowing for angular momentum
injection from a binary companion.  The handling of convective overshoot may also be relevant, but a full exploration
of its effects is beyond the scope of this paper.

\begin{figure}
\epsscale{1.25}
\plotone{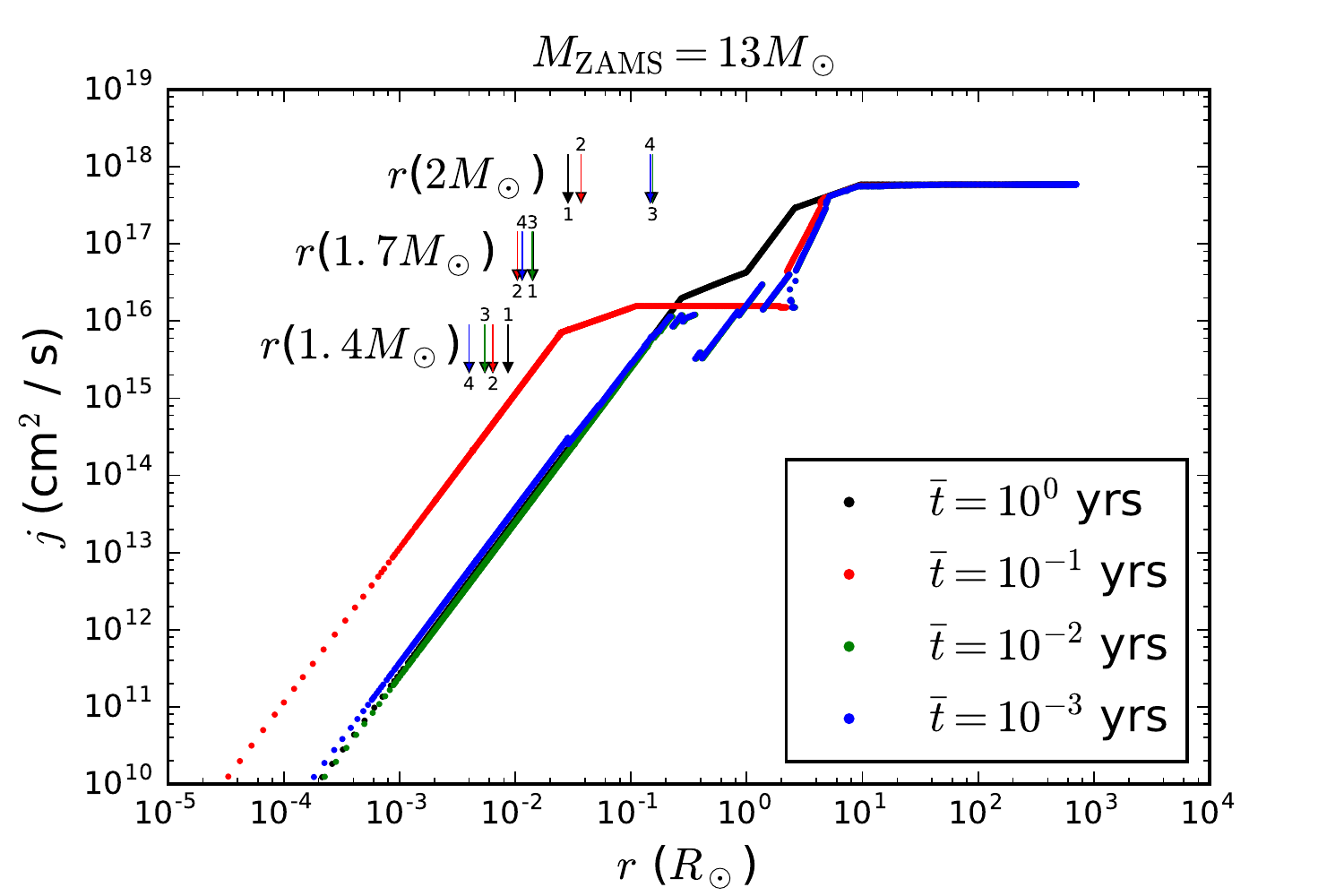}
\plotone{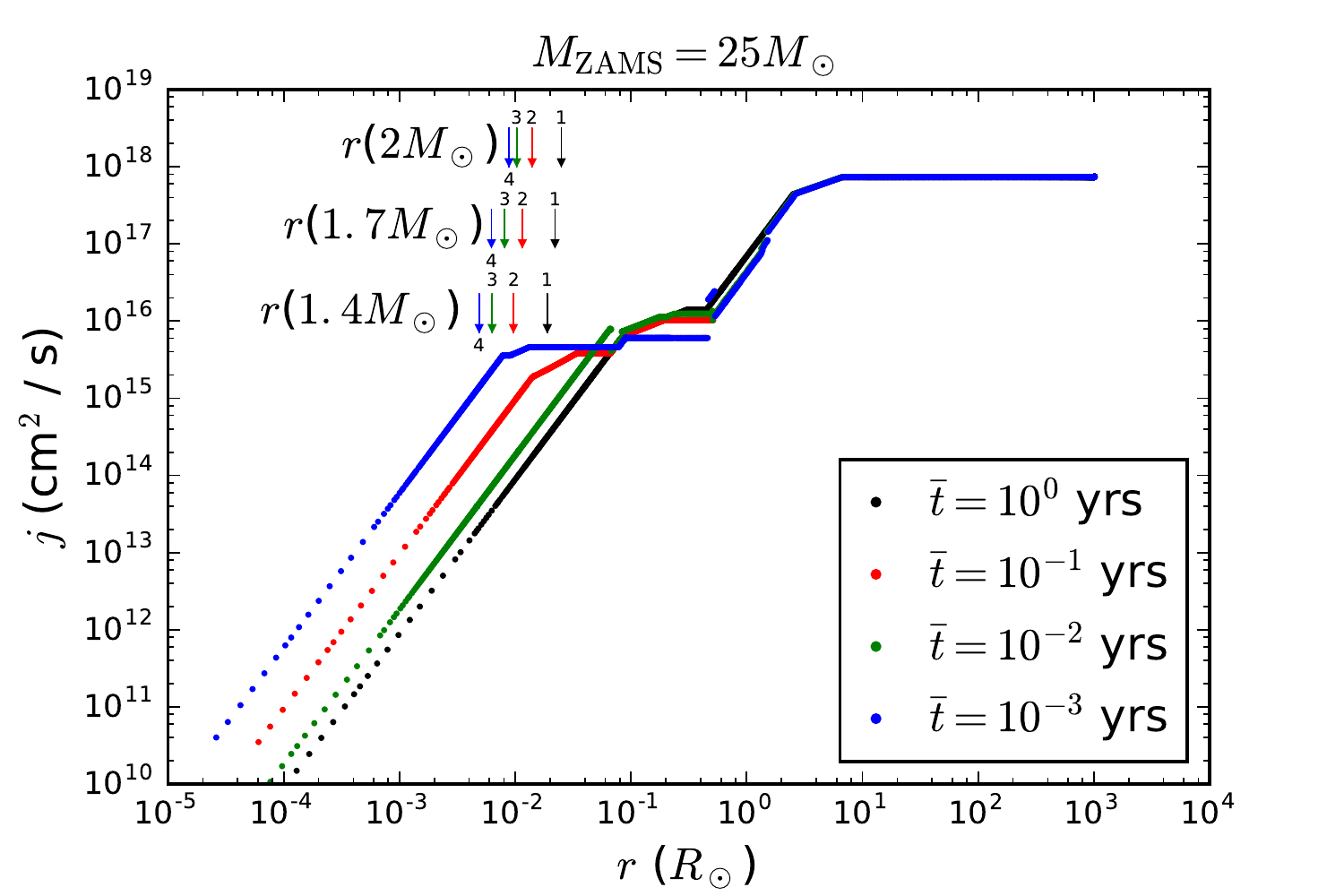}
\caption{{\it Top panel:}  Specific angular momentum profile in the 13$\,M_\odot$ model at four separate times 
$\bar{t} = 10^{0},10^{-1},10^{-2}, 10^{-3}$ yr before core collapse.   Flat parts of the curves correspond
to convective layers with ${\rm Co} < 1$, and quadratic parts ($j \propto r^2$) correspond to radiative layers
with nearly solid rotation.  Arrows (colored by the time to core collapse) mark the radii at
an enclosed mass $m=1.4$, 1.7 and 2$\,M_\odot$.  The spike in inner rotation seen at $\bar{t} = 10^{-1}$ yr
is caused by the development of a short-lived thick convection zone, at $m \sim 2$-$3.4\,M_\odot$.  (See for
comparison Figure \ref{fig:conv_zones_complete_13M_o}.)  By $\bar{t} = 10^{-2}$ yr this convection zone disappears,
and the inner rotation slows down.  {\it Bottom panel:} Corresponding figure for the 25$\,M_\odot$ model.  Now
the burning shells support much thicker and more persistent convective layers, as compared with the 13$\,M_\odot$
model.  The result is a significantly higher rotation rate in the inner core.}
\vskip .2in
\label{fig:j_profile}
\end{figure}

\subsection{Dependence on Stellar Mass} 

A significant difference in the rotation rate of the inner core is observed between the 13$\,M_\odot$ and 25$\,M_\odot$
models:  the center of the lower-mass model rotates much more slowly.   To explain this result, we show in Figure
\ref{fig:j_profile} the angular momentum profiles at different times preceding core collapse. 
The burning shells develop much thicker convective layers in the 25$\,M_\odot$ model.  They are slowly rotating enough
to be effective inward pumps of angular momentum, with a large ratio of inner to outer angular velocities.   These
convective layers correspond to the flat parts of the specific angular velocity curves in Figure \ref{fig:j_profile}.
As a guide, we mark off the radii of several enclosed mass coordinates ($m = 1.4$, 1.7 and 2$\,M_\odot$).
Close to core collapse, the specific angular momentum is about two orders of magnitude larger at $m=1.4\,M_\odot$ in the 
25$\,M_\odot$ model than in the 13$\,M_\odot$ model.

\subsection{Neutron Star Remnants: Rotation \\ and Dipole Magnetic Flux} \label{sec:NS_remnants}

Consider now the spin period $P_{\rm NS}$ of a NS that forms from a collapsed baryonic mass $M_{\rm col}$.
This depends on $M_{\rm col}$ as well as on the angular momentum profile of the pre-collapse core.   
A first estimate of $M_{\rm col}$ is obtained from the observation that the infall of the oxygen shell
causes an outward expansion of the post-collapse standing shock, which when combined with neutrino heating
can drive an explosion (e.g. \citealt{muller16b}).  
Taking $M_{\rm col}$ to be the mass enclosed by the silicon shell $+\,0.1\,M_\odot$, we obtain
$\sim 1.7\,M_\odot$ for the 13$\,M_\odot$ model and $\sim 2.0\,M_\odot$ for the 25$\,M_\odot$
model.  Figure \ref{fig:j_profile} shows that the lever arm is raised by a factor $\sim 2$, and the specific angular
momentum by a factor $\sim 4$, when moving outward from $m = 1.4$ to 1.7$\,M_\odot$ in the 13$\,M_\odot$ core
at a short interval ($\sim 10^{-3}$ yr) before core collapse.  Increasing the enclosed mass from 1.4 to 2$\,M_\odot$
raises the specific angular momentum by approximately the same factor in the 25$\,M_\odot$ pre-collapse core.

\begin{figure} 
\epsscale{1.2}
\plotone{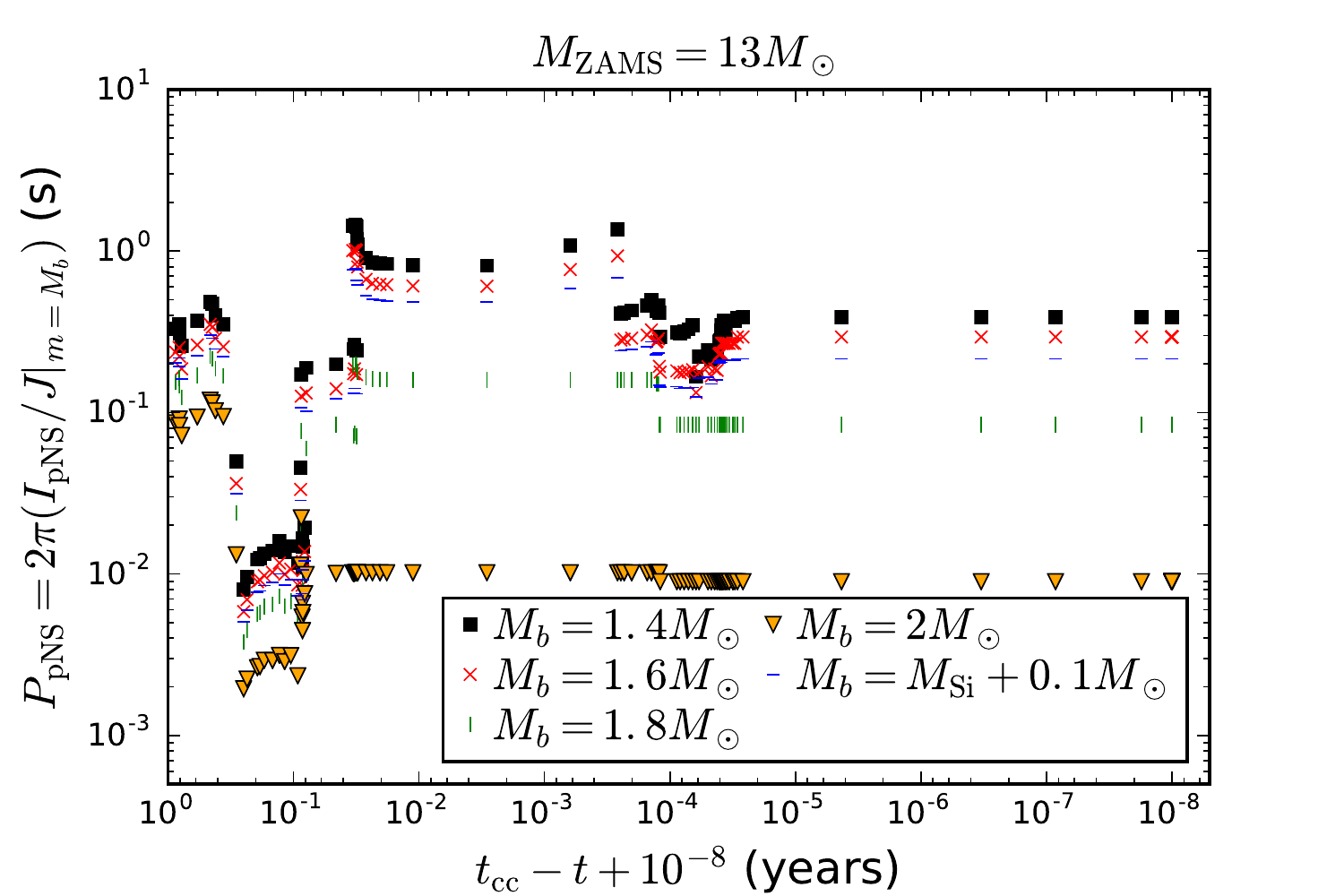}
\plotone{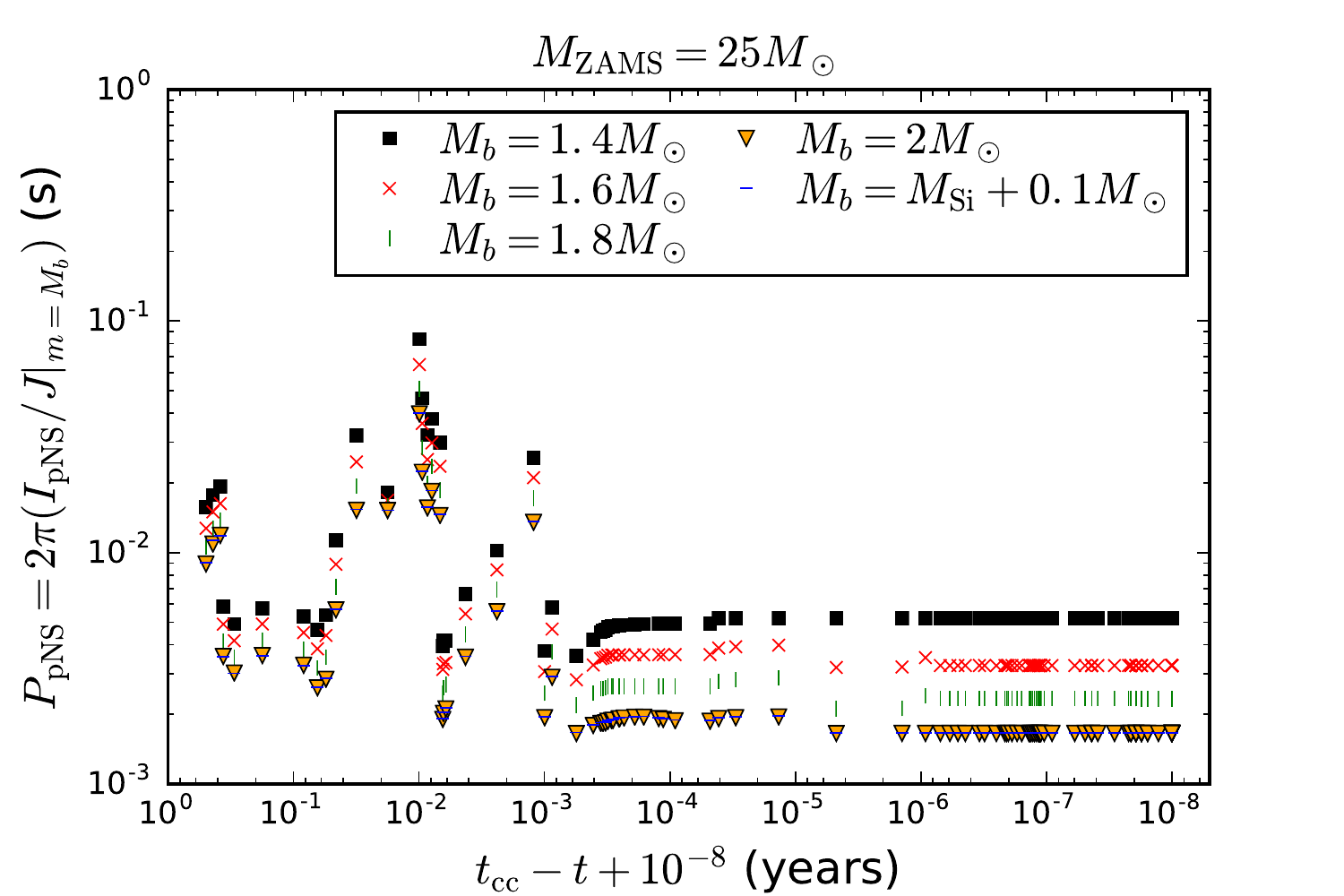}
\caption{{\it Top panel:}  Evolving rotation profile of the core of the 13$\,M_\odot$ model, 
as measured by the equivalent rotation period of a collapsed NS remnant, during the final year before core collapse.  
Points show different collapsed baryonic masses $M_b$ extending from 1.4$\,M_\odot$
to 2$\,M_\odot$.   Horizontal blue dashes show the mass $M_{\rm Si}$ enclosed by the Si burning shell $+ 0.1\,M_\odot$,
to represent the collapsed mass that might power a successful explosion.  Here $M_{\rm Si} + 0.1\,M_\odot \sim 1.7\,M_\odot$.
{\it Bottom panel:}  Equivalent results for the 25$\,M_\odot$ model (here $M_{\rm Si} + 0.1\,M_\odot \sim 2.0\,M_\odot$).}
\vskip .2in
\label{fig:P_pNS_13M_o}
\end{figure}

\begin{figure} 
\epsscale{1.2}
\plotone{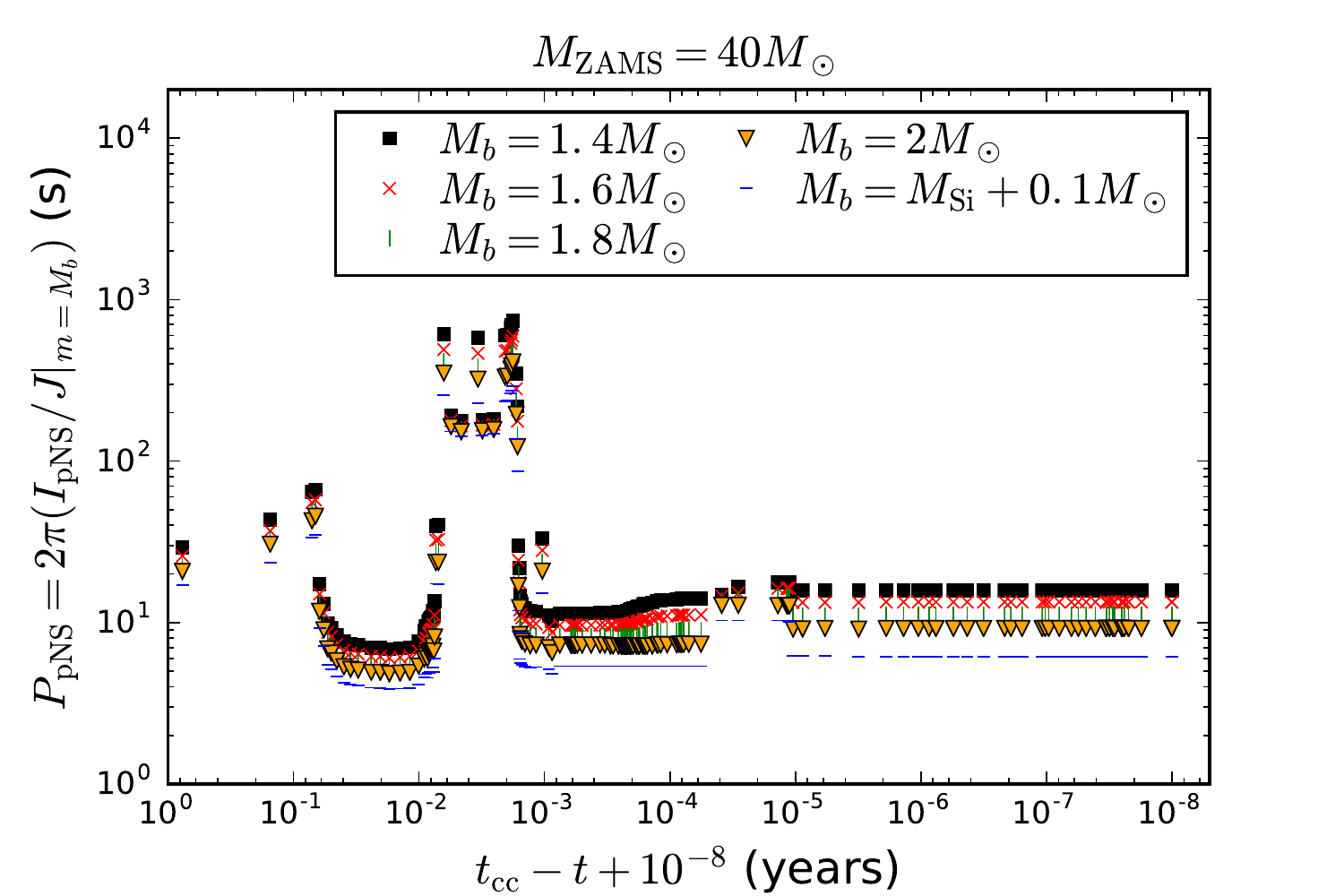}
\plotone{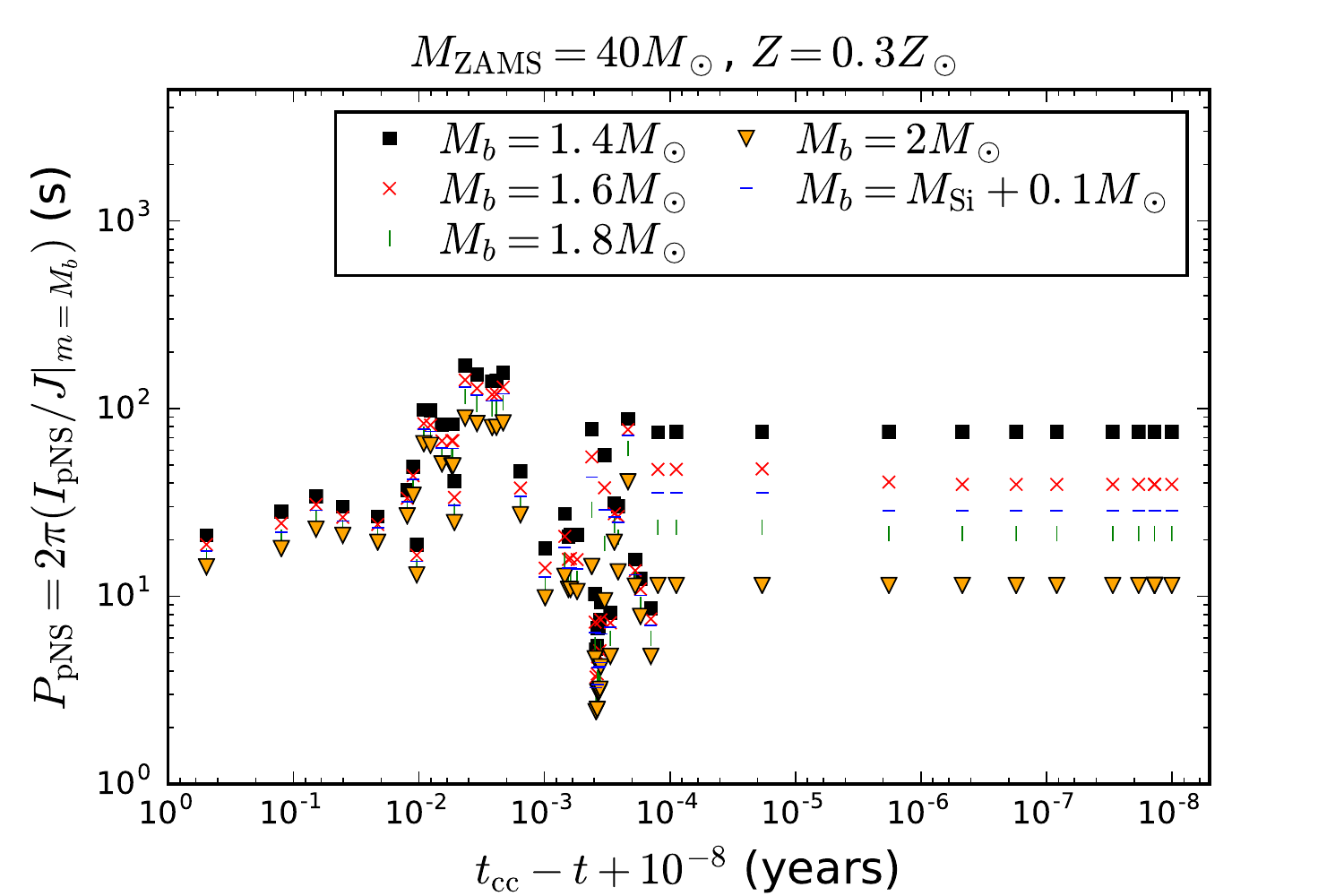}
\caption{Similar to figure \ref{fig:P_pNS_13M_o}, but for our 40$\,M_\odot$ models.   
$M_{\rm Si}+0.1\,M_\odot \sim 2.4\,M_\odot$ in the $Z_\odot$ model (upper panel) and $1.7\,M_\odot$ in the $0.3\,Z_\odot$ model (lower panel).  The solar metallicity star is expected to
collapse to a BH, but for ease of comparison we continue to use the NS rotation period as a probe of the core rotation
profile.  More realistically the points toward the right side represent the NS rotation at an intermediate stage
of the collapse to a BH (although in that case the proto-NS is 2-3 times larger than the 10 km radius assumed here,
and the spin period 4-9 times longer).  The greatly reduced rotation rate obtained in these models is a consequence
of the strong angular momentum loss to a wind.}
\vskip .2in
\label{fig:P_pNS_40M_o}
\end{figure}

The dependence of $P_{\rm NS}$ on $M_{\rm col}$ is shown in Figures \ref{fig:P_pNS_13M_o} and \ref{fig:P_pNS_40M_o}.
Here the evolving angular momentum profile is probed by assuming contraction to a cold neutron star
over a range of times in advance of the actual core collapse.  The angular momentum of the collapsed mass 
is equated with $2\pi P_{\rm NS}^{-1} I_{\rm NS}$.  The cold NS moment of inertia\footnote{The energy
lost to neutrinos is neglected here, meaning that the gravitational mass of the NS is slightly overestimated; but
in compensation our choice of $R_{\rm NS}$ is slightly smaller than most modern equations of state would suggest.}
is taken to be $I_{\rm NS} = 0.35M_{\rm col} R_{\rm NS}^2$, and the radius $R_{\rm NS} = 10$ km.  

The rotation of the 13$\,M_\odot$ remnant ($P_{\rm NS} \sim 0.2$ s for $M_{\rm col} = M_{\rm Si} + 0.1\,M_\odot$) 
is not atypical of pulsars:  a significant fraction of NSs appear to be born with spin periods in the range 
0.1-1 s \citep{PopoT2012}.  On the other hand, the 25$\,M_\odot$ model has a predicted spin period around 1.5 ms, 
making it a strong candidate for further dynamo amplification of the magnetic field post-collapse \citep{ThomD1993}.  
We also include, for completeness, the $P_{\rm NS}$ that would be obtained from the 40\,$M_\odot$ models if 
they were able to explode and leave behind a stable NS below the maximum mass.  As expected from the
greatly increased angular momentum loss in these models, some 2-4 orders of magnitude less angular momentum is stored 
in the core than in the lower-mass models (Figure \ref{fig:P_pNS_40M_o}).  The remnant spin is calculated
in Section \ref{sec:BH_mass_spin} in the more likely scenario of BH formation.

The blue lines in Figures \ref{fig:deposited_flux_13M_o} and \ref{fig:deposited_flux_25M_o} show the magnetic flux 
threading the hydrogen-depleted core just before collapse.  The thin red line represents the convective core on the ZAMS,
which has nearly the same amplitude but covers a wider range of mass; this demonstrates that the contribution from later
convective stages is small.   Here we assign a single number $\Phi_r = {\cal H}_{\rm con}^{1/2}$ for the polar flux, 
where ${\cal H}_{\rm con}$ is the magnetic helicity stored in the convective core.  This must overestimate the NS poloidal
flux:  only the inner $\sim 15$-$30\%$ of the core material is incorporated into the NS.   For example, if the core had a 
constant ratio of flux to mass before the collapse, then $\Phi_r$ should be corrected downward by a factor
$\sim 0.15^{2/3}$-$0.3^{2/3} \sim 0.3$-0.4.   

Applying this correction to the polar magnetic field $B_p \sim \Phi_r/\pi R_{\rm NS}^2 \sim 
{\cal H}_{\rm con}^{1/2}/\pi R_{\rm NS}^2$, we obtain $B_p \sim 2\times 10^{13}$ G in the cold NS.
This is in the upper part of the pulsar dipole field distribution:  the dipole fields of isolated
pulsars are typically $B = {1\over 2}B_p \sim (3 \times 10^{11}$ - $3 \times 10^{13}) 
R_{\rm NS,6}^{-3}$ G (\citealt{Bhatv1991}, with $R_{\rm NS,6}=R_{\rm NS}/10^6$ cm).
This translates into a hemispheric flux $\Phi_p \sim \pi B_p R_{\rm NS}^2  \sim (2 \times 10^{24}$ - $2 \times 10^{26})
R_{\rm NS,6}^{-1}$ Mx.  For comparison, \cite{HegeWS2005} used the dynamo model of \cite{Spru2002} and obtained 
a much lower flux, $\Phi_p \sim 1 \times 10^{22}$ Mx.  The implications for the origin of pulsar magnetism, including
the growth or decay of the dipole field post-collapse, are discussed further in Section \ref{s:dipole}.

\subsection{Black Hole Remnants: Mass and Spin} \label{sec:BH_mass_spin}

Black holes are expected to form during the collapse of the most massive stars, as represented by our solar-metallicity 40$\,M_\odot$ model, and possibly one or other of the 25$\,M_\odot$ model and the $0.3\,Z_\odot$, 40$\,M_\odot$ model 
(see Figure \ref{fig:explodability}).   The remnant of such a collapse may engulf a large part of the progenitor.
The fraction of the progenitor that is accreted is regulated by the spin up of the infalling material:
conserving angular momentum, it may become rotationally supported.   An energetic outflow from the black hole ergosphere
and the inner parts of an orbiting disk would expel outer mass shells which have not yet become strongly bound to the hole,
thereby limiting its mass.

As the black hole builds up by accretion, its mass $M_{\rm BH}$ and angular momentum $J_{\rm BH}$ are nearly equal
to those of the accreted precollapse core material (enclosed mass $m$).   Some loss of energy and angular momentum to
neutrinos occurs in the first part of the collapse, before the event horizon forms, but this represents a small
correction after the hole has grown to several solar masses.  We therefore neglect neutrino losses here.  The 
first mass shell that forms a centrifugally supported disk can be estimated by comparing the specific angular 
momentum $j_\star$ of the infalling material with that of the innermost stable circular orbit (ISCO).  The 
specific angular momentum in an orbit of radius $r$ is a function of $M_{\rm BH} = m$ and $J_{\rm BH} = J(m)$
\citep{fn98},
\begin{equation}
\frac{l(r)}{\sqrt{GM_{\rm BH}r}} = \frac{1 - 2j (R_g/r)^{3/2} + j^2 (R_g/r)^2}
{\sqrt{1 - 3(R_g/r) + 2j(R_g/r)^{3/2}}}.
\end{equation}
Here the spin parameter $j \equiv J_{\rm BH} c / GM_{\rm BH}^2$ and $R_g \equiv GM_{\rm BH}/c^2$.  We evaluate
$l_{\rm isco} \equiv l(R_{\rm isco})$ at the radius of the ISCO, 
\be
{R_{\rm isco}\over GM_{\rm BH}/c^2}  =  3 + Z - [(3-W)(3+W+2Z)]^{1/2},
\ee
where
\ba
    W &\equiv& 1 + (1-j^2)^{1/3} [ (1+j)^{1/3} + (1-j)^{1/3} ]\nn
    Z &\equiv& ( 3j^2 + W^2 )^{1/2}.
\ea

\begin{figure}
\epsscale{1.2}
\plotone{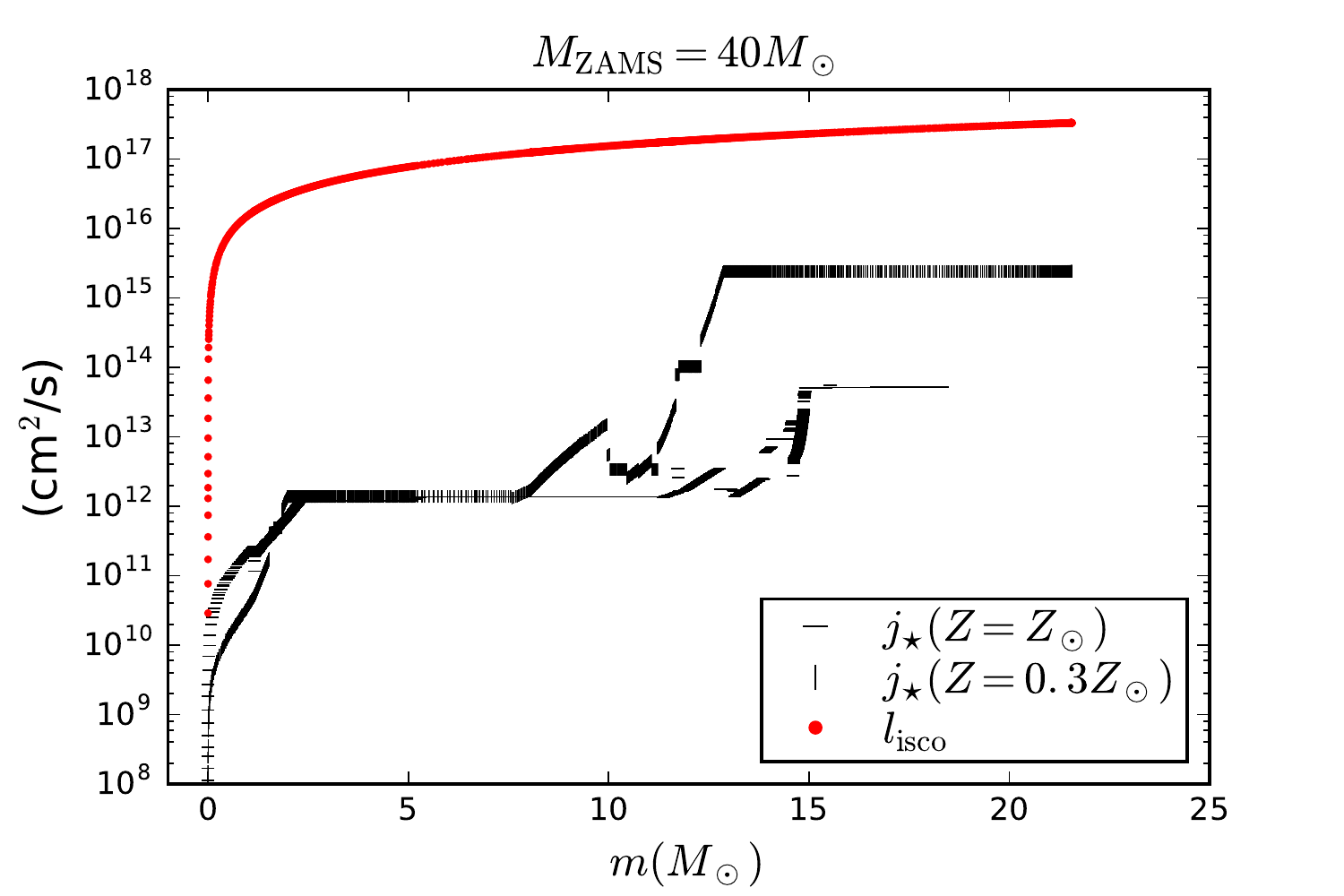}
\caption{Specific angular momentum profile in the two isolated 40$\,M_\odot$ models right before core collapse 
(red curves overlapping) compared with the angular momentum of the ISCO of a BH with the mass
and angular momentum of the enclosed material at the onset of core collapse (black curves).   
In this case the entire star will collapse into a BH, with rotation parameter 
$J_{\rm BH}c/GM_{\rm BH}^2 \sim 10^{-4}$ in the $Z_\odot$ model, and $\sim 10^{-2}$ in the 0.3 $Z_\odot$ model.}
\vskip .2in
\label{fig:specific_j_comparison_40M_o}
\end{figure}

Figure \ref{fig:specific_j_comparison_40M_o} compares $j_\star$ and 
$l_{\rm isco}$ as functions of $m$ in the 40$\,M_\odot$ models.  In both cases, so much angular momentum has been lost
to the wind that each star should almost entirely collapse into a BH (of mass 18.3$\,M_\odot$ in the $Z = Z_\odot$
model, versus 21.6$\,M_\odot$ for $Z = 0.3Z_\odot$).

\begin{figure}
\epsscale{1.2}
\plotone{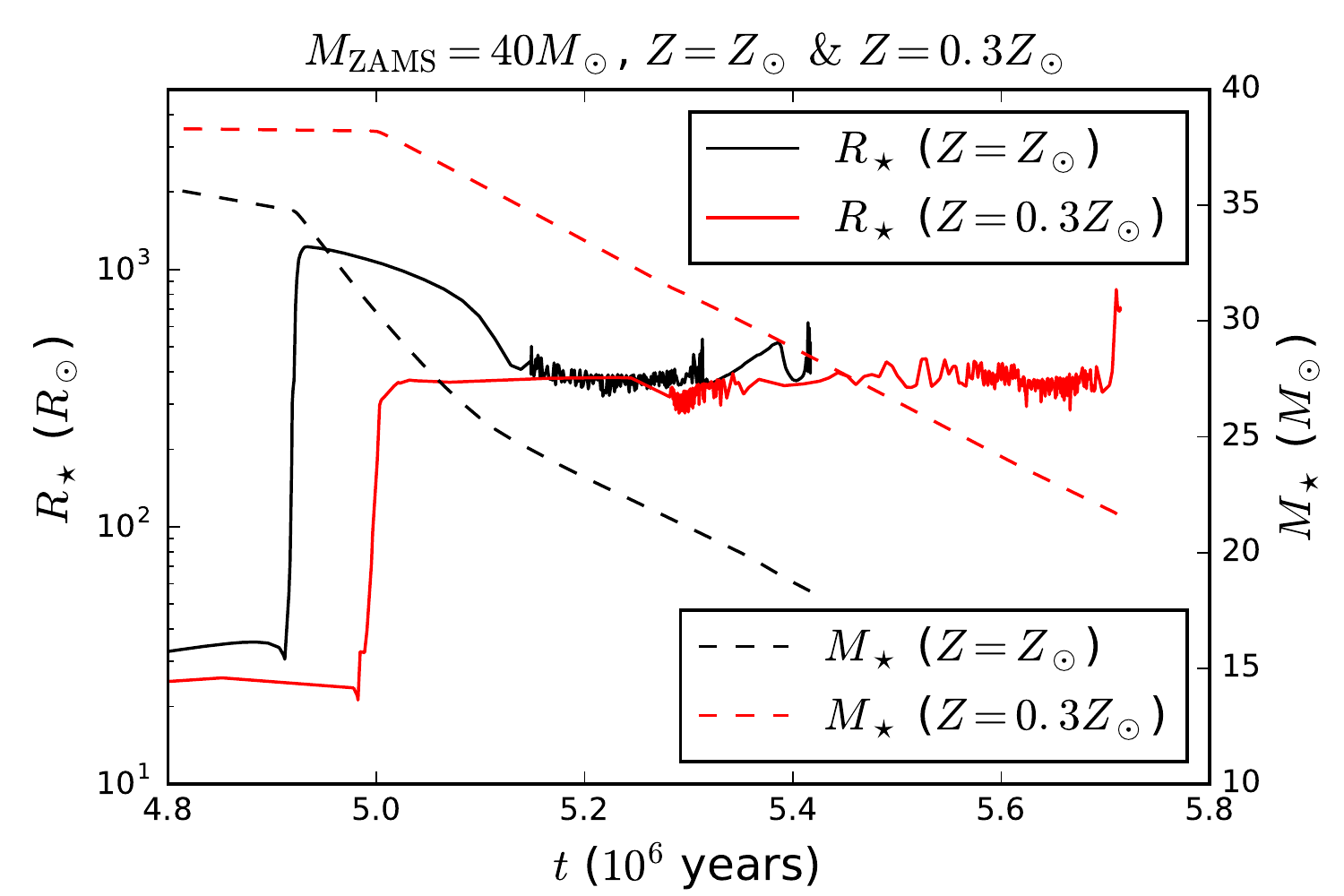}
\caption{Stellar radius and mass as a function of age for both 40$\,M_\odot$ models.  The $Z_\odot$ model 
contracts significantly (by a factor $\sim 4$ in radius) from its maximum expansion, whereas the 
$0.3\,Z_\odot$ model maintains a nearly constant size between the onset of strong mass loss and core collapse.  
In the first but not the second case, most of the angular momentum deposited from a binary companion
would be lost to the wind.}
\vskip .2in
\label{fig:R_star_M_star_vs_t_two_models_RGB_end}
\end{figure}

\begin{figure}
\epsscale{1.2}
\plotone{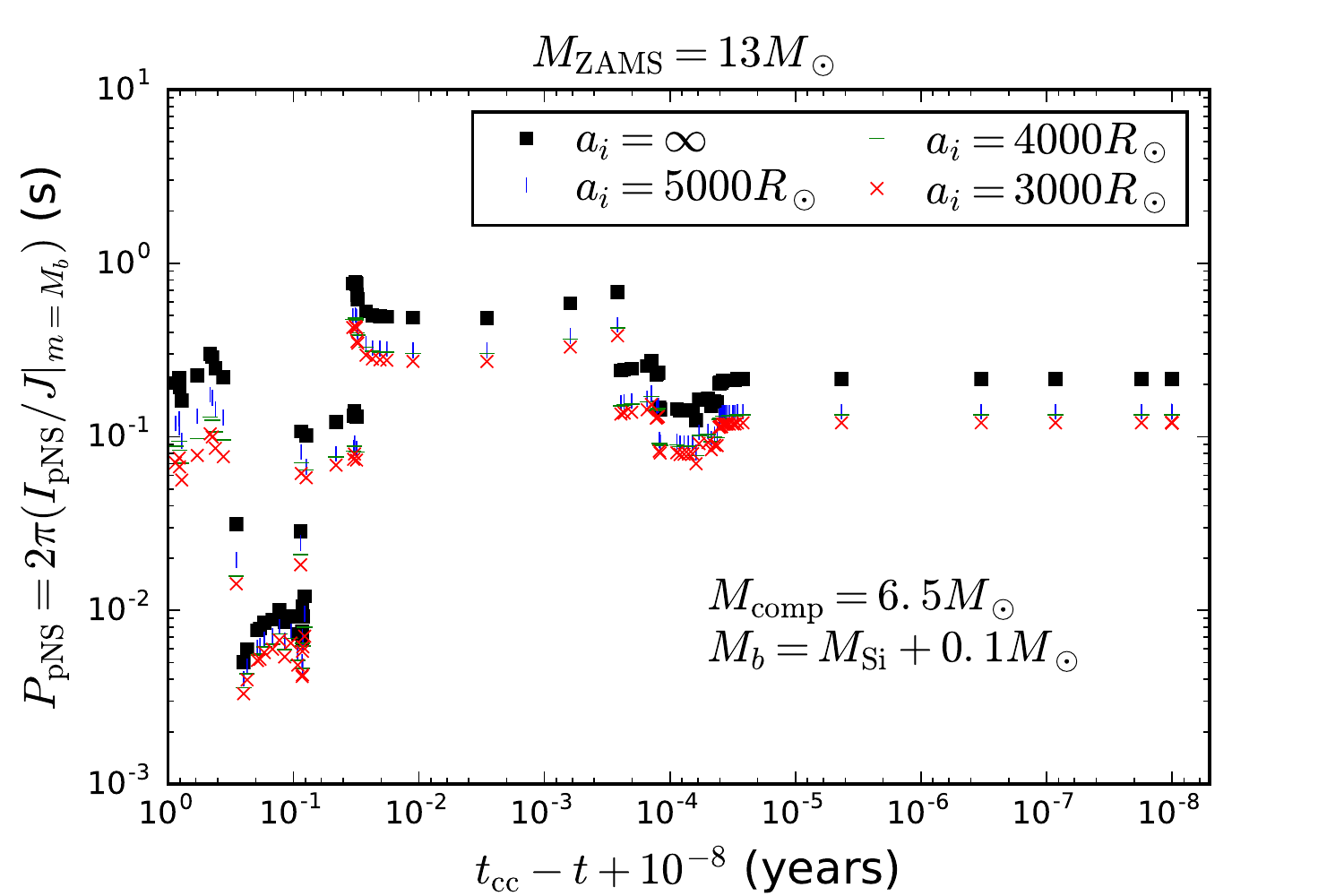}
\plotone{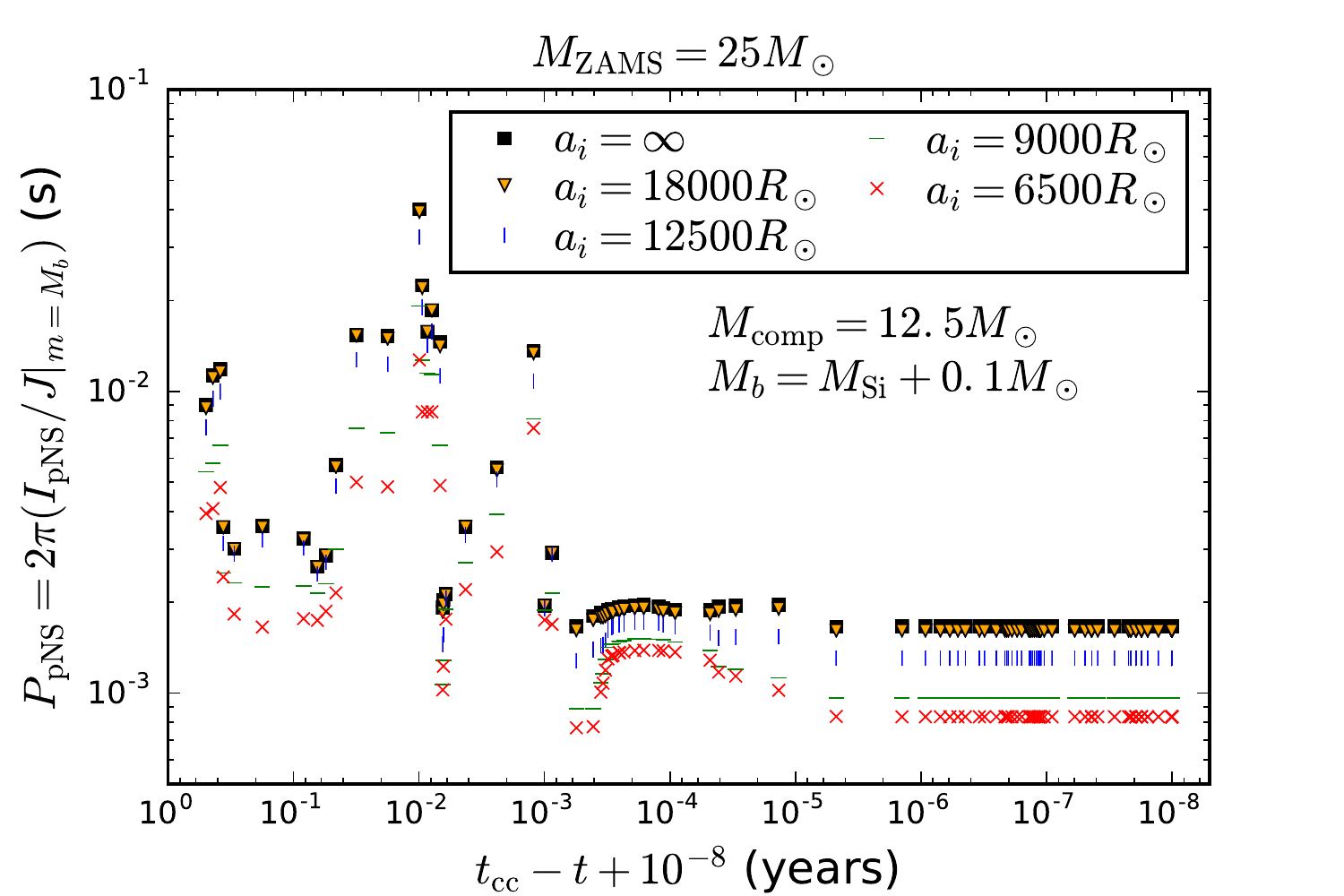}
\caption{{\it Top panel:}  Angular momentum stored in the evolving core of our 13$\,M_\odot$ model, now with an external
torque from a 6.5$\,M_\odot$ binary companion included.  As in Figures \ref{fig:P_pNS_13M_o}-\ref{fig:P_pNS_40M_o}, 
the core rotation is measured in terms of the rotation period of a collapsed neutron star. Here we focus on the results for the
baryonic mass $M_b = M_{\rm Si}+0.1\,M_\odot$.
Binary is initiated with a range of semi-major axes, with tidal evolution of the binary separation calculated 
self-consistently.  Closest binary separation corresponds to the onset of strong rotation in the convective envelope. 
{\it Bottom panel:}  Same, but for the 25$\,M_\odot$ with a binary companion of mass 12.5$\,M_\odot$.}
\vskip .2in
\label{fig:P_NS_post_SN_log_delta_t_end_diff_a_i_13M_o}
\end{figure}

\begin{figure}
\epsscale{1.2}
\plotone{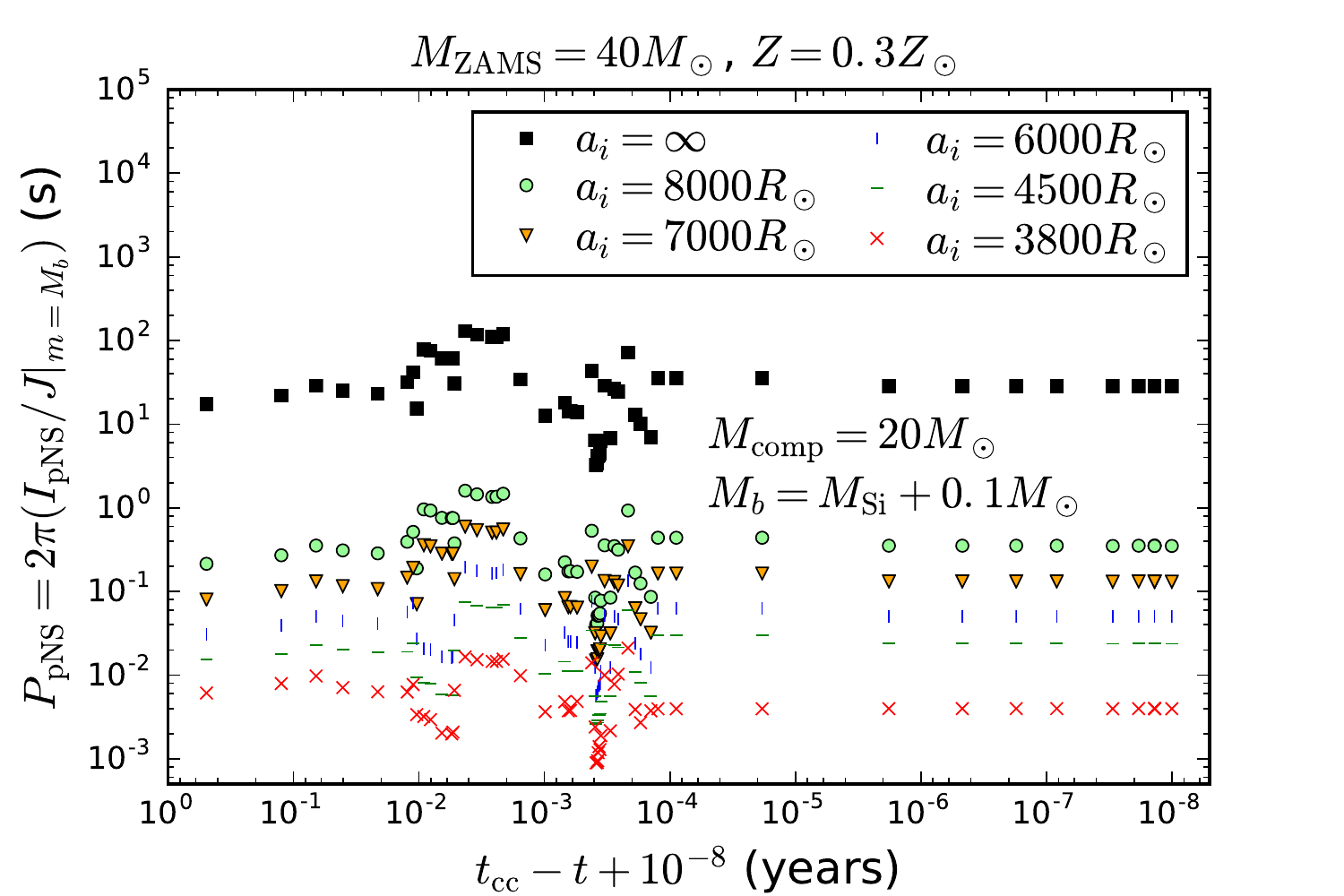}
\caption{Similar to figure \ref{fig:P_NS_post_SN_log_delta_t_end_diff_a_i_13M_o}, but for the 40$\,M_\odot$,
0.3$Z_\odot$ model and a companion mass 20$\,M_\odot$.}
\vskip .2in
\label{fig:P_NS_post_SN_log_delta_t_end_diff_a_i_40M_o_0.3Z_o}
\end{figure}

\begin{figure}
\epsscale{1.2}
\plotone{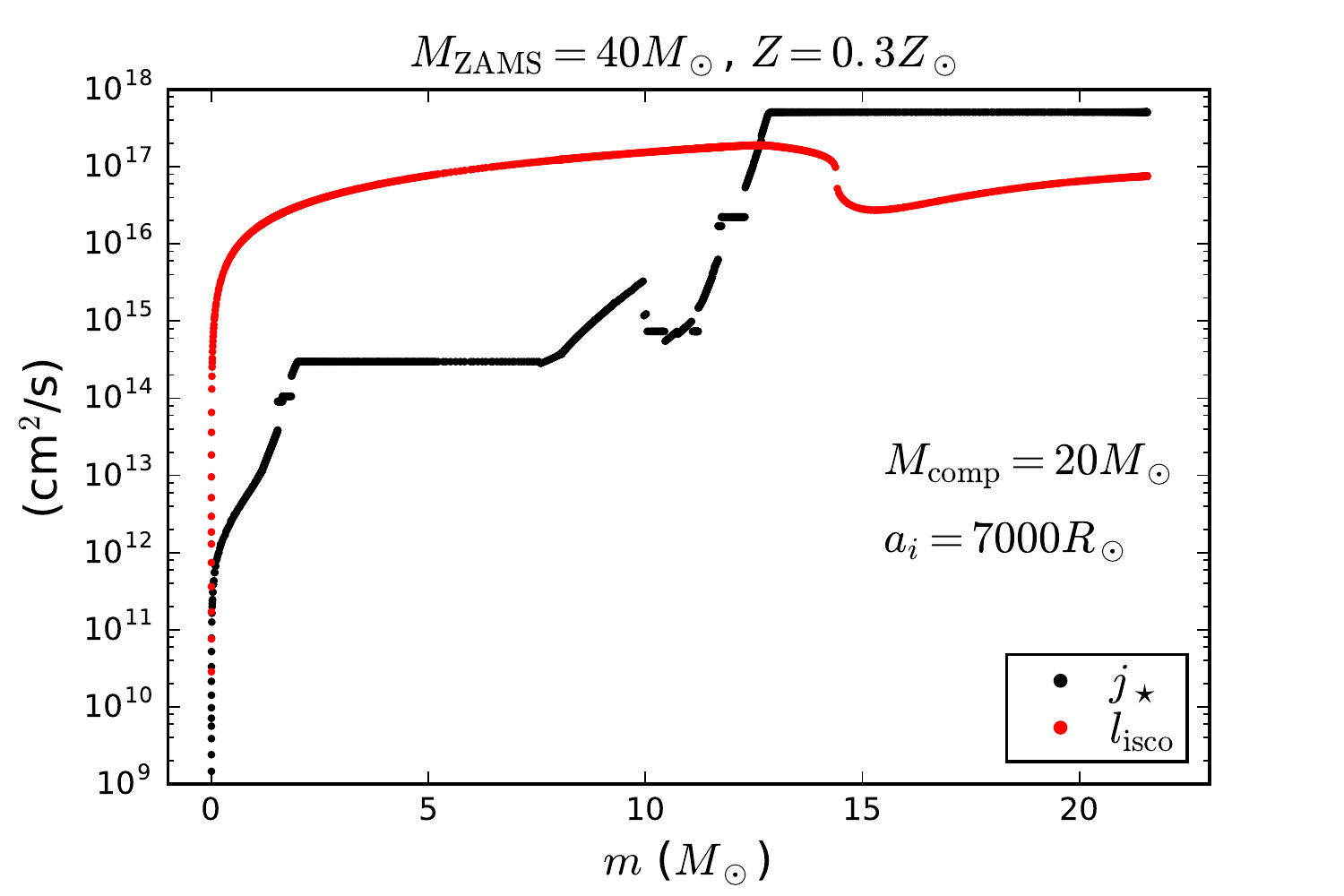}
\caption{Similar to figure \ref{fig:specific_j_comparison_40M_o}, but now including tidal angular momentum
deposition from a 20$\,M_\odot$ companion with initial separation $a_i = 7000R_\odot$.  The tidal interaction dominates when the
primary has nearly finished losing mass, allowing it to retain a significant fraction of the transferred angular momentum.
Now the core spins much more rapidly, and the collapsed mass outside 12.6$\,M_\odot$ is first accreted onto a disk.}
\vskip .2in
\label{fig:specific_j_comparison_40M_o_0.3Z_o_w_companion}
\end{figure}

\begin{figure} 
\epsscale{1.2}
\plotone{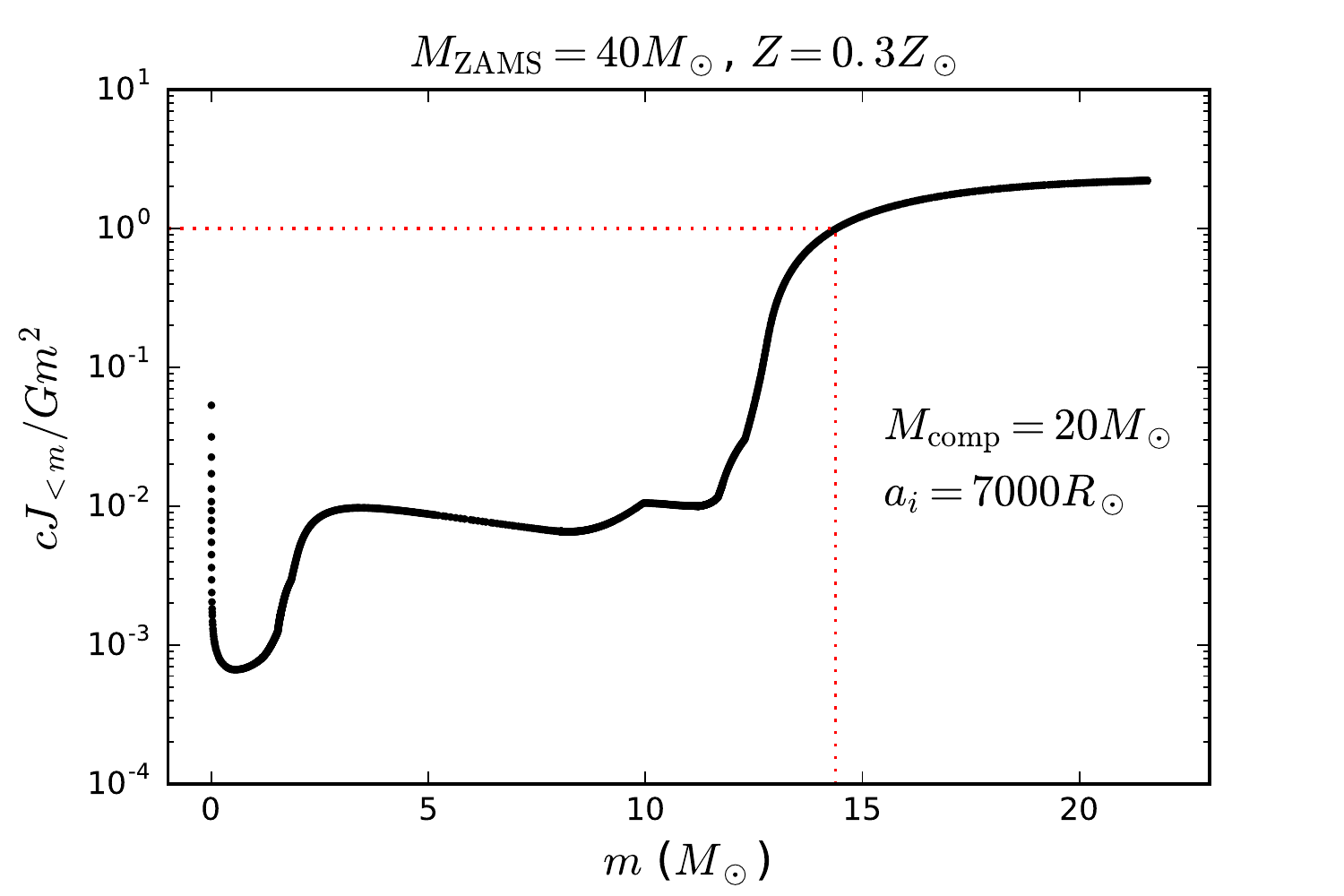}
\caption{Spin angular momentum of the collapsed BH in the 40$\,M_\odot$, $0.3\,Z_\odot$ model with binary
companion at initial separation $a_i = 7000\,R_\odot$ (Figure \ref{fig:specific_j_comparison_40M_o_0.3Z_o_w_companion}).
About 1.8$\,M_\odot$ of core material must accrete after the first centrifugally supported disk forms in order
for the BH to approach the extremal spin rate.}
\vskip .2in
\label{fig:BH_spin_parameter}
\end{figure}

\section{Angular Momentum Injection from a \\ Binary Companion} \label{sec:binary_interaction}

Most massive stars form in multiple systems:  for example, \cite{KobuF2007} find that the multiple 
fraction is greater than 80\%.  Interaction with a binary companion is therefore the rule rather than the exception.
Strong interactions involving conservative mass transfer, a merger, or a common envelope phase are beyond the 
scope of this paper.  

There is, however, a range of binary separations over which there is a gradual transfer of angular momentum by tides
to the spin of the more massive (primary) star during its supergiant phase.  This effect is implemented in our 
model stars by applying a positive torque to the hydrogen envelope.  
We assume a companion mass equal to ${1\over 2}$ the primary ZAMS mass, and consider a range of binary separations
with the orbital plane aligned with the initial rotation of the primary.

The orbital separation $a$ and angular momentum $L_{\rm orb}$ evolve according to \citep{Hut1981}
\be \label{eq:tidal_a_dot}
{da\over dt} = -6k_2 \frac{\tau_{\rm fric}}{\tau_{\rm dyn}^2} q(1+q) \left(\frac{a}{R_{\star,1}}\right)^{-8}a
\left(1 - \frac{\Omega_\star}{\Omega_{\rm orb}}\right)
\ee
and
\be \label{eq:tidal_L_dot}
{dL_{\rm orb}\over dt} = \frac{1}{2} \frac{M_{\star,1}M_{\star,2}}{M_{\star,1}+M_{\star,2}} 
\Omega_{\rm orb} a {da\over dt}.
\ee
Here the primary is labeled `1' and the secondary `2', the mass ratio is $q \equiv M_{\star,2}/M_{\star,1}$,
and $\Omega_{\rm orb}$ is the circular orbital frequency.  The tidal friction time is
\be
\tau_{\rm fric} = \left(\frac{M_{\rm env,1} R_{\star,1}^2}{L_{\star,1}}\right)^{1/3},
\ee
where $M_{\rm env,1}$ is the envelope mass of the primary, and we define
$\tau_{\rm dyn} = (R_{\star,1}^3/GM_{\star,1})^{1/2}$.
The Love number is obtained from an integral over the convective envelope,
\be
k_2 = 20.5 \alpha^{4/3} \frac{R_{\star,1}^{1/3} \tau_{\rm fric}}{M_{\rm env,1}}
      \int_{x_{\rm benv}}^1 x^{22/3} L_r^{1/3} \rho^{2/3} l_P dx.
\ee
Here $\alpha = 2$ is the convective mixing length parameter, $x = r/R_{\star,1}$, $x_{\rm benv}$ marks the
base of the convective envelope, and $L_r$ is the luminosity at a given radius.  As we integrate expressions 
(\ref{eq:tidal_a_dot}) and (\ref{eq:tidal_L_dot}) for $a$ and $L_{\rm orb}$, we linearly interpolate the various stellar 
parameters between MESA model snapshots.

The tidal interaction is concentrated near the maximum expansion of the primary, given the strong dependence of the 
torque on the aspect ratio $R_{\star,1}/a$.   This simplifies the calculation,
because the hydrogen envelope is mostly radiative during the early expansion, complicating the handling of the
tidal dissipation.  On the other hand, a deep outer convective envelope redistributes angular momentum rapidly,
allowing us to apply the rotation model described in Section \ref{sec:convective_transport}.

Here a significant difference emerges between the two 40$\,M_\odot$ models
(Figure \ref{fig:R_star_M_star_vs_t_two_models_RGB_end}).  The solar metallicity model reaches a maximum radius
and then contracts while mass loss continues.  This means that most of the angular momentum deposited by tides 
(or by a merger with a lower-mass companion) near the maximum expansion would be subsequently lost.   The situation
is different with the $0.3\,Z_\odot$ model, which maintains a nearly uniform radius as a supergiant, and then
experiences a late secondary expansion during which the tidal torque peaks but there is little time for additional
mass loss.

The binary separation is started with a range of values, but the integration is stopped (and the evolutionary track
discarded) if the internal rotation of the primary approaches breakup at any point within the convective envelope.
Applying this cut also allows us to discard binaries which merge.  

The resulting core rotation rate is shown in Figure \ref{fig:P_NS_post_SN_log_delta_t_end_diff_a_i_13M_o} 
for the 13$\,M_\odot$ and 25$\,M_\odot$ models.  In both cases, the maximum spin-up of the core, compared with
an isolated star, is about a factor $\sim 2$.

The lower metallicity 40$\,M_\odot$ model experiences a much greater relative spin-up (a factor $\sim 10^2$-$10^4$:
compare in Figure \ref{fig:P_NS_post_SN_log_delta_t_end_diff_a_i_40M_o_0.3Z_o} results for $a_i=\infty$ to the other results), a consequence of the strong angular momentum loss to a wind.  The consequences for the rotation and
accretion of a BH remnant are shown in Figure \ref{fig:specific_j_comparison_40M_o_0.3Z_o_w_companion}.  Now
the infalling material becomes centrifugally supported outside an enclosed mass $\sim 12.6\,M_\odot$.

An energetic outflow from the accretion torus that forms as a result could limit much further growth of 
$M_{\rm BH}$.  Here it is interesting to note that the outflow power is a strong increasing
function of the black hole spin \citep{tchek10}.   The accretion of an additional $\sim 1.8\,M_\odot$ of high
angular momentum core material (following the first appearance of the torus) is needed to bring $J_{\rm BH}$
close to the limiting value $GM_{\rm BH}^2/c$ (Figure \ref{fig:BH_spin_parameter}).  We also deduce 
that a successful magnetorotational explosion, leading to the formation of a NS, cannot be expected for these
binary parameters.

The observed properties of Galactic binary stellar-mass BHs are consistent with such a scenario.
Fitting of thermal X-ray spectra 
suggests in many cases a high spin rates for the hole, with $cJ_{\rm BH}/GM_{\rm BH}^2$ approaching 
unity (e.g. \citealt{McCletal2006,McCletal2011}).   Such a high spin could not have been gained by mass 
transfer from the companion star.  A similar effect should occur in closer binaries in which the primary
envelope is ejected during a common envelope phase.

\section{Conclusions}\label{s:conclusions}

We have investigated the evolving magnetism and rotation of massive stars, starting in the pre-MS accretion
phase and extending to the last stages of nuclear burning and the onset of core collapse.  Our focus is on 
two effects which have received little attention in models of stellar evolution:
i) inhomogeneous rotation in deep convective layers, especially slowly rotating
layers with a small Coriolis parameter $\Omega \tau_{\rm con}$;
and ii) the pumping of magnetic helicity into a growing radiative layer at a moving convective-radiative boundary.
The impact of these processes on the magnetism and rotation of lower-mass giant stars was considered previously
by \cite{KissT2015a,KissT2015b}.

The inward advection of even a small amount of angular momentum through a deep convective layer can have a 
major effect on the rotation rate of the stellar core and its collapsed remnant.
This effect is more important for red than blue supergiants, and for more massive stars with 
thicker convective burning shells.

We account for the limiting effect of a kink instability on the transport of 
angular momentum by the $r\phi$ Maxwell stress through radiative layers of a star. 
Our approach to magnetic field growth in radiative layers differs substantially from that of \cite{Spru2002}
in that it also incorporates a large-scale poloidal flux, which is stablized by buried magnetic twist.

We expand on previous efforts by tracking the growth of the magnetic field in
the stellar core during the pre-MS accretion phase, when the star acquires most of its mass.
This first phase of convective retreat turns out to be the dominant contributor to the magnetic
helicity stored in the core, yielding a poloidal flux density around $10^{13}$ G in the NS remnant.

A relatively simple rotation profile is found over much of the evolutionary history of our model stars,
corresponding to a significant redistribution of angular momentum between the inner and outer parts of the
star.  Radiative layers tend to rotate as solid bodies, and deep convective layers approach constant 
specific angular momentum, as motivated by anelastic calculations of deep and very slowly rotating
stellar envelopes \citep{BrunP2009}.  

To test these effects, we chose two MESA models which are likely to leave behind a NS remnant
(13$\,M_\odot$ and possibly the 25$\,M_\odot$, based on the explosion criterion of \citealt{ErtlJWSU2016} 
as implemented in the MESA output).  These were supplemented by the
two 40$\,M_\odot$ models, the more metal rich of which likely forms a BH.
The rotation of the collapsed remnant (either NS or BH) varies much more between these models than 
does the magnetic flux.  That is because the rotation is sensitive to the depth of the outer convective
envelope during later stages of mass loss, and because the most massive stars form thicker convective shells.
The predicted NS rotation period ranges from the millisecond range (in the 25$\,M_\odot$ model), up to $0.1-1$ s
(in the 13$\,M_\odot$ model).  
These cover the full range of the estimated initial rotation rates of pulsars (\citealt{PopoT2012}),
implying that radio pulsars can form from isolated massive stars.

Much longer spin periods are found in the highest mass models, which end up as blue supergiants and 
collapse to black holes.  This means that the entire star is likely to collapse through the event horizon 
of the BH, corresponding to a mass $M_{\rm BH} \sim 20\,M_\odot$ for our 40$\,M_\odot$ progenitors.

We find that tidal angular momentum exchange from a binary companion with semi-major axis 
3000-20000$\,R_\odot$ only spins up the core by a factor $\sim 2$ in the two 
lower-mass models.   On the other hand, enough angular momentum is deposited in the
40$\,M_\odot$, $0.3\,Z_\odot$ model star to produce a rapidly rotating BH:  the outer third of
the collapsing star forms a rotationally supported torus and may be expelled in a disk-driven outflow, 
leaving a remnant of mass $M_{\rm BH} \sim 13$-15$\,M_\odot$.

These calculations of the magnetism and rotation of the remnants of massive stars
bear a promising resemblance to observations.  It is possible that one or both of mechanisms 
investigated here is manifested more weakly in real stars.   Lower magnetization of the
radiative layers of a star could partly compensate weaker inward pumping of angular momentum through its 
convective layers.  Nonetheless, an intermediate level of differential rotation in the convective envelope 
would still have a profound effect on the core rotation rate during the supergiant phase.    And even 
a factor $\sim 10^{-2}$ reduction in the amplitude of the helicity flux (compared with our approach)
would leave behind a poloidal magnetic flux well within the range observed in radio pulsars.

\subsection{Comparison with Previous Models}

The most direct comparison is with the models of \cite{HegeWS2005} and \cite{wheeler15}.  Both models
include the effects of magnetic stresses as evolved according to the \cite{Spru2002} model, to which
the second adds mixing by the MRI excited near compositional boundaries.  
As was explained in Section \ref{s:intro}, the core magnetic flux preceding collapse cannot be usefully
extracted from these earlier calculations, because the  magnetic helicity is not computed, and the magnetic
instabilities considered have a high spatial wave number.  However, one does deduce that the poloidal
magnetic field is systematically weaker, by at least a few orders of magnitude, than is suggested by
pulsar magnetic fields, and is found in our calculations of helicity growth during pre-MS accretion.

We therefore focus on comparing core rotation rates.   Both the preceding calculations combine
an {\it assumption} of solid rotation in the deep convective envelope and convective hydrogen-depleted layers
of an evolved star, with an incomplete rotational coupling across radiative-convective boundaries.  Our
model has opposing properties, meaning that any near agreement between the results for central core rotation
should be viewed as fortuitous.   One can certainly imagine a sequence of rotational models 
with decreasing angular velocity gradient in the convective envelope, which is compensated by 
a growing (negative) angular velocity jump(s) across internal compositional boundaries.  The difficulty in constructing
such intermediate models lies in developing a deterministic approach to reconnecting the large-scale poloidal
magnetic field at compositional boundaries, and in prescribing the envelope angular velocity profile over
a wide range of Coriolis parameter.  A purely hydrodynamic approach to the latter problem was developed by
\cite{KissT2015a}, but the result probably also depends on magnetic feedback, especially from the MRI
where ${\rm Co} \gtrsim 1$.  Nonetheless, it should be kept in mind that solid rotation in the
envelope of a red supergiant would imply extremely low Co throughout the envelope, pushing conditions toward
the regime where hydrodynamic stresses may dominate.  

That being said, \cite{wheeler15} find a specific equatorial angular momentum $j \sim 3\times 10^{14}$
cm$^2$ s$^{-1}$ at an enclosed mass $\sim 1.8\,M_\odot$ in a 15 $M_\odot$ model, and $j \sim 3\times 10^{13}$ 
cm$^2$ s$^{-1}$ in a 20 $M_\odot$ model.  By comparison, Figure \ref{fig:j_profile} shows $j \sim 10^{14}$ cm$^2$ s$^{-1}$
at the same enclosed mass in our 13 $M_\odot$ model, increasing to $3\times 10^{15}$ cm$^2$ s$^{-1}$ in the 25 $M_\odot$
model.  Taking into account the signficant scatter in core compactness and mass that is expected with a changing
ZAMS mass, these results can at best be viewed as overlapping; and both show some promise in comparison with pulsar
rotation rates.   The calculations of \cite{HegeWS2005} do not include magnetic torques of any type at 
radiative-convective boundaries, and therefore typically yield stronger angular velocity jumps, and 
systematically faster core rotation.

\subsection{Systematic Uncertainties in the Model}\label{s:syst}

Much of
the systematic uncertaintly in our approach is related to the still developing understanding of 
differential rotation in deeply convective layers.   Here we explore how our predictions would be
altered in the presence of a slightly different pattern of differential rotation.

First, the inward pumping of angular momentum by deep convective plumes could result in a shallower
angular velocity profile than the one ($\Omega(r) \propto r^{-2}$) suggested by anelastic calculations
of slowing rotating envelopes \citep{BrunP2009}.  Supposing that the index softens to $-2 + \alpha$,
and taking into account that a large fraction of the stellar angular momentum is contained in the 
envelope, one finds that the core rotation period is increased by a factor $\simeq (1 + 4\alpha/3)^{-1/2} 
(R_\star / R_{\rm base})^\alpha \sim 8$ for $\Omega(r) \propto r^{-3/2}$ and 
$R_\star \sim 10^2 R_{\rm base}$.  (The coefficient here corresponds specifically to a density
profile $\rho(r) \propto r^{-3/2}$ in the envelope.)  Even with such a revision, the angular velocity
at the base of the convective envelope remains dramatically faster relative to the surface than in
existing evolution codes such as MESA.   

Second, we chose a strong latitudinal angular velocity gradient at the base of the convective layer,
again as observed in anelastic calculations of deeply convective layers.  A weaker level of differential
rotation would reduce the magnetic helicity flux in proportion to ${\cal H} \propto (\partial\Omega/\partial\theta)^2$
(following Equations (\ref{eq:Max}) and (\ref{eq:netH})), and the remnant hemispheric magnetic flux as
${\cal H}^{1/2} \propto \partial\Omega/\partial\theta$.  A reduction by an order of magnitude of the 
strength of differential rotation corresponds to a polar magnetic field $B_p \sim {\cal H}^{1/2}/\pi R_{\rm NS}^2
= 2\times 10^{12}$ G, which is still well within the range observed in young radio pulsars.   

Third, we explored in Section \ref{s:rim} how the introduction of RIM into one-dimensional MESA calculations
would modify the compactness and mass of the collapsing stellar core.   We estimated a $\sim 0.1\,M_\odot$
error in mass enclosed by the silicon layer from the dependence of $M_{\rm Si}$ on peak rotation rate (relative
to Keplerian) at the formation of the helium core (Figure \ref{fig:hecorespin}).   At a fixed progenitor
mass, the simplest measure $\xi_{2.5}$ of the core compactness is estimated to rise from $\sim 0.2$ to 
$\sim 0.35$.  However, much of these changes can be compensated by a modest 1-$2\,M_\odot$ adjustment of
the progenitor mass, as shown in Figure \ref{fig:xi25}. 

\subsection{Implications for Growth or Decay of the Magnetic Field Post-Collapse}\label{s:dipole}

The dipole magnetic fields of radio pulsars carry a minuscule fraction of the neutron star binding energy
(about $10^{-12} (B_p/10^{12}~{\rm G})^2$, where $B_p$ is the polar flux density).  This complicates our
understanding of their origin:  a variety of processes might contribute to such a relatively weak field.
The $\sim 10^{13}$ G NS magnetic field that our models produce (over a wide range of progenitor masses) is 
moderately stronger than the spindown field of most radio pulsars, but weaker than the magnetic fields 
of active magnetars.  Our next task is therefore to consider how the field may be modified post collapse.

Rapid neutrino-driven convection in the proto-NS, with an overturn time $\tau_{\rm con} \sim 3$ ms,
can have two competing effects on the magnetic field \citep{ThomD1993}.
First, the entrained field lines will diffuse across the surface of the star over a reasonably
short timescale, which depends on the initial rotation period.  A mixing together of radial fluxes
of opposing signs from the two magnetic hemispheres would reduce the external magnetic moment. 
We show that such a reduction is most feasible in slow rotators.

The convective motions do not extend fully to the surface of the proto-NS \citep{lattimer81}.  A stably
stratified layer of mass $M_{\rm rad} \sim 0.1\, M_\odot$ maintains significant inertia.  The
magnetic field lines which thread both the convective material and this more inert shell are
stretched in the non-radial direction near the interface between the two,
producing a strong horizontal field $B_h \gg B_r$.  

The shuffling motion is slow enough to be further impeded by the Coriolis force.  Integrating this
through the outer shell (column $\Sigma_{\rm rad} = M_{\rm rad}/4\pi R_{\rm NS}^2$)
gives an estimate of the speed $v_h$ of the shuffling motions, 
\be
{B_r B_h\over 4\pi} \sim \Sigma_{\rm rad} v_h \Omega_{\rm NS}.
\ee
The shuffling timescale is then
\ba\label{eq:tshuff}
{R_{\rm NS}\over v} &\sim&  {2\pi M_{\rm rad} \over P_{\rm NS} B_r B_h R_{\rm NS}}\nn
 &=& {10~{\rm s}\over B_{r,13} B_{h,14} R_{\rm NS,6}}\left({P_{\rm NS}\over 0.1~{\rm s}}\right)^{-1}
                     \left({M_{\rm rad}\over 0.1~M_\odot}\right).\nn
\ea
For the estimated post-collapse field $B_r \sim 2\times 10^{13}$ G, and for a spin period longer
than $\sim 0.1$ s, the timescale (\ref{eq:tshuff}) is comparable to the Kelvin-Helmholtz timescale of a few seconds.
Slower rotators can therefore experience greater dipole cancellation.

The second effect is encountered in the regime of faster rotation.  A proto-NS spinning with 
$P_{\rm NS} \lesssim 3$-10 ms would generate strong toroidal magnetic fields from the mean
poloidal field.  This happens preferentially after the bounce shock has expanded away and
the high-entropy mantle surrounding the star has collapsed (over perhaps the first $\sim 0.3$ s); 
but before the bulk of the interior has cooled.  During this phase, a positive radial angular 
velocity gradient develops in the outer parts of the star, driven by the collapse of the mantle.  
This shear is strong enough for the radial field to experience linear winding and the toroidal 
magnetic field to approach the dynamical limit, $B_\phi \sim (4\pi \rho)^{1/2} r\Delta\Omega = 
2\times 10^{16}\,\rho_{14}^{1/2} (\Delta\Omega/\Omega) (P_{\rm NS}/10~{\rm ms})^{-1}$ G, 
all before the NS cools.  A field this strong induces a strong temperature perturbation, 
which is rapidly erased by charged-current neutrino reactions, thereby generating buoyant motions 
and feedback on the seed poloidal field \citep{tm01}.   In this way a dynamo feedback loop becomes
possible.

\begin{appendix}

\section{MESA inlist} \label{app:MESA_inlist}

Below we include the inlist parameters we changed from their default values in our MESA runs:
\\

\&star\_job

      create\_pre\_main\_sequence\_model = .true.

      kappa\_file\_prefix = 'gs98'

      change\_v\_flag = .true.

      new\_v\_flag = .true.
      
      warn\_run\_star\_extras = .false.

/ ! end of star\_job namelist
\\

\&controls

         initial\_mass = 3

				 initial\_Y = 0.25

         initial\_Z = 2d-2 (6d-3 in the 40$\,M_\odot$, $0.3Z_\odot$ model)

         velocity\_logT\_lower\_bound = 7

         max\_dt\_yrs\_for\_velocity\_logT\_lower\_bound = 1

         mesh\_delta\_coeff\_for\_highT = 2

         okay\_to\_reduce\_gradT\_excess = .true. 

         cool\_wind\_RGB\_scheme = 'Dutch'

         cool\_wind\_AGB\_scheme = 'Dutch'

         Dutch\_scaling\_factor = 1

         cool\_wind\_full\_on\_T = 1d8

         cool\_wind\_full\_off\_T = 1.1d8

         include\_dmu\_dt\_in\_eps\_grav = .true.

         use\_Type2\_opacities = .true.

         Zbase = 2d-2 ! must set this in the main inlist

         mixing\_length\_alpha = 2

         MLT\_option = 'Henyey'

         use\_Ledoux\_criterion = .true.

         alpha\_semiconvection = 0.1

         thermohaline\_coeff = 2
   
         overshoot\_f\_*** = 1d-2      \ \ \  ! This applies to all variations of 'overshoot\_f' 

         overshoot\_f0\_*** = 5d-4    \ \ \ ! This applies to all variations of 'overshoot\_f0'

         min\_timestep\_limit = 1d-12

/ ! end of controls namelist

\section{Angular Momentum Transport by Winding a Poloidal Magnetic Field} \label{app:kink_instability}

Here we consider how the kinking of a wound-up magnetic field may limit the transport
of angular momentum through the radiative layers of a star.   The background state has finite magnetic helicity ${\cal H}$
and is threaded by large-scale toroidal and poloidal fields.  Each field component carries a
finite flux, which was deposited during a transition from a convective to the present radiative state (Section
\ref{sec:helicity}).  Neither the hemispheric poloidal flux $\Phi_r$ nor the toroidal flux $\Phi_\phi$ is directly
modified by differential rotation, even while the toroidal field energy may increase substantially.   

Magnetic twist is stored on loops of poloidal field that close within the radiative layer \citep{BraiS2004}.  
These structures act as barriers to the mixing and reconnection of open poloidal fluxes from the opposing hemispheres.
In the application to evolving stellar interiors, we posit an initial relaxation to
a roughly isotropic state with $\Phi_r \sim \Phi_\phi \sim {\cal H}^{1/2}$.  This maximizes the poloidal
flux for a given ${\cal H}$, as well as the rate of redistribution of angular momentum by poloidal torsional magnetic
waves.  

Winding by differential rotation leaves the magnetic field susceptible to an ideal hydromagnetic `kink' instability \citep{Tayler73}.
In contrast with the model analysed by \cite{Spru2002}, in which the mean poloidal field is absent, the kink is not the primary source of 
poloidal flux:  it only induces high-wavenumber distortions of the poloidal field.  
We therefore simplify the problem by assuming a fixed poloidal field, and consider the action of
the kink instability only on the wound-up component of the toroidal field.

Three characteristic timescales can be distinguished here:  i) the evolution time of the stellar mass profile,
$\tau_{\rm ev} = {\rm min}[l_P / |v_r|, (t_{\rm cc}-t)/3]$, which varies significantly between the inner and outer
parts of the star during post-MS evolution; ii) the growth time $\tau_{\rm kink}$ of the kink instability; and iii) the timescale $\tau_J$
for the redistribution of angular momentum between different shells within a slowly rotating radiative layer.
(Convective angular momentum transport is generally much faster, taking a few eddy overturns.)
We are interested in deducing the minimum $\Phi_r$ that will allow nearly solid rotation to be established,
corresponding to $\tau_J \sim \tau_{\rm ev}$.  

To do this, we first must address the growth of the kink.
Two inequalities which can be justified {\it ex post facto} are i) $\tau_{\rm kink} < \tau_{\rm ev}$ unless the rotation 
is extremely slow ($\Omega \lesssim \tau_{\rm ev}^{-1}$), meaning that the growth of the toroidal field will be
limited by kinking; and ii) $\tau_{\rm kink} > \Omega^{-1}$, so that the Coriolis force must be taken into account in
evaluating $\tau_{\rm kink}$.  Then the kink is associated with a hydromagnetic displacement of speed \citep{PittT1985}
\be\label{eq:vk}
v_{\rm kink} \sim {r\over \tau_{\rm kink}} \sim {B_\phi^2\over 8\pi \rho \Omega r}.
\ee
To obtain a relationship between $B_\phi$ and $B_r$ we use the induction equation,
\be
\frac{\partial B_\phi}{\partial t} \sim -r B_r \frac{\partial \Omega}{\partial r} - \frac{B_\phi}{\tau_{\rm kink}}.
\ee
Here we adopt a simplified analysis which focuses on the zone near the rotational equator.  Then
when $\tau_{\rm kink} < \tau_{\rm ev}$, one has
\be\label{eq:vaphi}
v_{A,\phi}^3 \sim 2\left|{d\ln\Omega\over d\ln r}\right| (\Omega r)^2 v_{A,r}.
\ee
Here $v_{A,\phi (r)} = B_{\phi (r)}/(4\pi\rho)^{1/2}$ are the toroidal (poloidal) Alfv\'en speeds.

The timescale for angular momentum transport is obtained by balancing the Maxwell torque exerted through
a lever arm $\sim r$ against the change in angular momentum,
\be\label{eq:torque}
{B_\phi B_r\over 4\pi} r  \sim  \rho l_P\,{r^2\Omega\over \tau_J}.
\ee
Rotational equilibrium corresponds to $\tau_J \sim \tau_{\rm ev}$, which we substitute into Equation
(\ref{eq:torque}) along with Equation (\ref{eq:vaphi}) to get an expression for $|d\ln\Omega/d\ln r|$.   Requiring
that the angular velocity gradient be weak, $|d\ln\Omega/d\ln r| \lesssim 1$, we obtain
\be\label{eq:varmin}
v_{A,r} > \left({rl_P^3\Omega\over 2\tau_{\rm ev}^3}\right)^{1/4}.
\ee
One can also substitute Equation (\ref{eq:vaphi}) into (\ref{eq:vk}) and show that
\be
{\tau_{\rm kink}\over \tau_{\rm ev}} \sim 2{v_{A,r}^2 \tau_{\rm ev}\over l_P^2\Omega} > 
                                    \left({2r\over \tau_{\rm ev}\Omega l_P}\right)^{1/2}.
\ee
The inequality is obtained from the threshold (\ref{eq:varmin}) for nearly solid rotation.   

One sees that the kink is excited self-consistently when this threshold is reached, unless the
rotation of the star is extremely slow, or the evolution time is very short, 
corresponding to $\Omega \lesssim \tau_{\rm ev}^{-1}$.  In this second regime, one can alternatively
substitute $B_\phi = -B_r |d\ln\Omega/d\ln r| \Omega \tau_{\rm ev}$ into the torque formula, and
obtain the threshold condition for angular momentum transport,
\be\label{eq:varminb}
v_{A,r} > {(rl_P)^{1/2}\over \tau_{\rm ev}}.
\ee
A more general condition is obtained by taking the minimum of the right-hand sides of Equations
(\ref{eq:varmin}) and (\ref{eq:varminb}).
The main point to take away from this analysis is that the kink only causes a slight reduction
in the Maxwell stress:  the threshold value of $v_{A,r}$ is lengthened compared with
$\sim l_P/\tau_{\rm ev}$ only by a factor $\sim (r\Omega \tau_{\rm ev}/2l_P)^{1/4}$.

\end{appendix}


\begin{thebibliography}{}

\bibitem[Ando (1983)]{Ando1983} {Ando}, H. \ 1983, \pasj, 35, 343

\bibitem[Augustson et al.(2016)]{auguston16} Augustson, K.~C., Brun, A.~S., \& Toomre, J.\ 2016, \apj, 829, 92 

\bibitem[Balbus \& Hawley(1994)]{BalbH1994} Balbus, S.~A., \& Hawley, J.~F.\ 1994, \mnras, 266, 769 

\bibitem[Balbus et al.(2009)]{BalbBLW2009} Balbus, S.~A., Bonart, J., Latter, H.~N., \& Weiss, N.~O.\ 2009, \mnras, 400, 176 

\bibitem[Bhattacharya \& van den Heuvel (1991)]{Bhatv1991}
{Bhattacharya}, D., \& {van den Heuvel}, E.~P.~J.\ 1991, \physrep, 203, 1 

\bibitem[Braithwaite \& Spruit(2004)]{BraiS2004} Braithwaite, J., \& Spruit, H.~C.\ 2004, \nat, 431, 819 


\bibitem[{{Brun} \& {Palacios}(2009)}]{BrunP2009}
{Brun}, A.~S. \& {Palacios}, A. 2009, \apj, 702, 1078

\bibitem[Couch \& Ott(2013)]{co13} Couch, S.~M., \& Ott, C.~D.\ 2013, \apjl, 778, L7 

\bibitem[Ekstr{\"o}m et al. (2012)]{Ekstetal2012}
{Ekstr{\"o}m}, S., {Georgy}, C., {Eggenberger}, P., 
	{Meynet}, G., {Mowlavi}, N., {Wyttenbach}, A., {Granada}, A., 
	{Decressin}, T., {Hirschi}, R., {Frischknecht}, U., 
	{Charbonnel}, C., \& {Maeder}, A. 2012, \aap, 537, 146

\bibitem[Endal \& Sofia (1976)]{EndlS1976}
{Endal}, A.~S., \& {Sofia}, S. \ 1976, \apj, 210, 184

\bibitem[Endal \& Sofia (1978)]{EndlS1978}
{Endal}, A.~S., \& {Sofia}, S. \ 1978, \apj, 220, 279

\bibitem[Ertl {et~al.} (2016)]{ErtlJWSU2016}
{Ertl}, T., Janka, H.~T., Woosley, S.~E., Sukhbold, T., \& Ugliano, M. 2016, \apj, 818, 124

\bibitem[Fricke (1968)]{Fric1968} {Fricke}, K., \ 1968, \zap, 68, 317

\bibitem[Frolov \& Novikov(1998)]{fn98} Frolov, V.~P., \& Novikov, I.~D.\ 1998, Black Hole Physics: Basic Concepts
and New Developments (Dordrecht : Kluwer Academic)


\bibitem[Fuller et al.(2015)]{FullCLQ2015} 
{Fuller}, J., {Cantiello}, M., {Lecoanet}, D., \& {Quataert}, E. \ 2015, \apj, 810, 101

\bibitem[Glebbeek et al.(2009)]{Glebetal2009} Glebbeek, E., Gaburov, E., de Mink, S.~E., 
Pols, O.~R., \& Portegies Zwart, S.~F. 2009, \aap, 397, 255

\bibitem[Goldreich \& Schubert (1967)]{GoldS1967} 
{Goldreich}, P., \& {Schubert}, G. \ 1967, \apj, 150, 571

\bibitem[Goldreich \& Kumar(1990)]{gk90} Goldreich, P., \& Kumar, P.\ 1990, \apj, 363, 694 

\bibitem[Heger et al. (2000)]{HegeLW2000}
{Heger}, A., {Langer}, N., \& {Woosley}, S.~E. \ 2000, \apj, 528, 368

\bibitem[Heger et al. (2005)]{HegeWS2005}
{Heger}, A., {Woosley}, S.~E., \& {Spruit}, H.~C. \ 2005, \apj, 626, 350

\bibitem[Herwig (2000)]{Herw2000}
{Herwig}, F. 2000, \aap, 360, 952H

\bibitem[Hillebrandt et al. (1987)]{HillHWT1987}
{Hillebrandt}, W., {Hoeflich}, P., {Weiss}, A., \& {Truran}, J.~W. 1987, \nat, 327, 597H

\bibitem[Hirata (2012)]{Hira2012}
{Hirata}, C. M. 2012, Lecture notes 

\bibitem[Huang et al. (2010)]{HuanGM2010}
{Huang}, W., {Gies}, D.~R., \& {McSwain}, M.~V. 2010, \apj, 722, 605H

\bibitem[Hut (1981)]{Hut1981}
{Hut}, P. 1981, \aap, 99, 126

\bibitem[de Jager et al. (1988)]{deJaNv1988}
{de Jager}, C., {Nieuwenhuijzen}, H., \& {van der Hucht}, K. A. 1988, \aaps, 72, 259

\bibitem[Kippenhahn et al. (1970)]{KippMT1970}
{Kippenhahn}, R., {Meyer-Hofmeister}, E., \& {Thomas}, H.~C. \ 1970, \aap, 5, 155

\bibitem[Kissin \& Thompson (2015a)]{KissT2015a}
{Kissin}, Y. \& {Thompson}, C. 2015, \apj, 808, 35

\bibitem[Kissin \& Thompson (2015b)]{KissT2015b}
{Kissin}, Y. \& {Thompson}, C. 2015, \apj, 809, 108

\bibitem[Klion \& Quataert(2017)]{kq17} Klion, H., \& Quataert, E.\ 2017, \mnras, 464, L16 

\bibitem[Kobulnicky \& Fryer (2007)]{KobuF2007}
{Kobulnicky}, H.~A., \& {Fryer} \ 2007, \apj, 670, 747

\bibitem[Kumar \& Quataert(1997)]{kq97} Kumar, P., \& Quataert, E.~J.\ 1997, \apjl, 475, L143 

\bibitem[Lai \& Goldreich(2000)]{lg00} Lai, D., \& Goldreich, P.\ 2000, \apj, 535, 402 

\bibitem[Lattimer \& Mazurek(1981)]{lattimer81}
Lattimer, J.~M., \& Mazurek, T.~J.\ 1981, \apj, 246, 955 

\bibitem[Lee et al. (2016)]{LeeMN2016}
{Lee}, U., {Mathis}, S., \& {Neiner}, C. \ 2016, \mnras, 457, 2445

\bibitem[Maeder \& Zahn (1998)]{MaedZ1998}
{Maeder}, A., \& {Zahn}, J.-P. \ 1998, \aap, 334, 1000

\bibitem[Maeder \& Meynet(2014)]{MM14} Maeder, A., \& Meynet, G.\ 2014, \apj, 793, 123 

\bibitem[McClintock et al. (2006)]{McCletal2006}
{McClintock}, J.~E., {Shafee}, R., {Narayan}, R., {Remillard}, R.~A., 
	{Davis}, S.~W., \& {Li}, L.-X. \ 2006, \apj, 652, 518 

\bibitem[McClintock et al. (2011)]{McCletal2011}
{McClintock}, J.~E., {Narayan}, R., {Davis}, S.~W., 
	{Gou}, L., {Kulkarni}, A., {Orosz}, J.~A., {Penna}, R.~F., 
	{Remillard}, R.~A., \& {Steiner}, J.~F. \ 2011, Classical and Quantum Gravity, 28, 114009

\bibitem[Menou et al.(2004)]{menou04} Menou, K., Balbus, S.~A., \& Spruit, H.~C.\ 2004, \apj, 607, 564 

\bibitem[Meynet \& Maeder (2000)]{MeynM2000}
{Meynet}, G., \& {Maeder}, A. \ 2000, \aap, 361, 101

\bibitem[Misner et al. (1973)]{MisnTW1973}
{Misner}, C.~W., {Thorne}, K.~S., \& {Wheeler}, J.~A. \ 1973, San Francisco: W.H.~Freeman and Co.

\bibitem[M{\"u}ller \& Janka(2015)]{muller15} M{\"u}ller, B., \& Janka, H.-T.\ 2015, \mnras, 448, 2141 

\bibitem[M{\"u}ller et al.(2016a)]{muller16a} M{\"u}ller, B., Heger, A., Liptai, D., \& Cameron, J.~B.\ 2016, \mnras, 460, 742 

\bibitem[M{\"u}ller et al.(2016b)]{muller16b} M{\"u}ller, B., Viallet, M., Heger, A., \& Janka, H.-T.\ 2016, \apj, 833, 124 

\bibitem[M{\"u}ller et al.(2017)]{muller17} M{\"u}ller, B., Melson, T., Heger, A., \& Janka, H.-T.\ 2017, arXiv:1705.00620

\bibitem[Murray \& Chang (2015)]{MurrC2015} 
Murray, N., \& {Chang}, P. 2015, \apj, 804, 44

\bibitem[Nordhaus et al.(2008)]{nordhaus08}
Nordhaus, J., Busso, M., Wasserburg, G.~J., Blackman, E.~G., \& Palmerini, S.\ 2008, \apjl, 684, L29 

\bibitem[Nugis \& Lamers (2000)]{NugiL2000}
{Nugis}, T. \& {Lamers}, H.~J.~G.~L.~M. 2000, \aap, 360, 227

\bibitem[O'Connor \& Ott (2011)]{OConO2011}
{O'Connor}, E., \& {Ott}, C.~D. \ 2011, \apj, 730, 700

\bibitem[Paxton et al.(2011)]{Paxtetal2011}
{Paxton}, B., {Bildsten}, L., {Dotter}, A., {Herwig}, F., {Lesaffre}, P., \&
  {Timmes}, F. 2011, \apjs, 192, 3

\bibitem[Paxton et al. (2013)]{Paxtetal2013}
{Paxton}, B., {Cantiello}, M., {Arras}, P., {Bildsten}, L., 
	{Brown}, E.~F., {Dotter}, A., {Mankovich}, C., {Montgomery}, M.~H., 
	{Stello}, D., {Timmes}, F.~X., \& {Townsend}, R. 2013, \apjs, 208, 4P

\bibitem[Peters {et~al.} (2011)]{PeteBKM2011}
Peters, T., Banerjee, R., Klessen, R.~S., \& Mac Low, M.~M. 2011, \apj, 729, 72

\bibitem[Pitts \& Tayler(1985)]{PittT1985}
Pitts, E., \& Tayler, R.~J.\ 1985, \mnras, 216, 139 

\bibitem[Popov \& Turolla (2012)]{PopoT2012} 
{Popov}, S.~B., \& {Turolla}, R. \ 2012, \apss, 341, 457

\bibitem[Ram{\'{\i}}rez-Agudelo et al. (2015)]{Ramietal2013} Ram{\'{\i}}rez-Agudelo, O.~H., 
Sim{\'o}n-D{\'{\i}}az, S., Sana, H., et al.\ 2013, \aap, 560, A29 

\bibitem[Sackmann \& Boothroyd(1991)]{SackB1991} 
Sackmann, I.-J., \& Boothroyd, A.~I.\ 1991, \apj, 366, 529

\bibitem[Spruit (2002)]{Spru2002}
Spruit, H.C. 2002, \aap, 381, 923

\bibitem[Spruit (2008)]{Spru2008}
Spruit, H.C. 2008, American Institute of Physics Conference Series, 983, 391

\bibitem[Spruit \& Phinney (1998)]{SpruP1998}
Spruit, H.C., \& {Phinney}, E.~S. \ 1998, \nat, 393, 139

\bibitem[Talon et al.(2002)]{tkz02}
Talon, S., Kumar, P., \& Zahn, J.-P.\ 2002, \apjl, 574, L175 

\bibitem[Tayler(1973)]{Tayler73}
Tayler, R.~J.\ 1973, \mnras, 161, 365 

\bibitem[Tchekhovskoy et al.(2010)]{tchek10}
Tchekhovskoy, A., Narayan, R., \& McKinney, J.~C.\ 2010, \apj, 711, 50 

\bibitem[Townsend(1958)]{Townsend58} Townsend, A.~A.\ 1958, Journal of Fluid Mechanics, 4, 361 

\bibitem[Thompson(2000)]{thompson00} Thompson, C.\ 2000, \apj, 534, 915 

\bibitem[Thompson \& Duncan(1993)]{ThomD1993}
Thompson, C., \& Duncan, R.~C.\ 1993, \apj, 408, 194 

\bibitem[Thompson \& Murray(2001)]{tm01} Thompson, C., \& Murray, N.\ 2001, \apj, 560, 339 

\bibitem[Triana et al.(2017)]{triana17} Triana, S.~A., Corsaro, E., De Ridder, J., et al.\ 2017, \aap, 602, A62 

\bibitem[Ugliano et al.(2012)]{UgliJMA2012}
Ugliano, M., Janka, H.-T., Marek, A., \& Arcones, A.\ 2012, \apj, 757, 69 

\bibitem[Vink et~al.(2001){Vink}, {de Koter}, \& {Lamers}]{VinkKL2001}
{Vink}, J.~S., {de Koter}, A., \& {Lamers}, H.~J.~G.~L.~M. 2001, \aap, 369, 574

\bibitem[White \& Malin (1987)]{WhitM1987}
{White}, G.~L., \& {Malin}, D.~F. 1987, \nat, 327, 36

\bibitem[Wheeler et al.(2015)]{wheeler15} Wheeler, J.~C., Kagan, D., \& Chatzopoulos, E.\ 2015, \apj, 799, 85 

\bibitem[Wickramasinghe et al.(2014)]{WickTF2014}
Wickramasinghe, D.~T., Tout, C.~A., \& Ferrario, L.\ 2014, \mnras, 437, 675 

\bibitem[Woosley(1993)]{woosley93}
Woosley, S.~E.\ 1993, \apj, 405, 273 

\bibitem[Woosley \& Weaver (1995)]{WoosW1995}
{Woosley}, S.~E., \& {Weaver}, T.~A. 1995, \apjs, 101, 181

\bibitem[Zahn (1992)]{Zahn1992}
{Zahn}, J.-P. 1992, \aap, 265, 115

\bibitem[Zahn et al.(1997)]{ztm97} Zahn, J.-P., Talon, S., \& Matias, J.\ 1997, \aap, 322, 320 


\end{thebibliography}
\end{document}